\documentclass{aa}
\usepackage{graphicx}
\usepackage[varg]{txfonts}
\usepackage{longtable}
\input{epsf}

\def\sun{\hbox{$\odot$}}

\def\lesssim{\mathrel{\hbox{\rlap{\hbox{\lower4pt\hbox{$\sim$}}}\hbox{$<$}}}}
\def\gtrsim{\mathrel{\hbox{\rlap{\hbox{\lower4pt\hbox{$\sim$}}}\hbox{$>$}}}}
\def\la{\mathrel{\hbox{\rlap{\hbox{\lower4pt\hbox{$\sim$}}}\hbox{$<$}}}}
\def\ga{\mathrel{\hbox{\rlap{\hbox{\lower4pt\hbox{$\sim$}}}\hbox{$>$}}}}

\begin{document}

\title{A2163: Merger events in the hottest Abell galaxy cluster}

\subtitle{I. 
Dynamical analysis from optical data\thanks{Based on data obtained with the 
European Southern Observatory, Chile (runs 073.A-0672 and 077.A-0813) 
and with the Canada France Hawaii Telescope.}}
\author{
S. Maurogordato\inst{1}, A. Cappi\inst{2,1}, C. Ferrari\inst{3}, C. 
Benoist\inst{1} , G. Mars,\inst{1}
G. Soucail\inst{4}, M. Arnaud \inst{5},  G.W.
Pratt\inst{6} , H. Bourdin \inst{7}, J-L., Sauvageot\inst{5}}

\offprints{Sophie Maurogordato, \email{sophie.maurogordato@oca.eu}}

\institute{
Laboratoire Cassiop\'ee, CNRS, UMR 6202,
Observatoire de la C\^ote d'Azur, BP4229, 06304 Nice Cedex 4, France
\and
INAF - Osservatorio Astronomico di Bologna, via Ranzani 1, I--40127 Bologna,
Italy 
\and
Institut f\"ur Astro- und Teilchenphysik, Innsbruck Universit\"at, 
Technikerstrasse 25/8,6020 Innsbruck, Austria
\and
Laboratoire d'Astrophysique de Toulouse-Tarbes, CNRS-UMR 5572 and
Universit\'e Paul Sabatier Toulouse III, 14 Avenue Belin,
31400 Toulouse, France
\and
Service d'Astrophysique, DAPNIA,CEA-CEN Saclay, 91191,
Gif-sur-Yvette, France
\and
Max-Planck-Institut f\"ur extraterrestriche Physik,
Giessenbachstrasse, 85748 Garching, Germany
\and
Dipartimento di Fisica, Universit\`a degli Studi di Roma "Tor
Vergata", via della Ricerca Scientifica, 1, I-00133 Roma, Italy
}
\date{Accepted~2007 November 6}

\abstract
{A2163 is among the richest and most distant Abell clusters, with outstanding
   properties in different wavelength domains. 
X-ray observations 
have revealed a distorted morphology of the gaz and 
strong features have been detected in the temperature map, 
suggesting that merging processes are important in this cluster. 
However, the merging scenario is not yet well--defined.}
{We therefore undertook a complementary optical analysis, 
aiming to understand the dynamics of the system, to 
constrain the
merging scenario and to test its effect on the properties of galaxies.}
{We present a detailed optical analysis of A2163 based
on new multicolor wide--field imaging and the medium--to--high resolution
spectroscopy of galaxies.}
{The projected galaxy density distribution shows strong subclustering
with two dominant structures: a main central component (A), and
a northern component (B), visible both in optical and in X-ray, while
other substructures are detected in the optical.
At magnitudes fainter than R=19, the galaxy distribution shows a clear
elongation approximately on the east--west axis 
extending over $~4 h_{70}^{-1}$ Mpc, while
a nearly perpendicular bridge of galaxies along the north--south axis
appears to connect (B) to (A).
The (A) component shows a bimodal morphology, and the positions
of its two density peaks depend on galaxy luminosity: when going
to magnitudes fainter than $R = 19$, 
the axis joining the peaks shows a counterclockwise
rotation (from NE/SW to E--W) centered on the position of the X-ray maximum.
Our final spectroscopic catalog of 512 objects includes
476 new galaxy redshifts. We have identified 361 galaxies as cluster
members; among them, 326 have high precision redshift measurements, which 
allow us to perform a detailed dynamical analysis of unprecedented accuracy. 
The cluster mean redshift and velocity dispersion are
respectively $z= 0.2005 \pm 0.0003$ and $1434 \pm 60$ km/s.
We spectroscopically confirm that the northern and western
components (A2163-B and A2163-C) belong to the A2163 complex. 
The velocity distribution shows multi-modality,
with a bimodal structure peaking at $\sim 59200$ km/s and $\sim 60500$ km/s.
A significant velocity gradient ($\sim 1250$ km/s) is detected along the 
NE/SW axis of the cluster, which partially explains the detected bimodality. 
A2163 appears to be exceptionally massive:
the cluster virial mass is  
$M_{vir} = 3.8 \pm 0.4 \times 10^{15} \ M_\odot h_{70}^{-1}$.
}
{Our analysis of the optical data, combined with
the available information from X-ray observations 
and predictions of numerical simulations,
supports a scenario in which A2163-A has
undergone  a recent ($t\sim 0.5 ~$Gyr) merger along a a NE/SW 
(or E--W) 
axis, and A2163-B is connected to the main complex, probably
infalling on A2163-A.}

\authorrunning{Maurogordato et al.}
\titlerunning{A2163: Merger events in the hottest Abell galaxy cluster}

\maketitle

\section{Introduction}

A2163 is among the richest (richness class 2 as Coma) and most distant
Abell clusters ($z=0.20$) and presents
outstanding properties at various wavelengths. 
Extensive X--ray observations have shown that A2163 
is the hottest Abell cluster (temperature estimates
vary between 11.5 and 14.6 keV; Arnaud et al. 1992; Elbaz et
al. 1995; Markevitch et al. 1996; Markevitch \& Vikhlinin 2001; Pratt,
Arnaud \& Aghanim 2001) 
and among the most luminous ones ($L_X[2-10 {\rm keV}]=6.0\times10^{45}$ 
ergs/s, Ginga+ROSAT/PSPC, Elbaz et al. 1995).
The clear subclustering in the ICM density
map (Elbaz et al. 1995), in particular the presence of a secondary 
cluster in the North (A2163B in Elbaz et al. 1995) and the rotation 
of the main cluster isophotes with radius, together with strong 
temperature variations in the central region 
(Markevitch and Vikhlinin 2001, Bourdin et al. 2001, 
Govoni et al. 2004) are typical signatures of merging processes. 
Nevertheless, the surface brightness profile of the cluster is well 
fitted by a $\beta$-model (Elbaz et al., 1995; Pratt et al. 2001) and 
the temperature profile was shown to be relatively flat outside the 
very center (Pratt et al. 2001). The binding mass, derived from X-ray 
observations under the isothermal hypothesis, is very large: 
$4.6^{+0.4} _{-1.5} \times 10^{15} M_{\sun}$ within a radius of $4.6$ Mpc 
(Elbaz et al. 1995, assuming a cosmology 
with $H_0=50$ km s$^{-1}$ Mpc$^{-1}$, $\Omega_0=1$).

The galaxy density and mass distribution in the central region of this
cluster have also been  
determined by weak gravitational lensing (Squires et al. 1997, Cypriano et
al. 2004). These analyses show very similar mass and galaxy
distributions, with two coincident maxima and a flat shape elongated
in the E-W direction, but the weak lensing signal is surprisingly
faint in comparison to what could be expected from the cluster X-ray 
properties.
However, these detailed studies are limited to
the inner 8'$\times$8' region of the cluster, and do not include the
peripheral clumps such as A2163-B. La Barbera et
al. (2004) estimated the photometric redshifts of galaxies in A2163-B,
showing that this structure lies at the typical redshift of the main
cluster ($z = 0.215\pm 0.0125$).

The available optical and X-ray results suggest that A2163 is not a
completely relaxed cluster, as shown both by signatures of a
merging event in the central region, and by the presence of A2163-B, which is
a possible interacting subcluster in the North, at $\sim$7 arcmin from
the peak of X-ray emission. Radio observations (Feretti et al. 2001) 
revealed both an elongated and diffuse source (a possible relic) in the N-E
peripheral region of A2163, and, above all, one of the most powerful
and extended halos ever detected, which is quite regular in shape and 
elongated in the E-W direction, similarly to the X-ray emission.
Further analysis has shown that the spectral index
map of the halo is flatter in a region elongated along the N--S direction
and crossing the cluster centre, 
and at the northern and southern boundaries of the halo
(Feretti et al. 2004). While the physical processes generating radio
halos are still not clarified, several studies have shown 
that these objects are preferentially found in rich X-ray bright clusters
(Giovannini et al. 2002), and the mechanism of formation is probably
connected to cluster mergers (Brunetti et al. 2003, Govoni et al. 
2004, Feretti 2006, Ferrari et al. 2008). A better understanding of the merging scenario in
A2163 will hopefully also shed some light on the
formation of its radio halo and on the origins of its clumpy spectral map.

This work is part of a more general program aiming at reconstructing
the merging scenario(s) of Abell 2163 through a combined large-scale
optical/X-ray analysis, and test the relation between the merging
event and the properties of the galaxies and of the gas. Recent analyses of
merging clusters combining a high number of redshifts to X-ray and/or radio
data have brought significant insight in the understanding of these
complex systems (for a non exhaustive list, 
see for instance Quintana et al. 1996, Biviano et al. 1996, Maurogordato et
al. 2000, Arnaud et al. 2000, Flores
et al. 2000, Bardelli et al. 2001,  Czoske et al. 2002, Ferrari et al. 2003,
Ferrari et al. 2005, Miller
et al. 2004, Ledlow et al. 2005, Miller et al. 2006, Boschin et al. 2004,
Girardi et al. 2005, Boschin et al. 2006, Girardi et al. 2006, 
Barrena et al. 2007).
In this paper
we present the results obtained from the analysis of our optical
observations carried out in 2004 and 2006 at ESO, including wide-field 
multi-band imaging with the Wide Field Imager on the 2.2m telescope,
high resolution  (R $\sim 2000$) spectroscopy obtained with the VIMOS
Spectrograph at the VLT/UT3 telescope, and former spectroscopy obtained at
CFHT with the MOS instrument at lower resolution.
This paper will be followed by an
analysis of the X-ray observations with XMM and Chandra focused on 
the signature of the merging process in the temperature
maps, and by the study of the star formation history of cluster
members with respect to the merging event.

In Section 2 we present the data and the reduction
procedure. Section 3 focuses on the projected density distributions of the
galaxies and of the gas and their comparison through the determination
of isodensity maps and projected density profiles. We analyse
the velocity distribution and the dynamics for
the whole cluster in Section 4, and 
the existence of subclustering both in density and
velocity space in Section 5. In Section 6 we derive an estimate of
the cluster mass and compare it to independent 
X-ray and weak-lensing estimates. 
Discussion and conclusions are presented in Section 7.

In the following, we adopt the cosmological parameters of a
$\Lambda$CDM model with $\Omega_M$=0.3,
$\Omega_{\Lambda}$=0.7 and $H_0= 70\rm ~km ~s^{-1} ~Mpc^{-1}$. With
these parameters, at $z = 0.2$ one degree 
corresponds to a physical length of $11.9$ Mpc.

\section{Observations and data reduction}

\subsection{Data acquisition and reduction}

Our observations were carried out at ESO in 2004 (run 073.1-0672) and 2006 
(run 077.A-0813). We also used additional unpublished data
from spectroscopic observations at lower resolution obtained at CFHT 
by Soucail and collaborators in 1996 (see below).
A2163 was observed with the Wide Field Imager at the ESO 2.2m telescope 
and with the VIMOS (VIsible
Multi--Object Spectrograph; Le F\`evre et al. 2000) at the Melipal UT3
telescope of the ESO VLT.  
Imaging of the central 30'x30' field has been
performed in the R (filter ESO/878) and V (filter ESO/843) bands. For
each filter eight dithered images were obtained, leading to a total exposure
time of 40~mn. 
The seeing FWHM values of the R and V band images amount to 1.25 and 
1.40 arcsec, respectively.

These sets of images were reduced and combined using
the ESO/MVM package (Alambic, Vandame et al. 2002), and the galaxy catalogs
in the two passbands were extracted with SExtractor 
(Bertin and Arnouts, 1996). These catalogs include half-light radii 
and magnitudes (we adopted the MAG$\_$AUTO estimate), and 
the resulting magnitude--half light radius diagram
was used to classify stars and galaxies, up to a magnitude of 21 
in both bands. The limiting magnitudes, 
defined as $5 \sigma$ detections within an aperture of 
twice the seeing FWHM,  are 22.5 and 21.5 in the R- and V-bands, respectively. 
Finally, the V- and R-band catalogs were matched by adopting a matching radius 
of 1.5 arcsec.

\begin{figure*}
\begin{center}
\hspace{6mm}
\epsfxsize=15.0cm \epsfverbosetrue \epsfbox{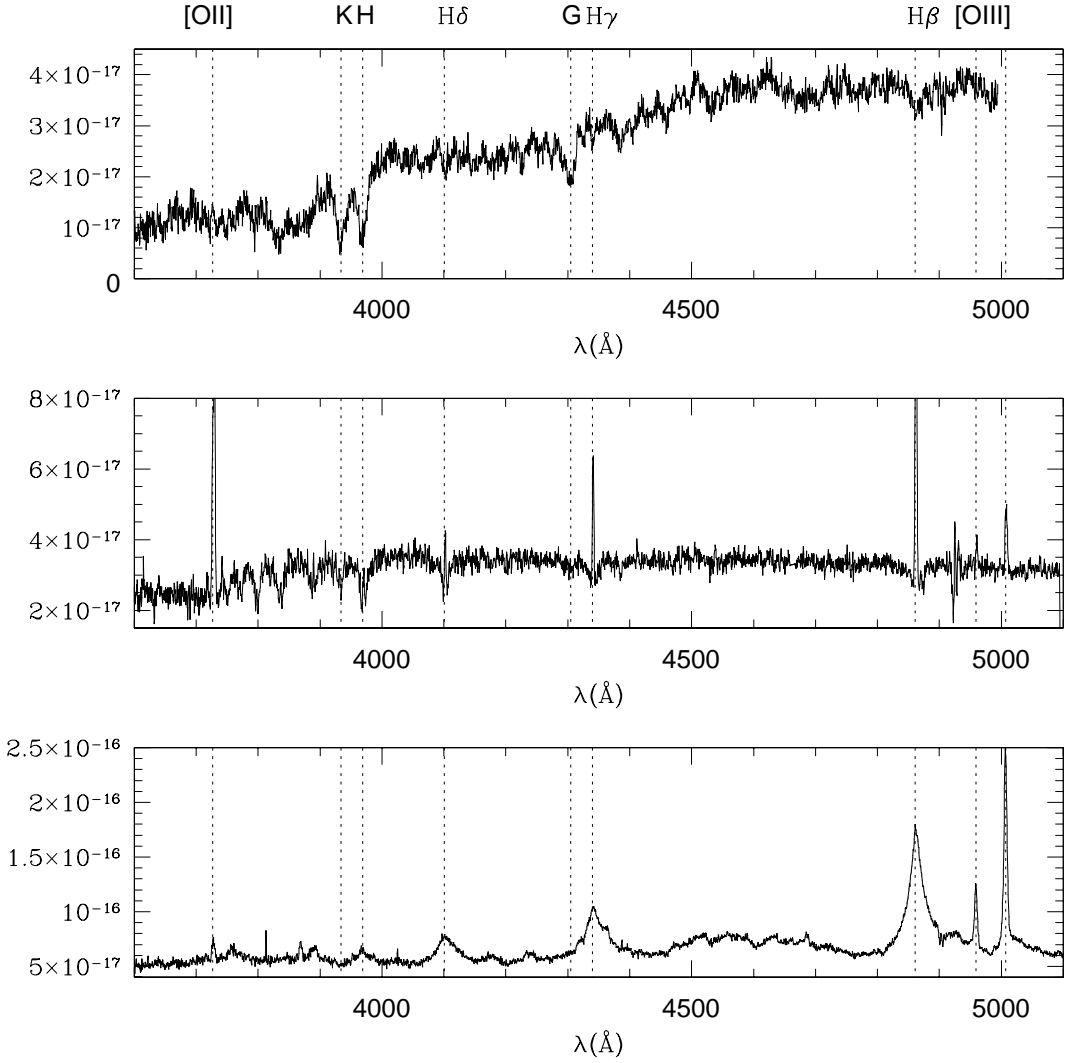}
\end{center}
\caption[]
{\sl Examples of spectra obtained with VIMOS in HR mode which are
representative of our Flag~0 sample. Spectra have been deredshifted, and
the rest wavelengths of various lines ([OII], Calcium K and H, $H\delta$, 
Gband, $H\gamma$, $H\beta$, [OIII]$_{a}$ and [OIII]$_{b}$) 
are shown with dotted lines. Intensity values are in arbitrary units. 
From top to bottom: a) spectrum of the brightest cluster member BCG-1 
(see Table \ref{Table_BCG}), with the absorption lines typical of an old 
population;
b) spectrum of a typical emission line galaxy,
with Balmer lines both in absorption and emission and oxygen emission lines;
c) spectrum of the galaxy with X-ray emission north of A2163-B with 
broad hydrogen lines, which is a type I AGN.}
\label{spectra}
\end{figure*}

For multi--object spectroscopy, masks were prepared through the
VMMPS (VIMOS Mask Preparation Software, Bottini et al. 2005).
Spectra were obtained in the 30'x30' field covered by WFI, 
with VIMOS in Multi-Object Spectroscopy mode. We used the High Resolution
Blue grism, with a slit of 1'' width, which leads to a resolution in the 
range [2050-2550] and covers the wavelength range  
$4200-6200$\AA~ for a centered slit. 
At the cluster redshift this range includes 
[OII], K and H, G band, $H_\delta$, $H_\gamma$,$H_\beta$, [OIII]a\&b 
among the main spectral lines.
About $100$ slits per pointing were assigned to targets, and 
the field was covered with six pointings 
(four obtained in 2004 and two in 2006) following 
a dithering pattern, in order to fill the 
cross-shaped gap (2' wide) among the 4 VIMOS quadrants. 
For each pointing, we obtained four exposures of 45~mn.
The prereduction of spectroscopic data was performed
with the dedicated VIPGI package (Scodeggio et al. 2005).
Finally, redshifts were measured through
the standard cross-correlation package {\em rvsao} in 
IRAF (Tonry and Davis, 1981),
using three stellar template spectra obtained during our observing runs
and a set of synthetic spectra derived from
the population synthesis library of Bruzual \& Charlot (2003).
All the spectra were checked visually for the 
presence of [OII], [OIII], and Balmer lines in emission 
with equivalent widths larger than 5\AA, and in case of positive
detection the emission lines redshifts were measured.
For galaxies where both absorption and emission redshifts were
available, we selected the value with lower error.
Redshifts are heliocentric (in our case the heliocentric correction
is small, amounting to $+3$ km/s).

Objects were classified in two categories according to the 
quality of the spectra, and thereby to the
precision of the derived redshifts:
a) the ``high precision'' sample (Flag 0), including
galaxies having high S/N ratio spectra 
($\sim 10$ per pixel or higher)
and typical redshift errors of $\sim 20$ km/s; 
b) the ``medium precision'' sample (Flag 1), 
including galaxies having lower S/N ratio spectra but
still reliable redshifts with larger errors ($\sim 100$ km/s). 
 Examples of flag 0 spectra are displayed in Fig.\ref{spectra}.
The dynamical analysis was generally performed on the high precision 
spectroscopic sample; the total sample was used 
when the redshift precision was not critical 
(e.g. when associating galaxies to the cluster in the color-magnitude
diagram for the best determination of the red sequence).

As we mentioned above, we dispose also of supplementary data
coming from multi-object spectroscopy
obtained at CFHT with the MOS instrument (PI: G. Soucail),
where the grism had a dispersion of 300 \AA /mm, giving a
wavelength scale of 7 \AA~ per pixel on the CCD in a binned mode and a
maximum wavelength range of 4400 to 8000 \AA. Due to the lower spectral
resolution of CFHT observations, these objects have  
Flag 1.
CFHT observations provided 108 spectra 
and 79 galaxies belonging to the cluster were identified: 51 of them
were also observed with VIMOS. 
After applying a 3$\sigma$ clipping,
the mean offset in velocity is
$-90 \pm 42$ km/s, which is within the mean errors from the CFHT measurements. 
When both
measurements were available, we chose the higher precision VIMOS measurement.
At the end, 28 objects were added from the CFHT observations.
Therefore our final spectroscopic catalog includes 
512 spectra with a successful radial velocity
measurement. 476 of them are galaxies: 430 have a ``high precision'' 
and 46 have a ``medium precision'' redshift.
 
The catalog is shown in Table \ref{Table_redshifts}, where 
colums are as follow: 
1) identification number; 
2) and 3) right ascension and declination (J2000.0);
4) radial velocity; 
5) velocity error; 
6) $R_{TR}$ parameter (Tonry and Davis 1981; when $R_{TR} > 3$ 
the cross-correlation redshift can be considered as reliable);
7) quality flag for the redshift (0: high precision, 1: medium precision); 
8) emission lines flag (0: no emission, 1: with emission); 
9) instrument (1: VIMOS at ESO/VLT; 2: MOS at CFHT).

\onllongtab{1}{
\begin{longtable}{ccccccccc}
\caption{\sl Radial velocities measurements
in the field of Abell 2163.
   The full catalogue (512 redshifts) is
   available in electronic form at {\rm www.edpsciences.org}} \\
\label{Table_redshifts} \\
\hline
\hline
Galaxy & Right Ascension(J2000)  & Declination (J2000)  & v (km~$s^{-1}$) & 
$\epsilon$ (km~$s^{-1}$) & $R_{TR}$ & Flag & Emission & Run \\
(1)  & (2) & (3) & (4) & (5) & (6) & (7) & (8) & (9)\\
\hline
 1  & 16 14 53.88 & -06 09 00.4 & 100970 & 50 &  5.1 & 0 & 1 & 1   \\ 
 2  & 16 14 54.15 & -06 06 08.3 &  63576 & 32 &  8.6 & 0 & 1 & 1   \\ 
 3  & 16 14 58.18 & -06 00 44.9 & 126845 & 71 &  3.4 & 0 & 1 & 1   \\ 
 4  & 16 14 58.59 & -06 08 01.8 &  62075 &  8 & 18.2 & 0 & 0 & 1   \\ 
 5  & 16 14 58.65 & -06 14 34.4 &  62851 & 14 &  9.7 & 0 & 0 & 1   \\ 
 6  & 16 14 59.21 & -06 05 55.4 &  59068 & 29 &  6.0 & 0 & 1 & 1   \\ 
 7  & 16 15 00.26 & -06 15 16.5 &  57805 & 11 & 14.5 & 0 & 0 & 1   \\ 
 8  & 16 15 00.84 & -06 18 59.4 &  60983 & 14 &  8.9 & 0 & 0 & 1   \\ 
 9  & 16 15 00.87 & -06 07 02.0 &  52493 & 59 &  6.5 & 0 & 1 & 1   \\ 
10  & 16 15 01.03 & -06 17 46.4 &  59140 & 11 & 20.6 & 0 & 0 & 1   \\ 
11  & 16 15 01.86 & -06 02 47.1 &    -89 &  8 & 14.8 & 0 & 0 & 1   \\ 
12  & 16 15 02.18 & -06 07 57.4 &    -86 & 14 & 11.7 & 0 & 0 & 1   \\ 
13  & 16 15 03.34 & -06 10 32.9 &  37716 & 11 & 14.3 & 0 & 0 & 1   \\ 
14  & 16 15 03.38 & -06 09 55.8 &  75970 & 17 & 19.8 & 0 & 1 & 1   \\ 
15  & 16 15 03.54 & -06 01 25.6 &    -71 & 11 & 11.9 & 0 & 0 & 1   \\ 
16  & 16 15 03.65 & -06 05 16.7 &  73835 & 14 & 11.4 & 0 & 0 & 1   \\ 
17  & 16 15 03.67 & -06 20 15.0 & 100610 & 20 &  8.7 & 0 & 0 & 1   \\ 
18  & 16 15 03.91 & -06 05 40.4 &  73697 & 17 & 13.6 & 0 & 1 & 1   \\ 
19  & 16 15 03.99 & -06 02 29.1 &  61148 & 11 & 18.0 & 0 & 0 & 1   \\ 
20  & 16 15 04.02 & -06 15 46.9 &  62845 & 128&  3.2 & 1 & 1 & 1   \\ 
21  & 16 15 04.05 & -06 14 19.9 &  61928 & 11 & 13.1 & 0 & 0 & 1   \\ 
22  & 16 15 04.92 & -06 13 53.8 &  58984 &  5 & 27.4 & 0 & 0 & 1   \\ 
23  & 16 15 05.41 & -06 02 46.6 &    -23 & 17 & 10.1 & 0 & 0 & 1   \\ 
24  & 16 15 05.50 & -06 09 23.9 &  49627 & 47 &  6.5 & 0 & 1 & 1   \\ 
25  & 16 15 05.95 & -06 11 17.3 &  60414 &  8 & 23.9 & 0 & 0 & 1   \\ 
26  & 16 15 06.14 & -06 00 33.9 &  79684 & 32 &  6.0 & 0 & 1 & 1   \\ 
27  & 16 15 06.56 & -05 59 46.8 &  79606 & 47 &  5.6 & 0 & 1 & 1   \\ 
28  & 16 15 06.61 & -06 00 28.7 &  79534 & 23 &  8.9 & 0 & 1 & 1   \\ 
29  & 16 15 07.06 & -06 03 12.6 &  34296 & 38 &  8.5 & 0 & 1 & 1   \\ 
30  & 16 15 08.11 & -06 14 14.9 &     80 & 11 & 14.8 & 0 & 0 & 1   \\ 
31  & 16 15 08.48 & -06 14 00.3 &  58012 & 20 & 13.3 & 0 & 0 & 1   \\ 
32  & 16 15 09.02 & -06 19 13.8 &  58984 & 11 & 12.7 & 0 & 0 & 1   \\ 
33  & 16 15 09.17 & -06 11 13.7 &  76042 & 38 & 10.8 & 0 & 1 & 1   \\ 
34  & 16 15 09.17 & -06 15 27.6 &  57680 & 11 & 12.3 & 0 & 0 & 1   \\ 
35  & 16 15 09.19 & -06 00 50.1 &    -74 & 11 & 14.2 & 0 & 0 & 1   \\ 
36  & 16 15 09.30 & -06 17 28.4 &  59038 &  8 & 23.4 & 0 & 0 & 1   \\ 
37  & 16 15 09.69 & -06 01 09.6 &   9173 & 62 &  3.8 & 1 & 0 & 1   \\ 
38  & 16 15 09.80 & -06 10 32.7 &  60806 & 11 & 18.8 & 0 & 0 & 1   \\ 
39  & 16 15 09.84 & -06 11 23.7 &  62093 & 14 & 10.4 & 0 & 0 & 1   \\ 
40  & 16 15 09.90 & -06 06 49.0 &  59098 & 20 &  8.9 & 0 & 0 & 1   \\ 
41  & 16 15 10.79 & -06 02 38.1 &  73754 & 20 & 12.1 & 0 & 1 & 1   \\ 
42  & 16 15 10.79 & -06 01 46.2 &  73230 & 29 &  6.2 & 0 & 0 & 1   \\ 
43  & 16 15 10.94 & -06 05 03.7 & 129285 & 17 & 11.3 & 0 & 0 & 1   \\ 
44  & 16 15 11.84 & -05 59 29.1 &  46527 & 29 &  2.7 & 0 & 1 & 1   \\ 
45  & 16 15 12.03 & -06 00 19.9 &  74042 & 23 &  9.3 & 0 & 1 & 1   \\ 
46  & 16 15 12.07 & -06 05 27.6 &  60423 & 20 &  8.5 & 0 & 0 & 1   \\ 
47  & 16 15 12.09 & -06 17 20.1 &  83306 & 11 & 15.7 & 0 & 0 & 1   \\ 
48  & 16 15 12.12 & -06 08 32.5 &  60063 &  8 & 23.4 & 0 & 0 & 1   \\ 
49  & 16 15 12.18 & -06 04 03.4 &  61769 &  8 & 27.6 & 0 & 0 & 1   \\ 
50  & 16 15 12.44 & -06 07 25.2 &  59224 & 53 &  5.2 & 0 & 1 & 1   \\ 
\hline
\hline
\end{longtable}
} 

\begin{figure*}
\begin{center}
\hspace{6mm}
\epsfxsize=17.0cm \epsfverbosetrue \epsfbox{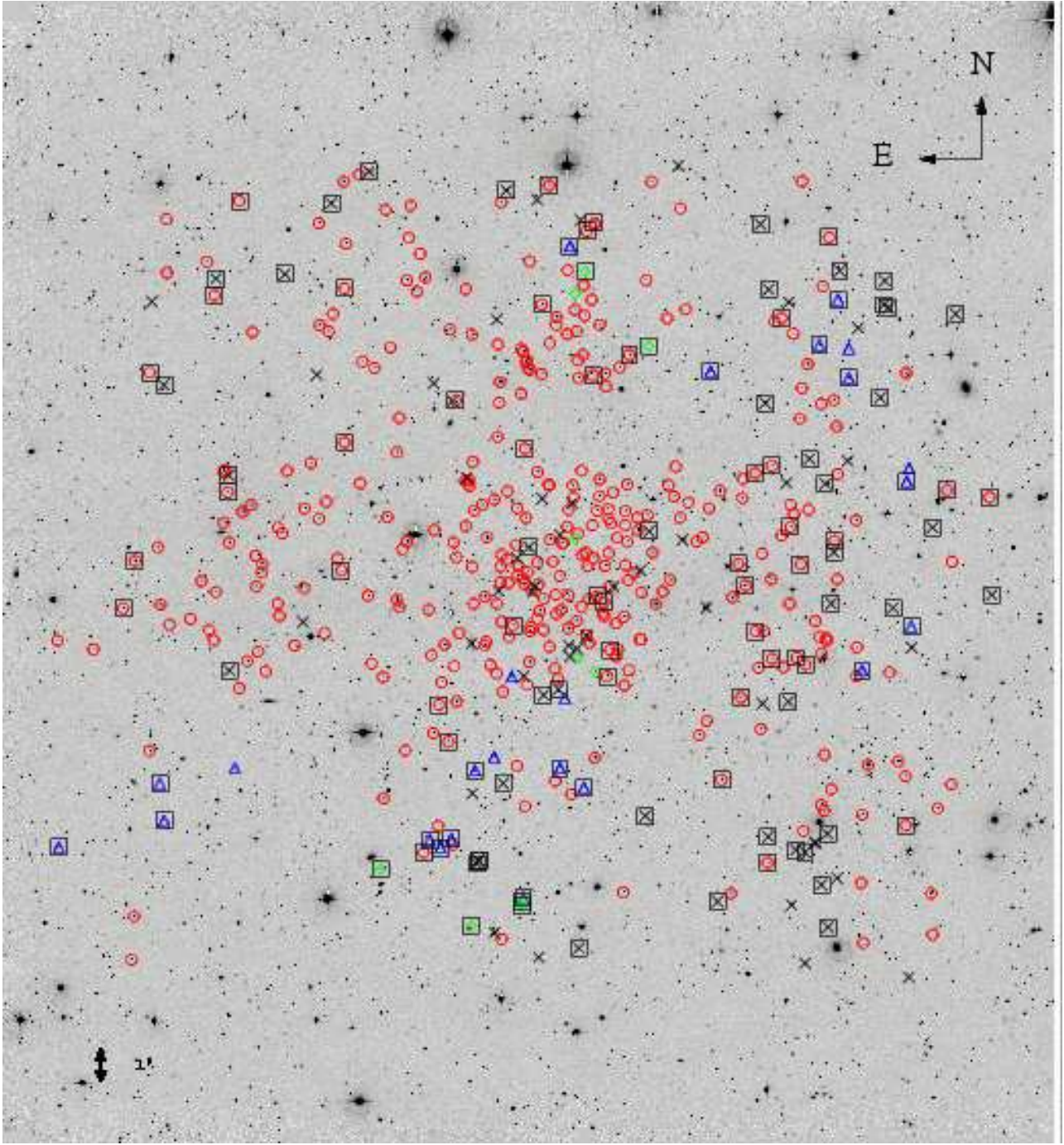}
\end{center}
\caption[]
{\sl WFI R--band image of the A2163 field.
Galaxies for
which successful spectra have been obtained (Flag 0 and 1) are
marked by different symbols. 
Red circles: velocity range [54000,66000] km/s corresponding to
expected cluster members. 
Blue triangles: velocity range [71000,77000] km/s corresponding to the first
background overdensity in the velocity histogram.
Green diamonds: velocity range  [87000-93000] km/s, corresponding
to the second background overdensity.
Black crosses: other redshifts. 
Squares indicate emission line galaxies.
There is a high number of emission line galaxies belonging to
the cluster on the West side. 
The spatial scale of the image is shown in the lower-left corner.}
\label{A2163_spectro}
\end{figure*}

\subsection {Completeness}

Apparent magnitudes were transformed to absolute ones following: $M = m
- 25 - 5log_{10}(D_L) - A - K(z)$, where $D_L$ is the luminosity
distance in Mpc, $A$ is the galactic extinction,
and $K(z)$ is the K-correction. 
We note that A2163 is
in a region with strong local variations of the extinction, and that 
the estimate of the extinction by Schlegel et al. (1998) differs from 
that of Burstein and Heiles (1984). For this region
La Barbera et al. (2004) have found a value for E(B-V) of ~0.41, 
in good agreement with that of Schlegel et
al. (1998), which we adopted in our work.
We computed the K-correction corresponding to an elliptical galaxy at $z=0.2$
for the set of adopted filters; in the $R$ (AB) passband,
assuming a Schechter $M_{R}^*=-21.97$ (Popesso et al. 2005), we expect
$m_{R}^*$ = 18.1. As the limiting magnitude of our catalog is $R=22.5$,
this means that we observe galaxies which are 4.4 magnitudes 
fainter than $M^*$. 
Moreover, the total field covered by the WFI camera (30'x30')
corresponds to $6$x$6$Mpc$^2$ at the mean redshift of the cluster.
Thus we have both a deep and wide sampling of the galaxy population in
A2163.

\begin{figure}
\begin{center}
\hspace{6mm}
\epsfxsize=8.0cm \epsfverbosetrue \epsfbox{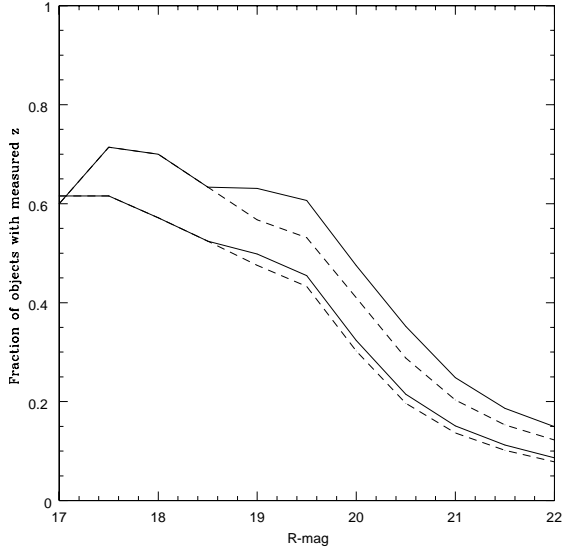}
\end{center}
\caption[]
{\sl The spectroscopic completeness of our observations as a
function of R-band magnitude.
Solid lines: Flag 0 + Flag 1 sample; dashed lines: Flag 0 sample.
Upper and lower lines refer respectively to the $8'x8'$ central field 
and to the $20'x20'$ wide field.}
\label{fig:compz}
\end{figure}

\begin{figure}
\centering
\resizebox{6cm}{!}{\includegraphics{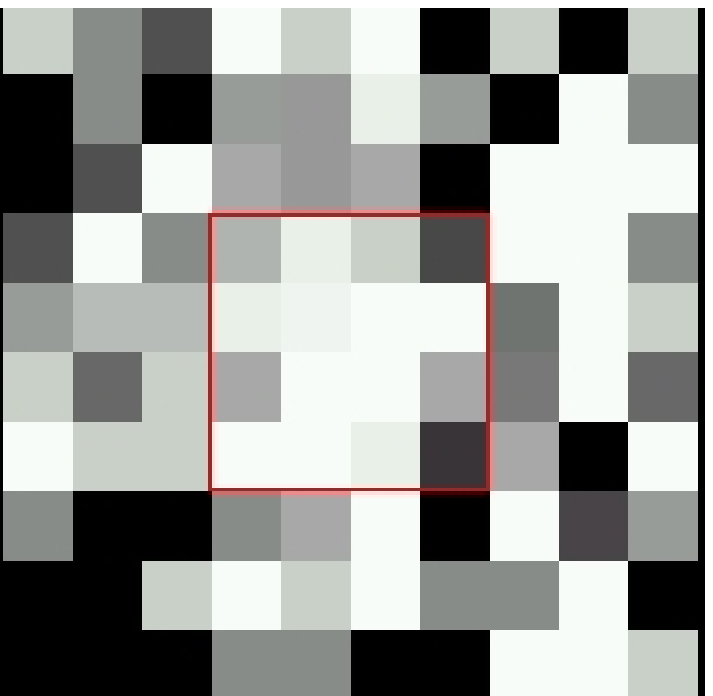}}
\resizebox{6cm}{!}{\includegraphics{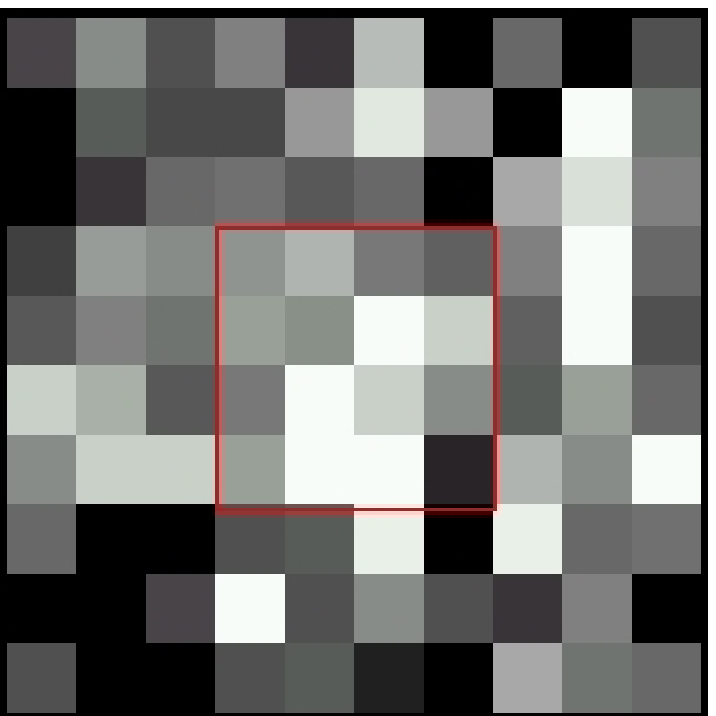}}
\caption[]
{\sl The spectroscopic completeness for our high precision velocity
   sample in the inner 20'$\times$20'. 
Top panel: galaxies brighter than $R = 19.5$;
bottom panel: galaxies brighter than $R = 20$. 
Individual cells are 2'x2' wide. 
The completeness ratio increases from black to
white. The threshold value is 0.1 for black pixels and 0.6 
for white pixels. The inner field (8'x8') is displayed as a red square.}
\label{sampling}
\end{figure}

The original scope of the program was to obtain spectra of all the galaxies
brighter than $R = 20$ in the 20'x20' region centered on
A2163, in order to sample the cluster population to luminosities 
$\sim L^* +2$, and to a radius of $\sim 2 h_{70} ^{-1}Mpc$. 
Due to our dithering pattern and to the field of VIMOS (16'x18'),
the total field covered by spectroscopy is slightly wider ($\sim 25$' x 35'),
but the spectroscopic completeness in the external parts is low. 
Therefore we limit the completeness analysis to the original 20'x20' region.
Fig. \ref{A2163_spectro} shows the the WFI R-band image of A2163 
with symbols identifying the galaxies which have
a measured redshift (Flag0+ Flag1). 
We have computed the spectroscopic completeness 
$f(R)= N_z(R)/N_{2D}(R)$, where  $N_z(R)$ is the number of galaxies 
with a measured redshift and magnitude brighter than $R$,
and $N_{2D}(R)$ is the number of galaxies in our
photometric catalog with magnitude brighter than R.
Fig. \ref{fig:compz} shows $f(R)$ for our two
spectroscopic catalogs, the high precision and total sample, 
corresponding respectively to Flag 0 and Flag 0 + Flag 1 objects in Table
\ref{Table_redshifts}. Completeness has been computed in two fields:
the inner (8'x8') and the wider (20'x20'). As a general trend, the wide-field
completeness is stable up to $R\leq19.5$
and rapidly drops at higher magnitudes. In particular,
the central field is highly sampled: the completeness ratio at $ R \le 19.5$
is between 0.63 and 0.75 for the total (Flag 0 + Flag 1) sample 
and between 0.55 and 0.62 for the high precision (Flag 0) sample.

Fig.\ref{sampling} shows how the completeness at  $R \leq 19.5$ and
$R \leq 20.0$ varies in the 20'x20' central field;
the completeness is computed in cells of 2'x2'. 
It is clear that at both magnitudes some regions in the periphery are 
very poorly, or even not sampled 
(such as for instance the South-East and North-East corners).
The main reason is due to 
technical problems during observations in service mode.
As we have previously explained, we covered the ``blind cross'' 
among the 4 VIMOS quadrants through a dithering pattern: because of the
technical problems this pattern was not completed, as apparent from 
the dark pixels in the completeness map.
On the other hand, some regions are very well sampled,
such as the central 8'x8' core (except in its North-West and
South-West corners).
Sampling is not only higher but also  more homogeneous at $R\leq19.5$ than at 
$R\leq 20$, in particular in the central region.

\section{Density distribution of galaxies}
\subsection{Projected isodensity maps}

\begin{figure*}
\centering
\resizebox{15cm}{!}{\includegraphics{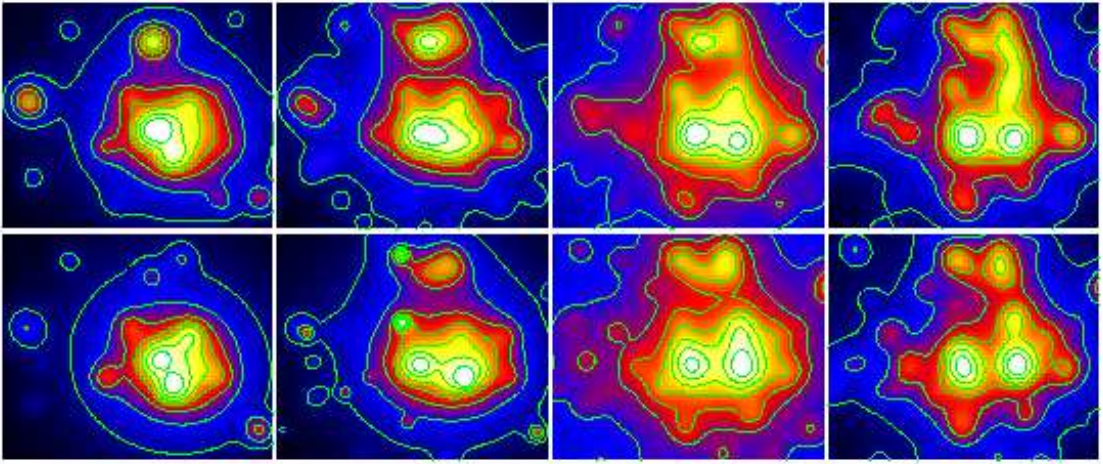}}
\caption[]{\sl Projected galaxy density maps for different magnitude cuts
(from left to right: R$<19$, R$<20$, R$<21$, and $19<R<21$). 
In the top panels no
color selection has been applied, while in the bottom panels only
red-sequence galaxies 
have been used to compute the density
maps (see text). The field displayed in each panel corresponds to the central 
$15'\times20'$. As usual, North is on the top and East on the left.
}
\label{isodens_mag}
\end{figure*}

\begin{figure*}
\begin{center}
\hspace{6mm}
\epsfxsize=17.0cm \epsfverbosetrue \epsfbox{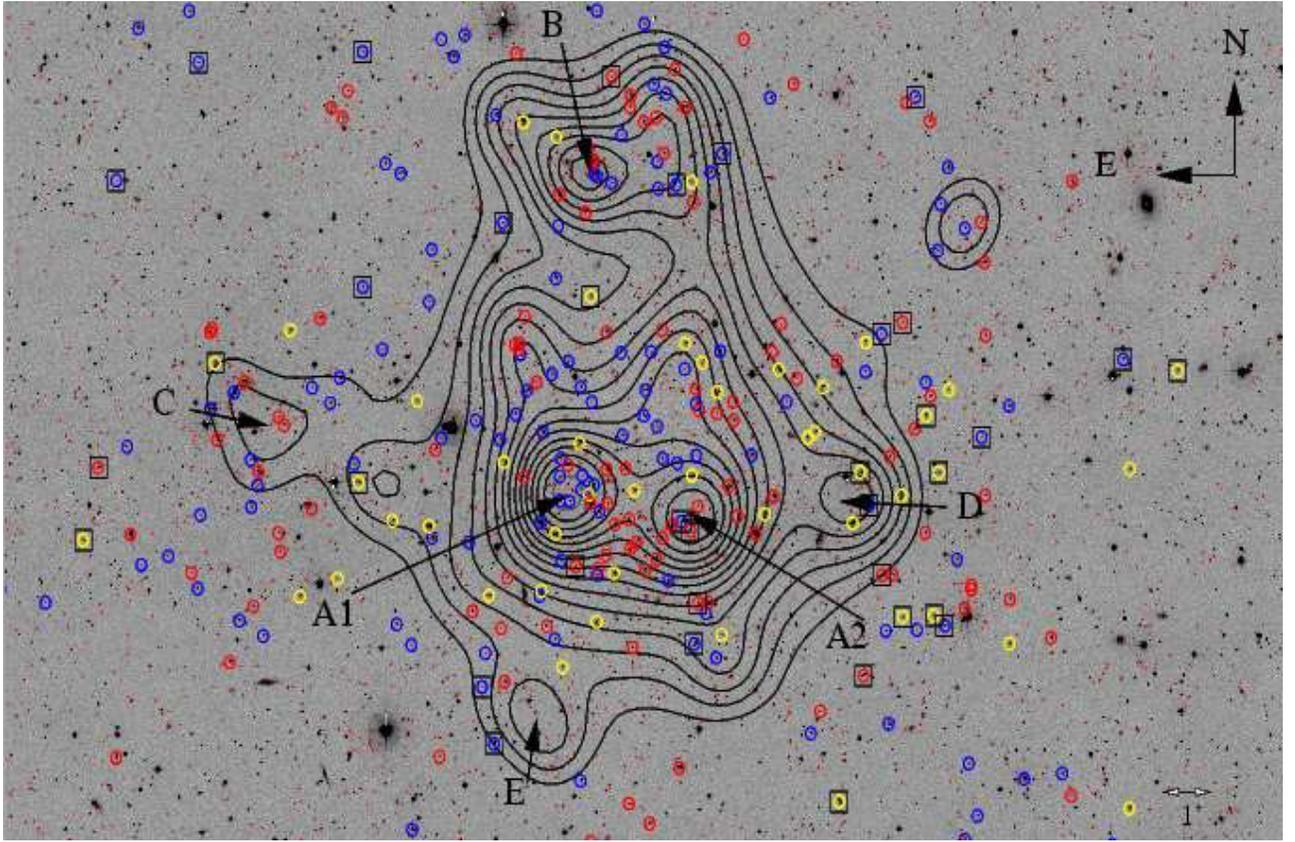}
\end{center}
\caption[]{\sl 
Projected galaxy density maps for $R<21$ superimposed on the WFI R-band image.
Substructures significant at more than $3 \sigma$ levels are identified with
capital letters. 
Cluster members in the three velocity bins corresponding to the KMM partitions 
are marked with circles of different colors. 
Blue: KMM1; red: KMM2; yellow: KMM3.
Emission line galaxies are marked with a black square.
}
\label{4slices}
\end{figure*}

The projected density distribution of galaxies has been computed from
the photometric galaxy catalog in the R band through the
multi-scale algorithm of Slezak et al. (in prep., see also Fadda,
Slezak and Bijaoui 1998 and Ferrari et al. 2005 for a description of the
algorithm). Density maps for different cuts in the R
magnitude (from $ R<19$ to $R<21$, and for $19<R<21$) are shown 
in Fig.\ref{isodens_mag} (top panel), where at each scale only structures 
which are $3 \sigma$ above the background are visualized. 
The $R<21$ density map is superimposed to the WFI $R$ 
image in Fig.\ref{4slices}, where the most significant substructures 
are identified. 
A visual inspection of the isodensity maps at the different
magnitudes cut-offs reveals a main, central component (A), while at
larger distances there are a variety of substructures, the most
significant one lying at $\sim 7$ arcmin North (B), and another
one, less prominent, located at $\sim 9$ arcmin East of the
center (C). These three components are visible at all
magnitude cut-offs. At $R<20$ and fainter cut-offs, new substructures are
visible, such as (D) which is $\sim 6$ arcmin to the west of 
the centroid, 
and (E) which is $\sim 4$ arcmin to the south of it.
Other smaller substructures in the nearby periphery of the
main subcluster (A) depend on the magnitude cut-off and might be
spurious.
substructures.
We note however the presence of two substructures, more prominent 
at fainter
magnitudes, to the northern periphery of the main subcluster (A). 
At faint magnitudes ($19<R<21$), a bridge of faint
galaxies connects the main subcluster (A) to the northern subcluster (B). 
There is also a long, low density structure extending in the
E--W direction, and including D, the main subcluster A and C.

The inner structure of A changes significantly with luminosity: 
it has an elongated shape and is bimodal, with two main components 
(A1 and A2 in Fig.~\ref{isodens_mag}) visible at all magnitudes. However, when
including fainter objects, the position of A1 and A2
changes, with a counterclockwise twisting of the axis
joining them: 
at R$<$19 the two density peaks are aligned along a NE/SW direction, 
but at fainter limiting magnitudes
(R$<20$ and R$<21$), A1 and A2 are lying on the (E--W) large-scale
axis of the cluster.
The angular separation of
the two peaks is 2 arcmin (3.5 arcmin) in the case of bright (faint)
objects. The coordinates of these maxima as a
function of magnitude are listed in Table~\ref{Table_subclusters}.

The projected density distribution of the second main component, A2163-B, 
also varies with magnitude. At $R<19$ it has a regular,
circular shape but at fainter magnitudes
it becomes more elongated,with two tails
extending  in the NE and NW directions, respectively.

On large scales, two main directions clearly appear at fainter
magnitudes: an E--W axis, corresponding to the large-scale orientation of the 
main component A, and including also the C and D substructures, 
and a N--S axis, joining the northern subcluster B, 
the main component A, and the E substructure.

\begin{table}
\begin{center}
\small
\begin{tabular}{cccc}
\hline
\hline
subcluster &Limiting Rmag & R.A.(J2000)  &Dec(J2000)\\
\hline
A1 &R $\le$ 19 &16~15~47.8  &-06~08~11\\
A2 &R$\le$19 &16~15~42.5  &-06~09~53\\
B  &R$\le$19  &16~15~48.8  &-06~02~21\\
C  &R$\le$19  &16~16~22.8  &-06~06~17\\
D  &R$\le$19.5  &16~15~27.9  &-06~09~18\\
A1 &R$\le$21 &16~15~50.9  &-06~08~29\\
A2 &R$\le$21 &16~15~39.3  &-06~09~15\\
X-ray (main) &- &16~15~46.0 &-06~08~55\\
X-ray (North)&- &16~15~48.0 &-06~02~25\\
\hline
\hline
\end{tabular}
\caption{\sl Coordinates of the maxima in the projected galaxy
  isodensity maps and in the X-ray ROSAT image (Elbaz et al. 1995) of
  A2163. As the positions of the maxima in the central subcluster A2163-A, A1
  and A2, depend on the limiting R-band magnitude, two sets of values are
  listed, corresponding to R $\leq$ 19 and R $\leq$ 21, respectively.}
\label{Table_subclusters}
\end{center}
\end{table}

\begin{table*}
\begin{center}
\small
\begin{tabular}{ccccccc}
\hline
\hline
Name & R.A.(J2000)  & Dec(J2000)  &R   &vel(km/s) &error(km/s)      &PA\\
\hline
BCG-1  & 16~15~48.9 &-06~08~41  &16.86  &60361     &10   &5.\\
BCG-2  & 16~15~33.5 &-06~09~16  &16.45  &60006     &13   &-9.\\
\hline
\hline
\end{tabular}
\caption{\sl The properties of the two brightest elliptical galaxies 
in the central  region of A2163,
BCG-1 and BCG-2. Column 1) Identificator; 
Column 2) and 3) Right ascension and
declination (J2000.0); Column 4) R-band magnitude; Column 5)
and 6) Radial velocity and its error; Column 7) Position
   Angle of the major axis (starting from the E--W axis, anti-clockwise)}
\label{Table_BCG}
\end{center}
\end{table*}

\begin{figure}
\centering
\resizebox{8cm}{!}{\includegraphics{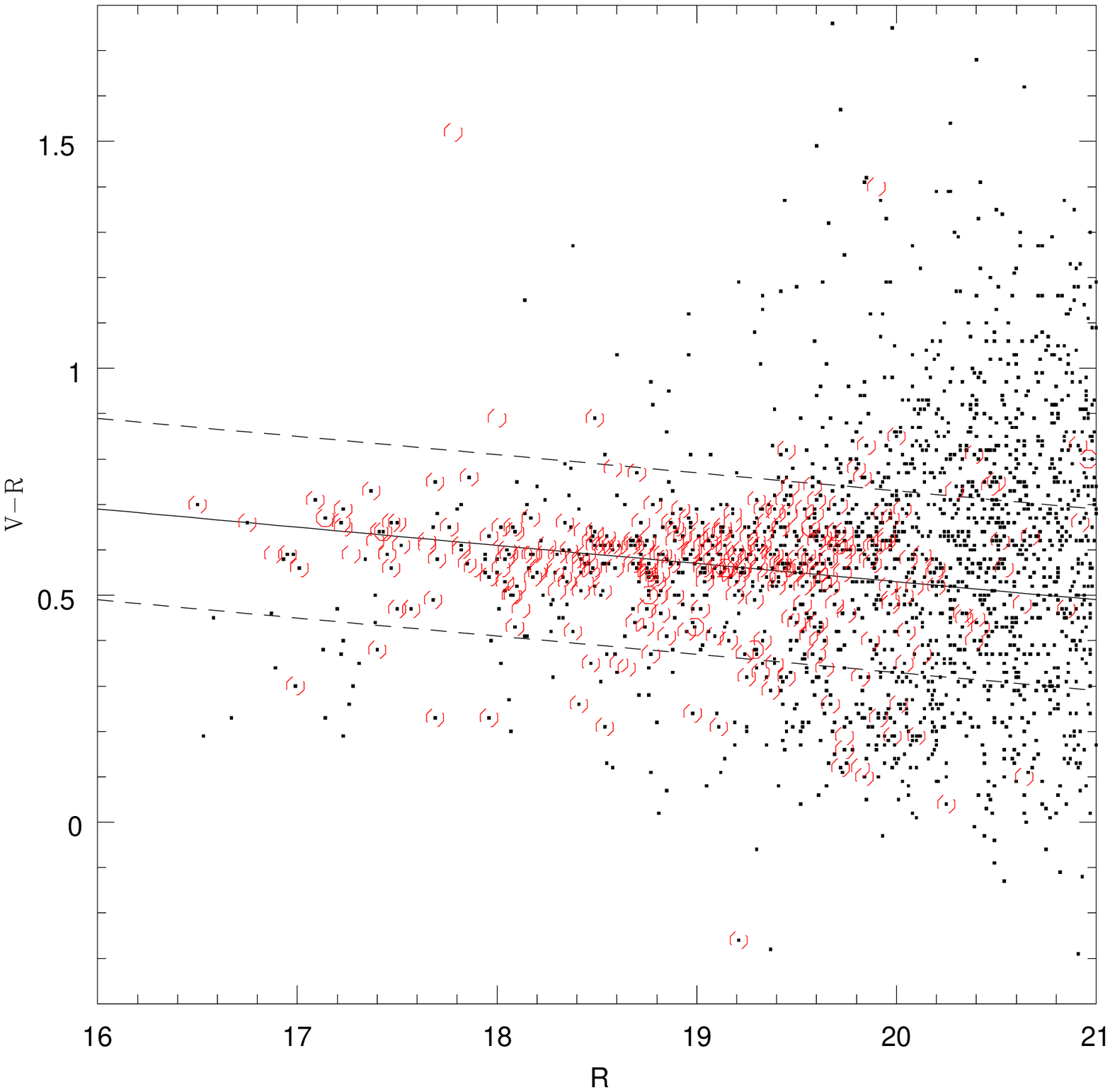}}
\resizebox{8cm}{!}{\includegraphics{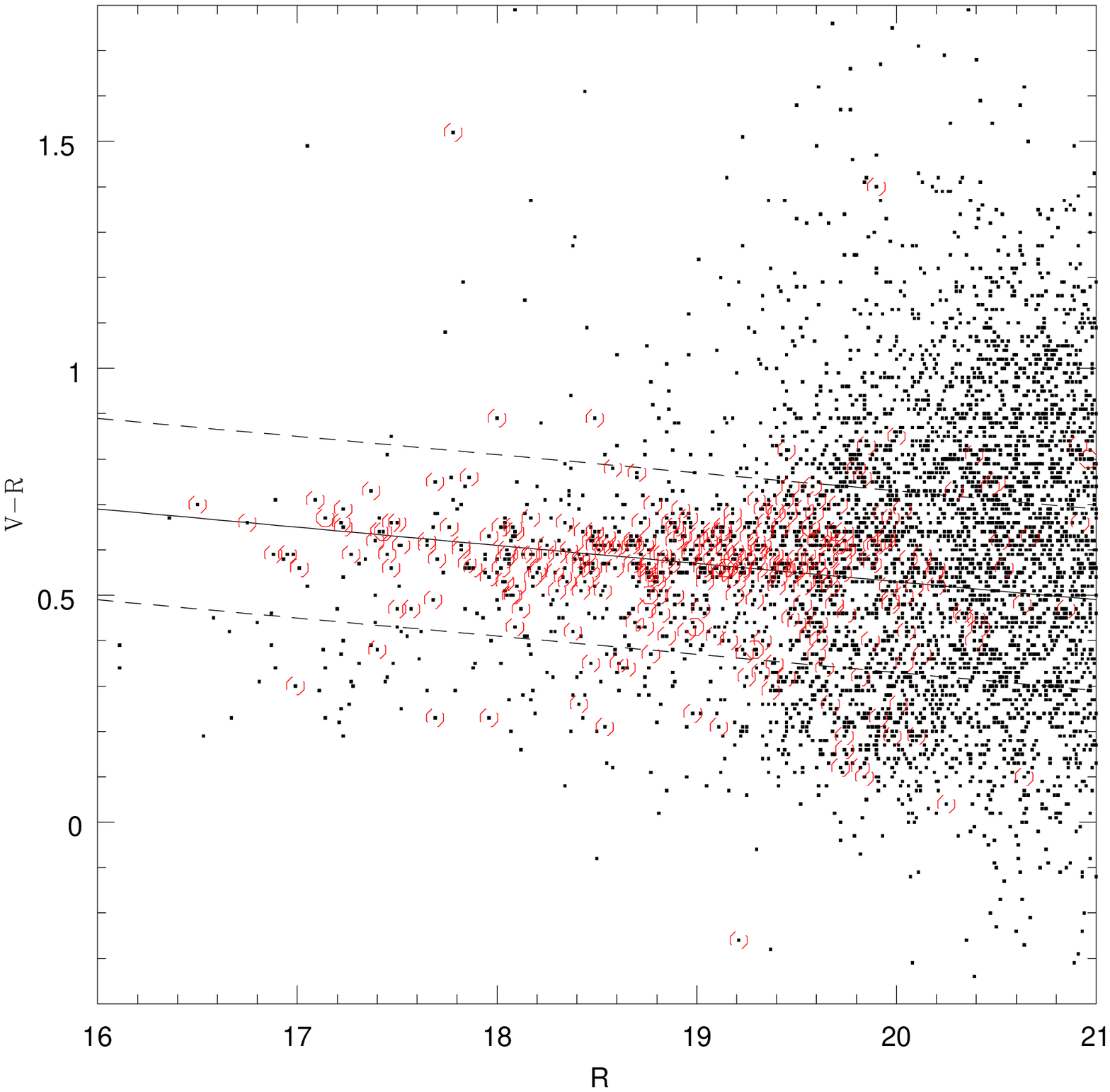}}

\caption{\sl 
Color-magnitude diagram in the field of A2163. Top panel: the central
   10'x10' field. Bottom panel: the whole 30'x30' field covered by WFI.
Black dots: galaxies in the photometric catalogs; 
red circles: spectroscopically confirmed cluster members.
The solid line represents the Red Sequence best fit, while dashed lines 
indicate the $\pm 2 \sigma$ limits from the best fit.}
\label{RS_all}

\end{figure}

\begin{figure}
\centering
\resizebox{8cm}{!}{\includegraphics{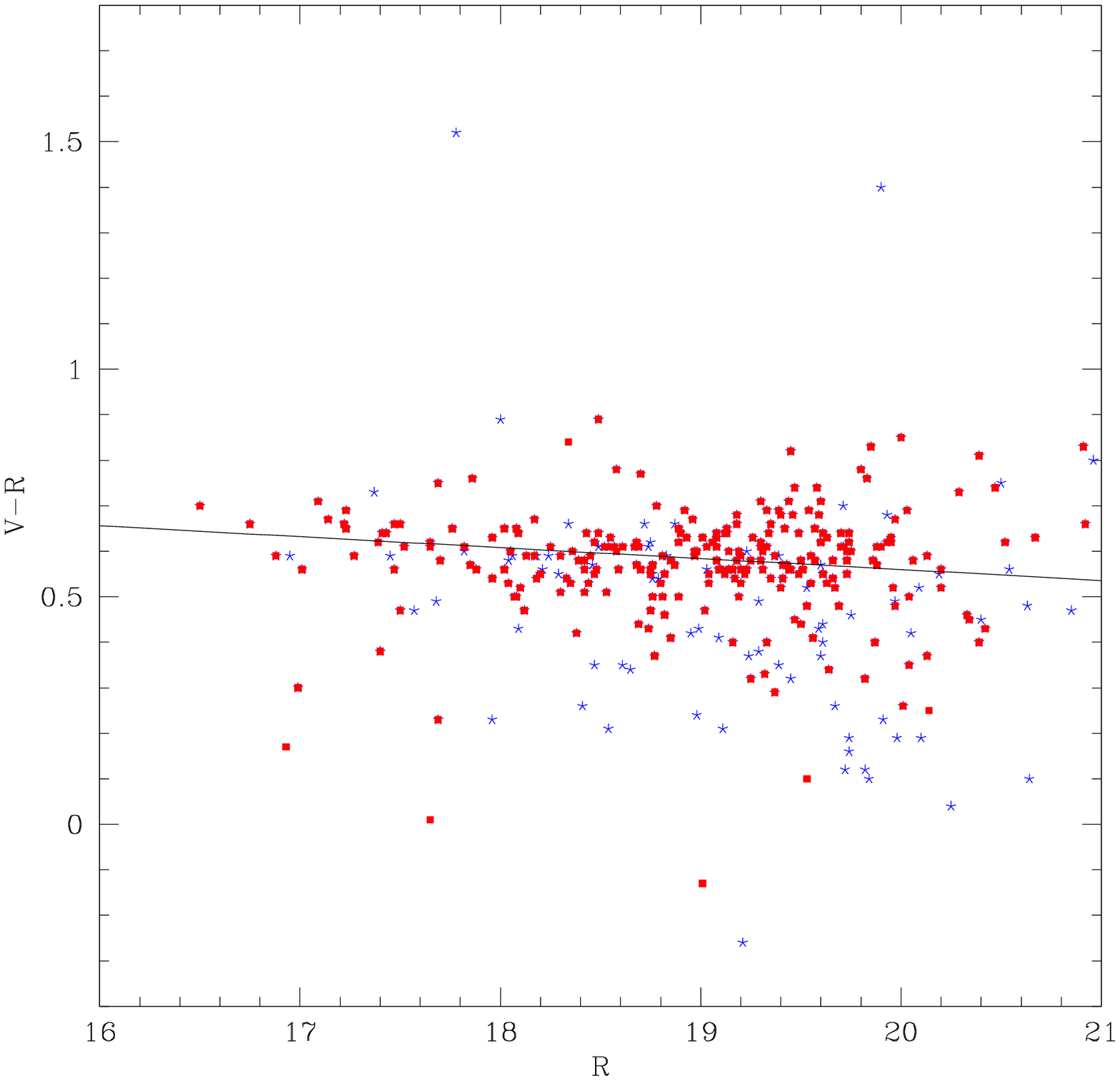}}
\caption{\sl 
Color-magnitude diagram for cluster members identified by
   spectroscopy. Red symbols indicate spectra with no emission lines while blue
   symbols correspond to spectra with emission lines. 
The solid line is the best--fit obtained  
   applying the robust method of Lopez--Cruz
   et al. (2004) to the subsample of galaxies without emission lines.} 
\label{RS_zcl}
\end{figure}

In order to better isolate the cluster population of
early type galaxies, we used our photometric
data to construct the color-magnitude diagram and to identify the red 
sequence of A2163. 
The resulting density map has the advantage to
be mostly decontaminated from projection effects, and it can be considered as
more representative of the cluster density map (it will obviously miss 
eventual concentrations dominated by objects with late--type or 
peculiar colors).
We have plotted  the (V-R) versus R diagram of galaxies:  
a) in the central $10' \times 10'$ region
(Fig.\ref{RS_all}, top panel); b) in
the whole $30' \times 30'$ field covered by WFI 
(Fig.\ref{RS_all}, bottom panel); 
c) with velocities within the cluster range (Fig.\ref{RS_zcl}). 
In all diagrams the red sequence is clearly visible at 
$R < 20$ (with an obvious higher dispersion 
when including the whole field). 
In order to parametrize the red sequence (hereafter RS), we 
have used the robust fitting
method introduced by Lopez-Cruz et al. (2004).
We have fixed the magnitude cut-off at R=20, which corresponds 
to $M^*+2$ at the
cluster redshift.
We have found that the values found by the minimization for the slope 
and the intercept for the different subsamples are very stable.

We have finally fitted the RS selecting only the ``redshift confirmed'' 
cluster members as defined in section \ref{sec:vel}.
The color--magnitude diagram of this sample is shown in Fig.\ref{RS_zcl}, 
both for emission line (blue symbols) and no emission line (red symbols) 
galaxies. Emission line galaxies were also 
excluded from the fit as they are not expected to populate the RS.
In conclusion, the best fit parameters for the RS
relation  $V - R = a + bR $ 
of this subsample are: $a=1.04$ and $b=-0.024$, with a $\chi^2$ value of 0.10.

In order to construct the projected density distribution of the cluster,
we selected all galaxies (with and without a measured redshift)
within $\pm 2 \sigma$ from the RS best fit, assuming that they are
early--type cluster members (see Fig.\ref{RS_all}).

The density maps are displayed in Fig. \ref{isodens_mag} (bottom panels).
The overall morphology of the cluster is 
globally unchanged, showing the main central cluster A2163-A and 
the northern subcluster A2163-B. The density contrast of
A2163-A is enhanced in the red sequence (hereafter RS) density maps. 
However, this is not the case for the smaller clumps: while A2163-E 
is still visible, A2163-C and A2163-D are hardly distinguishable. 
We have indeed verified that several galaxies in C and D
still follow the cluster red sequence, but with a somewhat 
larger dispersion than that found for galaxies in the cluster central region. 
We conclude
that the clumps C, D and E all belong to the A2163 complex, and
that their galaxy population includes
both bright early type RS galaxies and
fainter objects with colours outside the red sequence.

\begin{figure}
\centering
\resizebox{8cm}{!}{\includegraphics[angle=-90]{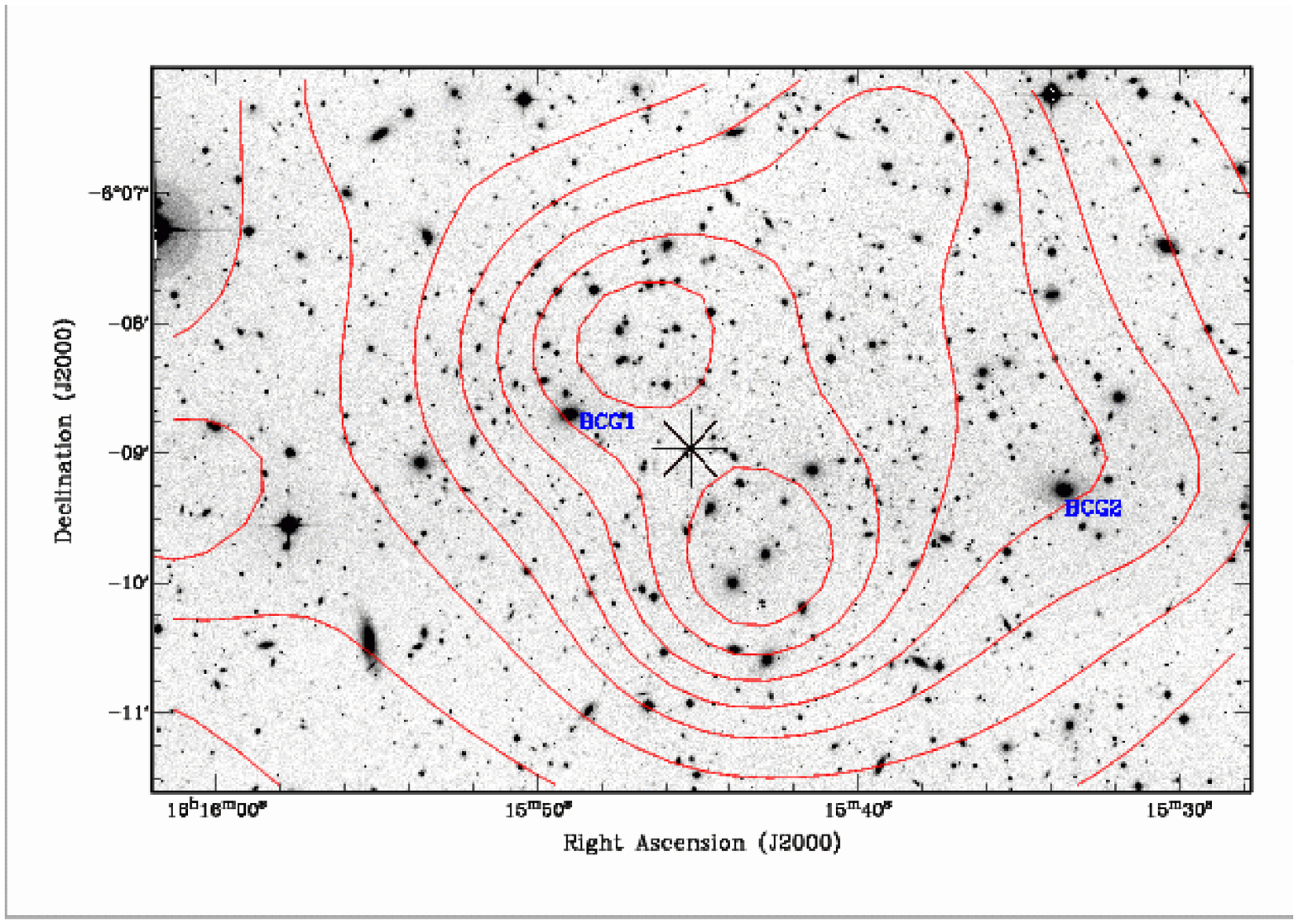}}
\resizebox{8cm}{!}{\includegraphics[angle=-90]{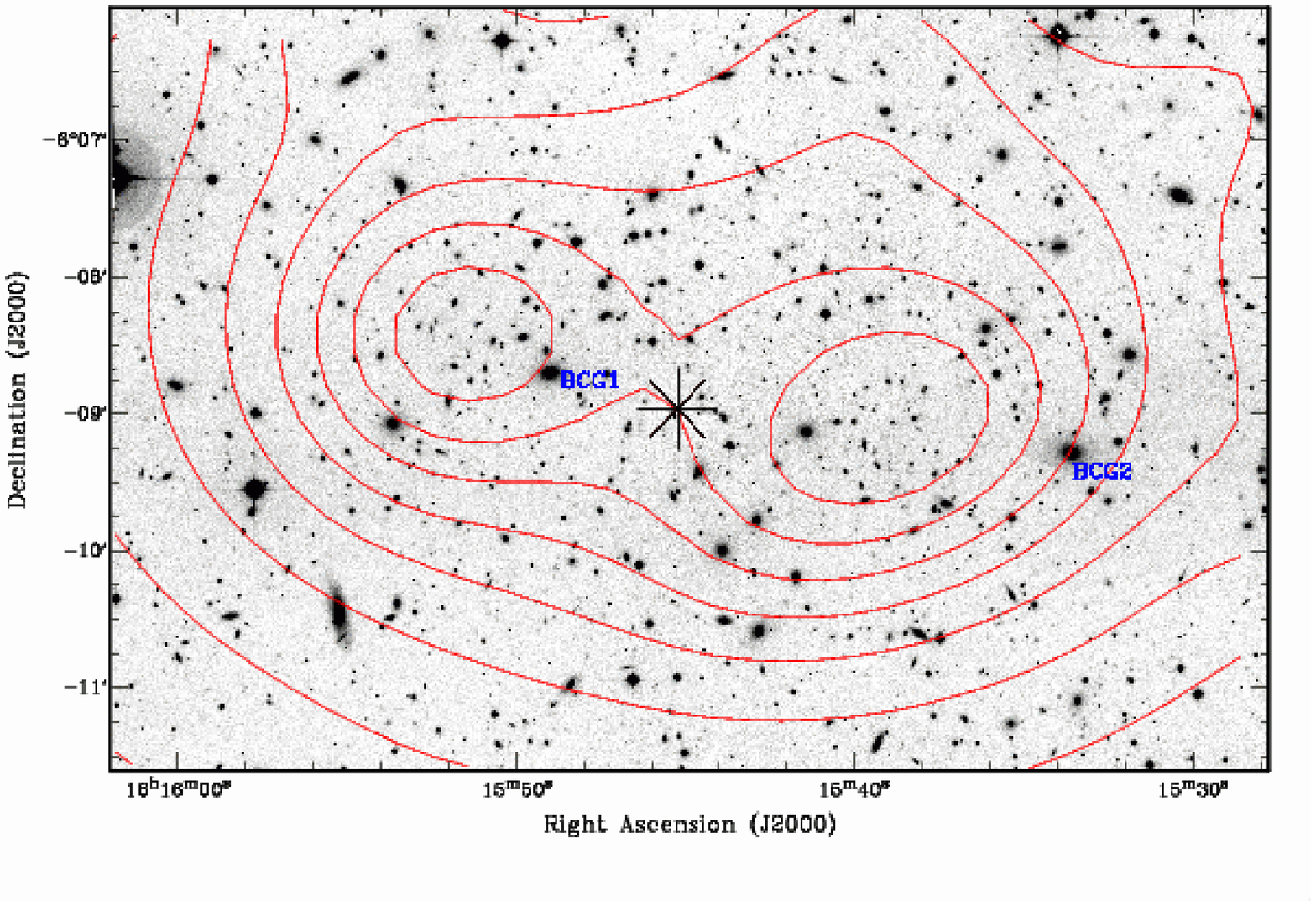}}
\caption{\sl R-band image of the central field of A2163, hosting the two
BCGs (marked in figure). Contours correspond to the projected
density distribution of Red Sequence galaxies with R$<$19 (top panel) and
R$<$20 (bottom panel). The position of the X-ray emission peak derived 
from XMM/EPIC observation is
indicated by an asterisk.}
\label{fig:BCGs}
\end{figure}

In Fig.~\ref{fig:BCGs}, the optical isodensity contours for
the RS galaxies with R$<$19 (top) and R$<$20 (bottom) are
superimposed on the R-band image of the cluster central region. 
The presence of  bimodality  (clumps A1 and A2 in
Fig.\ref{isodens_mag}) is striking, as well as the 
twist of the axis joining the two density peaks as a function of
the magnitude cut, while in both cases the peak of X-ray emission 
is located between the two central maxima 
(see the
angular coordinates of the X-ray and optical subclusters listed in 
Table \ref {Table_subclusters}).
It is important to stress
that the two Brightest Cluster Galaxies (BCG1 and BCG2 in
Fig.~\ref{fig:BCGs} and Table \ref {Table_BCG}) are both located
in the central clump A, but their positions do not
coincide with the maxima of the density peaks. However, 
while BCG1 is near A1 (at bright and faint magnitudes), BCG2 has
a significant offset to the west of A2. 
The axes of these two giant
galaxies and the line joining them are 
aligned along the same direction of the large--scale cluster
density distribution, i.e. the E--W direction
(see Table \ref{Table_BCG}). We note the presence of a bright
galaxy in proximity (east) of A2, which is also
on the E-W axis.

To summarize, the galaxy density distribution of A2163 shows:
\begin{itemize}

\item a main cluster component (A) elongated along an 
E--W axis on large scales, 
and a northern subcluster (B), visible both in optical and in X-ray.

\item several peripheral clumps; the two most significant ones 
(C) and (D) are located east and west of the cluster centroid 
respectively, on the same E-W axis defined by
  the large scale distribution of A2163-A. At faint magnitudes,
a southern clump (E) appears on the N--S axis joining (A) and (B). 

\item{} a bimodal morphology (clumps A1 and A2) in the central
(10'$\times$10') cluster field;

\item{} a significant counterclockwise twisting of the axis joining the 
highest density peaks of the
central clumps A1 and A2 when including fainter magnitude objects;

\item{} both the major axes of BCG1 and BCG2 and the axis joining them
lie on the same E--W axis of the cluster
large scale structure, which is seen both in the optical and X-ray maps;

\item{} a clear offset between a) the positions of the BCGs, b) the peaks in
the projected galaxy density maps and c) the X-ray intensity peak.

\end{itemize}

\subsection{Comparison of galaxy/gas density distribution}

The comparison between the galaxy and gas projected density
is important to characterize the dynamical state of the cluster,
 as the relative distribution of the collisional versus
non-collisional components of the cluster is indicative of its merger 
stage (Roettiger et al. 1997). 

\begin{figure*}
\begin{center}
\hspace{6mm}
\epsfxsize=16.0cm \epsfverbosetrue \epsfbox{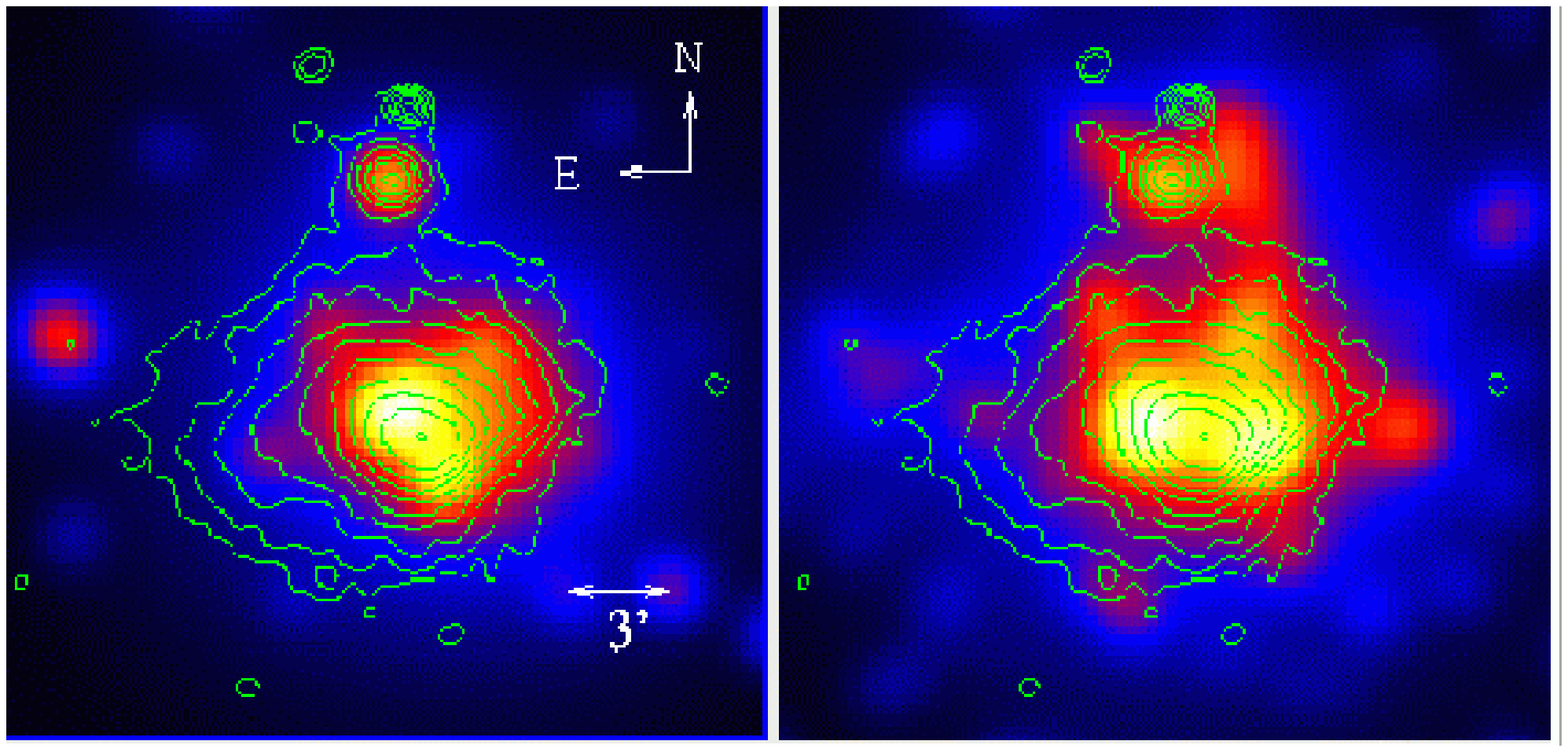}
\end{center}
\caption[]
{\sl 
X-ray isocontours superimposed on the projected density
maps of galaxies with $R < 19$ (left panel) and $R < 21$ 
(right panel). The X-ray contours derive from smoothed [0.5-2.keV] 
vignetted corrected XMM/EPIC images, and are logarithmically spaced 
by 0.2 dex, with the lowest contour at 4.65E-3 ct/s/arcmin$^2$ and the
highest one at 1.86E-1 ct/s/arcmin$^2$ }
\label{isodens_X}
\end{figure*}

As the projected galaxy density distribution changes with magnitude, 
we compare the two density maps limited at $R < 19$ and $R < 21$ 
(previously shown in Fig. \ref{isodens_mag}) 
to the XMM/EPIC images. Using archival XMM data 
(ID-0112230601, ID-0112230701, ID-0112230801, ID-0112230901 
and ID-0112231001) we have built a mosaic image  of the cluster 
in the [0.5-2.] keV energy band (all instruments and pointings summed 
together). The isocontours of the X-ray image superimposed on the 
galaxy density distribution are shown in Fig. \ref{isodens_X}.

While A2163-A and A2163-B are detected both in optical and in X-ray, 
the western and eastern clumps A2163-C and A2163-D are
not detected in X-ray. At both magnitude limits, the galaxy density
distribution in A2163-A is elongated and bimodal; 
the gas distribution is centrally peaked and  more regular, 
but it exhibits the same E--W elongation seen in 
the optical data at large scales.

In the case of bright galaxies ($R< 19$), the projected galaxy 
density distribution in the central regon is nearly aligned with the 
inner major axis of the X-ray map (NE/SW), and perpendicular to the 
compressed X-ray contours in the SW region which
corresponds to the position of the secondary optical clump A2.
At large scales, the gas distribution presents the same E--W 
elongation as the galaxy distribution at faint
magnitudes ($R < 21$) (with the E--W structure embedding
A2163-A, A2163-C and A2163-D).
As we have seen, faint galaxies follow the same E--W orientation also
in the cluster central region. 

Independently of the magnitude cutoff, the X-ray center position 
does not correspond to any of the two optical maxima, but is 
located between them; moreover, also the positions of the two BCGs do
not correspond to the maxima of the density distribution, neither of the 
galaxies nor of the gas.
These facts suggest that the gas is in a more relaxed state than the 
galaxies and that we are witnessing a post merger event in A2163-A. 
On the contrary, the distributions of bright galaxies and of gas 
in the northern component A2163-B are comparable, a fact which might suggest 
that A2163-B has not yet 
collided with A2163-A. We will discuss the merging scenario in section 7.

\subsection{Density profiles}

\begin{figure}
\begin{center}
\hspace{6mm}
\epsfxsize=8.0cm \epsfverbosetrue \epsfbox{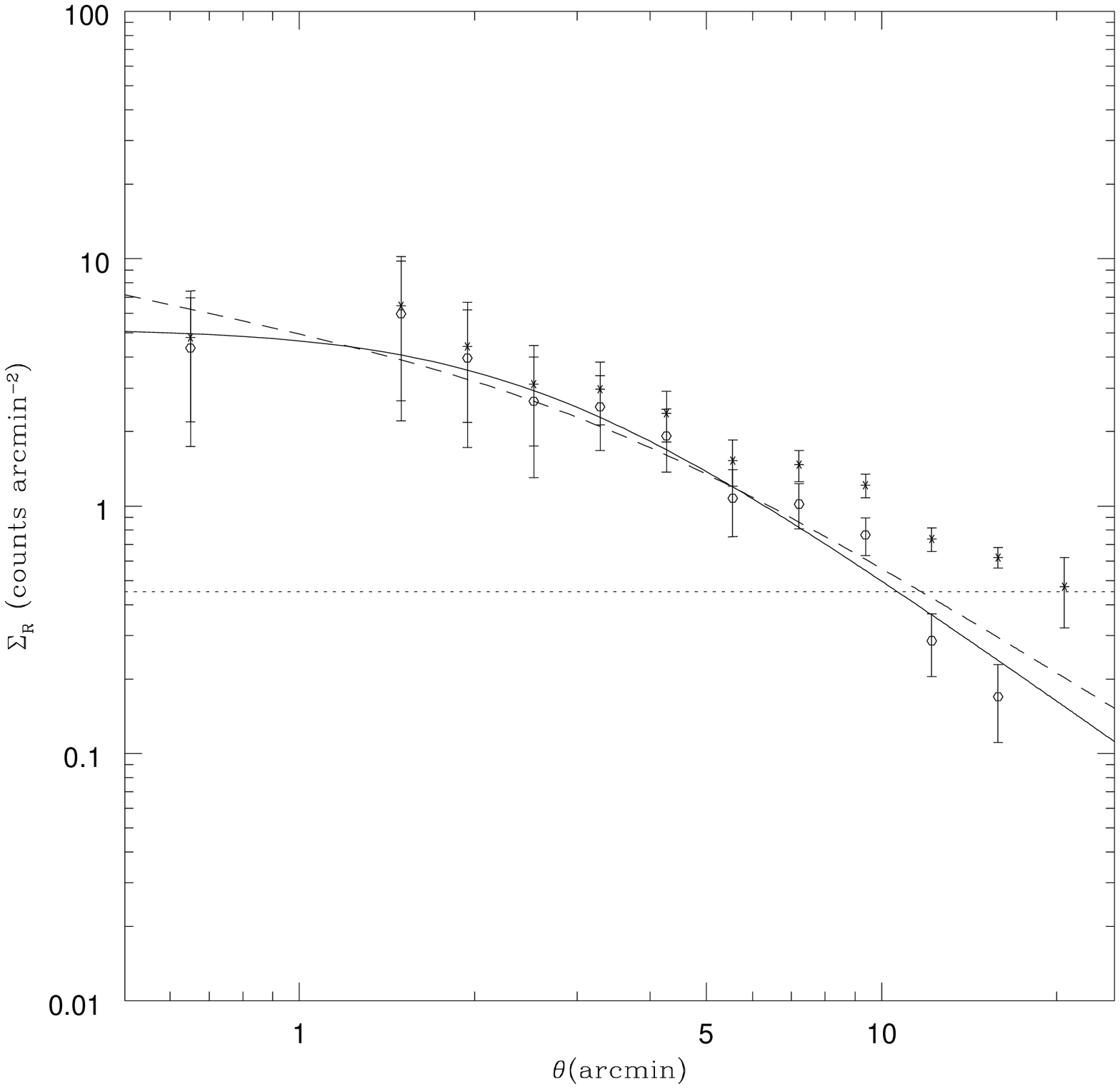}
\end{center}
\caption[]
{\sl Projected galaxy density profile in the central field of A2163.
Galaxies with $R\leq20$ belonging to the Red Sequence of the cluster
have been selected.
The raw profile (stars) and the background-subtracted
profile (open circles) are plotted. 
The horizontal short dashed line marks the estimated background value. 
The best-fit function with generalized beta and cusped
models are also shown with solid and dashed lines respectively.}
\label{dens_prof}
\end{figure}

We have estimated the projected radial density profile in the
$30'\times30'$ WFI field centered on the cluster. 
In order to correct for the areas which have been
masked due to the presence of bright saturated stars, 
we have generated a random catalog reproducing the same geometry as
the observed one. 

We have calculated the number of objects in concentric circles of radius
$\theta$ in the galaxy catalog, $N_{gal}(\theta)$, and in the random catalog, 
$N_{rand}(\theta)$, deriving the projected density profile $\sigma(\theta)$:

\begin{equation}
\sigma(\theta)=\frac{N_{gal}(\theta)}{N_{rand}(\theta)} 
\frac{N_{rand}^{tot}}{N_{gal}^{tot}}
\end{equation}

where $ N_{gal}^{tot}$ and $N_{rand}^{tot}$ are  the total number
of objects in the galaxy catalog and  in the random
catalog, respectively. We fixed $N_{rand}^{tot}= 200000$, a value
high enough to avoid introducing additive noise in the profiles.

We have fixed the cluster center at the position of
the X-ray centroid. 
The X-ray center is located between the local maxima A1
and A2, and roughly coincides with the centroid of the galaxy distribution
when smoothing over large scales. 
The profile was calculated for RS galaxies with $R < 20.0$, 
in order to minimize the contribution of the background.
The estimate of the background level (critical on large scales)
was done within a circular annulus comprised between 15 and 20 arcmin from
the cluster center. Apparently the background is reached near the edge of our
field.

The background subtracted profile was fitted with
a generalized beta model:
$\sigma(r) = \sigma_0 /(1+ (r/r_c)^2)^\beta$
and a generalized cusped profile:
$\sigma(r) = \sigma_0 /(r/r_c(1+ (r/r_c)^2))^\beta$, 
as defined in Adami et al. (2001),
using a Levenverg-Marquardt $\chi^2$ algorithm, which 
gives simultaneously the three parameters $\sigma_0$, $r_c$ and $\beta$.

The raw and the background subtracted radial profile with its best fits
are plotted in Fig.\ref{dens_prof}.
 Both the beta and cusped models provide a good fit to
the data up to $\sim 10$ arcmin, 
with $\chi^2$ values of 0.25 and 0.35, respectively (the most deviating
points --at $1 \sigma$ level-- are the last ones, 
where the uncertainty due to the background subtraction is larger).
For the beta model we have $\beta =0.83\pm 0.15$
and $r_c=2.5\pm 0.5$ arcmin, corresponding to 
$r_c=0.5 \pm 0.1$ Mpc (in our cosmology), 
which is a high value.
In the case of a cusped model, the best fit values are  $\beta =0.49\pm 0.10$
and $r_c=3.4\pm 0.3$ arcmin ($0.72 \pm 0.07$ Mpc). 
The values of $\beta$ are, within the errors, in agreement with those found by
Adami et al. 2001 on a sample of nearby  clusters 
($\beta_{beta~model}=1.0\pm 0.02 $ and $\beta_{cusped}=0.56\pm 0.01$), 
while the values of $r_c$ are significantly higher  
(the values in Adami et al. 2001, when converted to our cosmology, are
$r_c = 0.13 \pm 0.07$  Mpc for the beta model
  and $r_c=0.45\pm 0.05$ for the cusped model).
For comparison, the values for the gas profile are 
$r_c=1.2$ arcmin and $\beta=0.2$  (Elbaz et al. 1995). 
As expected from the density and isocontours maps (Fig. \ref{isodens_X}), 
the galaxy profile is less centrally 
peaked than the gas profile. The larger core radius found for the 
galaxies reflects the elongated, bimodal galaxy distribution in the 
center.

\section{Velocity distribution}\label{sec:vel}

\subsection{General behavior}

\begin{figure}
\begin{center}
\hspace{6mm}
\epsfxsize=8.0cm \epsfverbosetrue \epsfbox{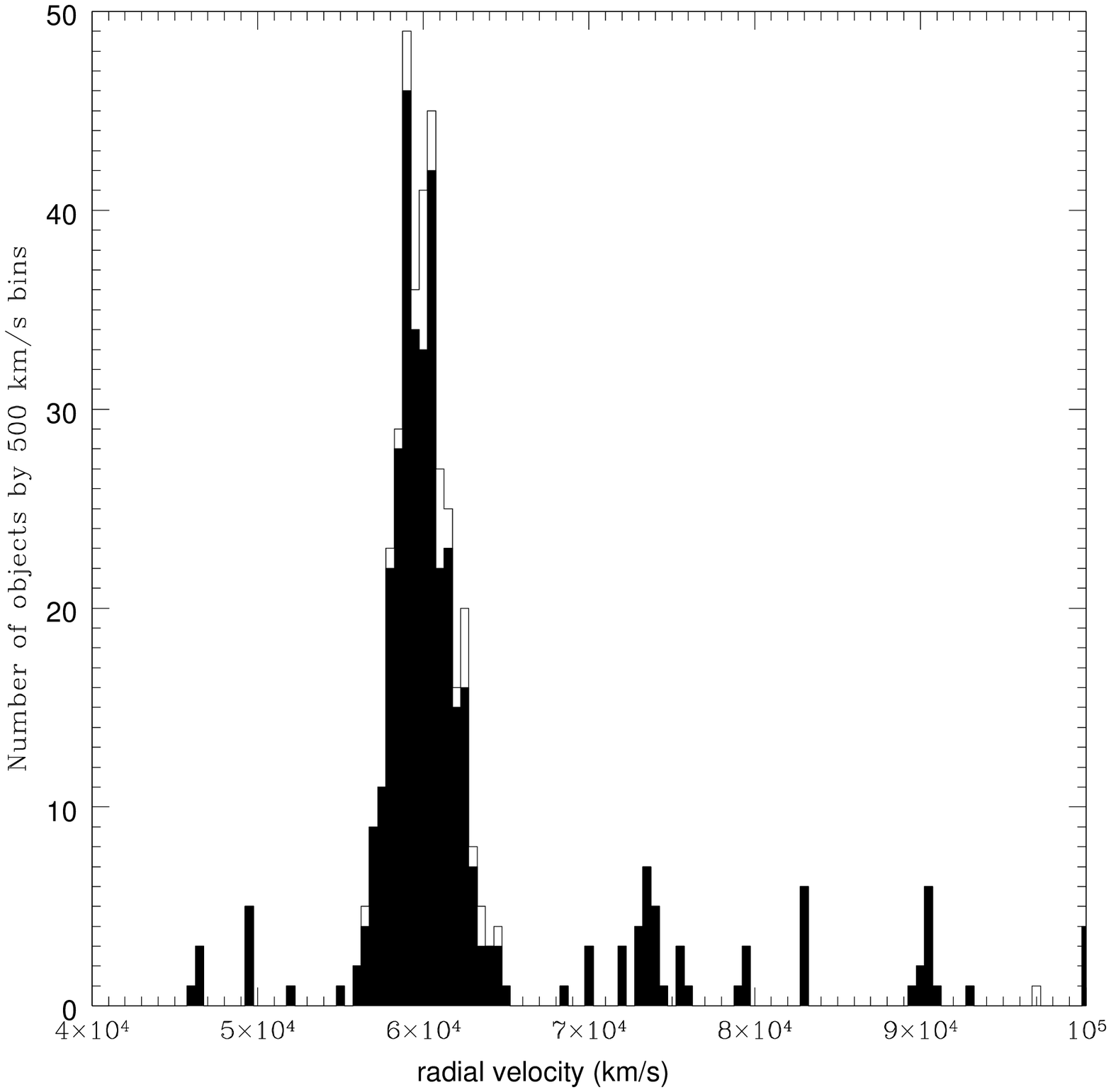}
\end{center}
\caption[]
{\sl Velocity distribution histogram in the WFI field centered on A2163. 
Filled histogram: high precision VIMOS spectroscopic
   catalog (430 galaxies). Empty histogram: total sample including 
also redshifts from CFHT spectra and low S/N VIMOS spectra (476 galaxies).}
\label{histobrut}
\end{figure}

In this section we analyze the spectroscopic high precision
 (Flag 0, 430 galaxies) and the total sample (Flags 0 and 1, 476 galaxies).
A visual inspection of the velocity histogram (Fig. \ref{histobrut})
clearly shows the main component of the cluster at
$\sim 60000$ km/s, while background overdensities are detected
at $\sim 74000$ km/s (18 objects), and more marginally at $\sim 90000$ km/s
(11 objects). See also the field image in Fig. \ref{A2163_spectro}, where 
galaxies in the velocity bins corresponding to
these three peaks are marked by different symbols.

\begin{figure*}
\centering
\resizebox{8cm}{!}{\includegraphics{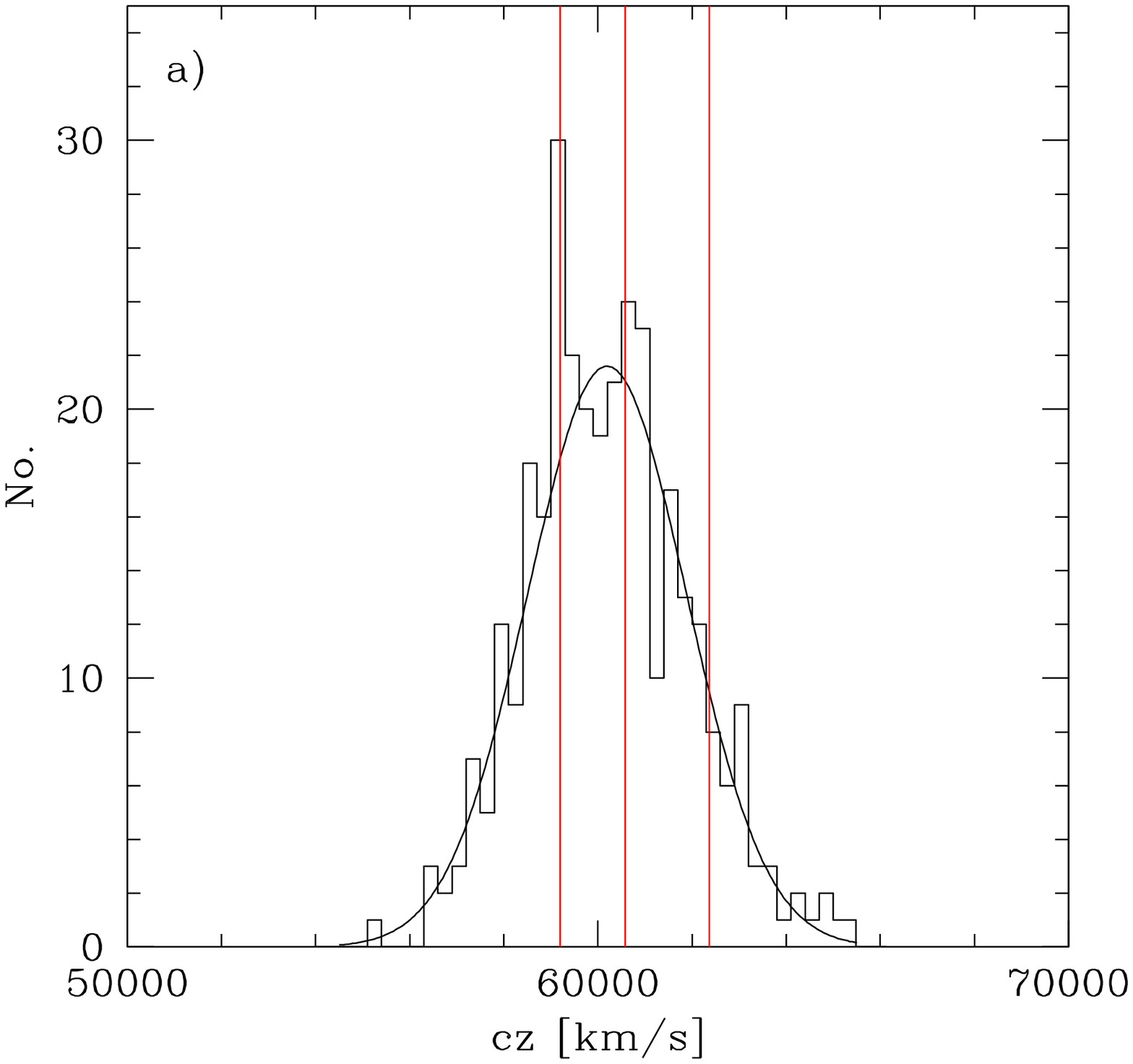}}
\resizebox{8cm}{!}{\includegraphics{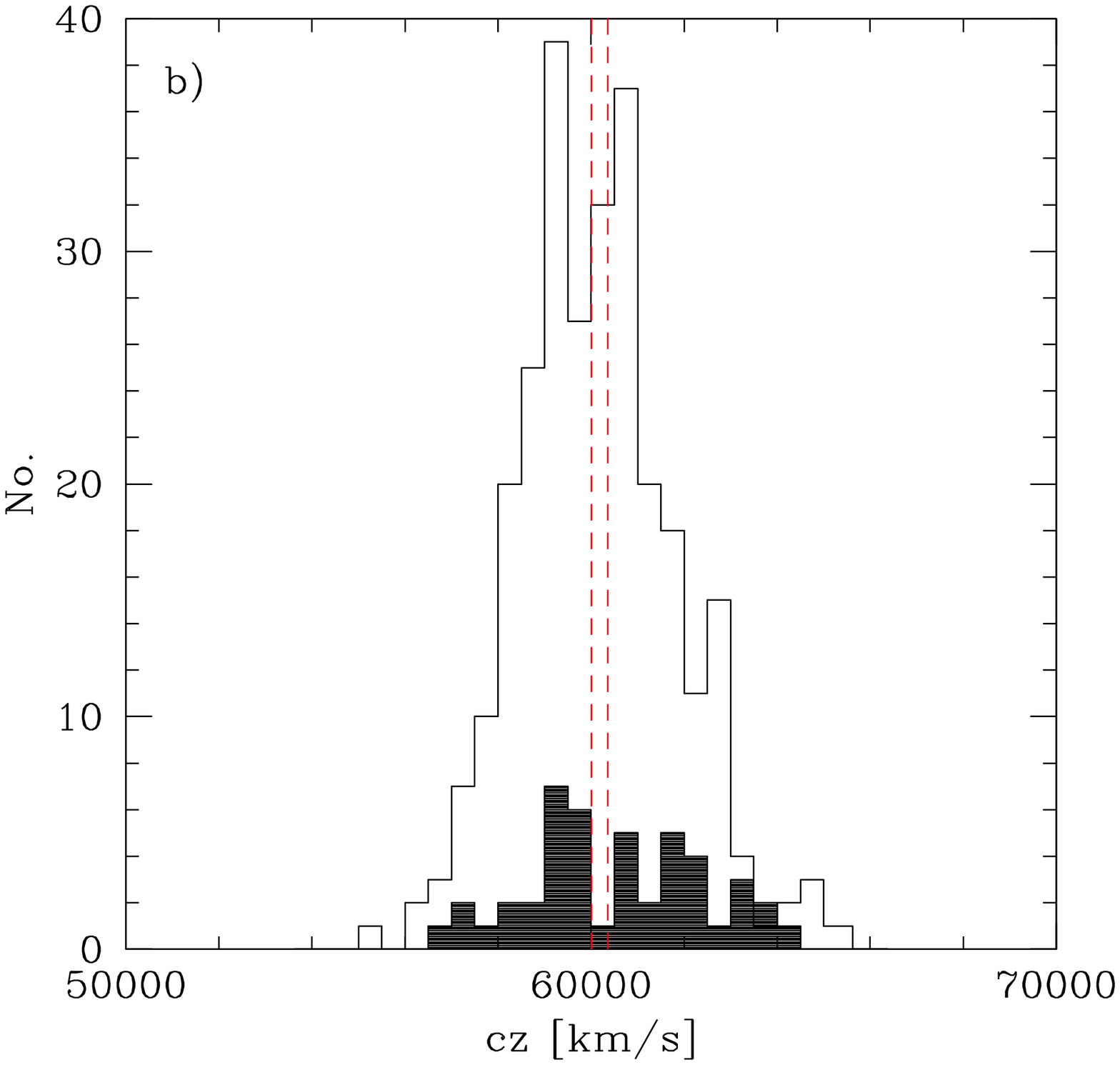}}
\parbox[b]{17cm}{
\caption{\sl a) Velocity histogram of the
   confirmed cluster galaxies in our high precision
   redshift sample (326 members), with the 
   Gaussian best--fit.  
   Bins of 300 km/s have been used.
   The thick solid lines correspond to the
   estimated mean velocities of the three groups identified by KMM. 
   b) Same sample as in a), but divided into
   galaxies without emission lines (empty histogram), and with
   emission lines (filled histogram). Bins of 500 km/s have been used. 
   The velocities of the two BCGs have been plotted as 
   red dashed lines.}
\label{histoV}
}
\end{figure*}
A critical point is the identification of cluster members and the exclusion
of interlopers (see e.g. den Hartog \& Katgert 1996). 
As A2163 is a rich cluster, we have to take into account its broad velocity
distribution.
Within $\pm 6000$ km/s from the main peak of the velocity distribution
at $\sim 60000$ km/s, we find 326 galaxies from the high precision (flag 0)
catalog.
This sample satisfies the criterion of
a maximum velocity gap of $1000$ km/s between adjacent members in velocity
space.
An alternative way to select cluster members is the classical 
three-sigma clipping method (Yahil \& Vidal 1977). 
We have applied it to all flag 0 galaxies
within the redshift range $0.18 \le z \le 0.22$, obtaining again the same 
final sample of 326 galaxies. We have also checked the velocity of these 
galaxies as a function of their projected distance from the cluster center, 
finding no significant outliers.
Applying the same analysis to the total (flag 0+1) sample,
we find 361 cluster members.
These are our reference catalogs of cluster members, which were 
used to study the kinematical and dynamical properties of A2163

Applying the program ROSTAT (Beers et al. 1990),
we find very stable values for the cluster mean velocity and velocity 
dispersion calculated with the different methods.
In Table \ref{ROST_Tab} we give the biweight estimates of the
location ${C}_{\rm BI}$ and scale ${S}_{\rm BI}$, 
which are better estimators than the classical mean and standard deviation, 
for different subsamples.
The total cluster sample with 361 galaxies and the high-precision one with
326 galaxies have comparable values of location and
scale, with a huge value of velocity dispersion ($\sim 1430$ km/s). 
The subsample including only emission line galaxies shows 
--not surprisingly-- a higher velocity dispersion, 
with a value of $1564^{+152}_{-152}$ km/s;
its mean velocity is marginally 
higher than the mean velocity of the no emission and total cluster samples,
with a velocity offset 
$\Delta V = 400\pm 280$ km/s.
The velocity distribution of these objects (see Fig. \ref{histoV}) is
very broad. Their projected spatial distribution (Fig.\ref{4slices}) 
shows that most of them lie in the cluster outskirts.
These results are expected for late-type galaxies
infalling on the main cluster. However, several emission line 
galaxies are found within the cluster core, which is quite unusual.  
We have examined in more detail the position 
of the emission line galaxies in the field, by dividing the 30'x30'
field in individual cells of 10'x10',
and estimating the fraction of emission line cluster members 
in each cell. The mean fraction in the whole field 
is $\sim 13 \%$. There is a systematic
increase of the fraction from the central cell, where it has the minimum value
of $\sim 6.5 \%$, to the lateral cells where the mean
value is $\sim 11 \%$. However, in the western region of the cluster it
reaches the high value of $\sim 30 \%$. This corresponds to the excess shown 
in Fig. \ref{4slices} in the immediate surroundings of A2163-D.

\begin{figure*}
\centering
\resizebox{16cm}{!}{\includegraphics{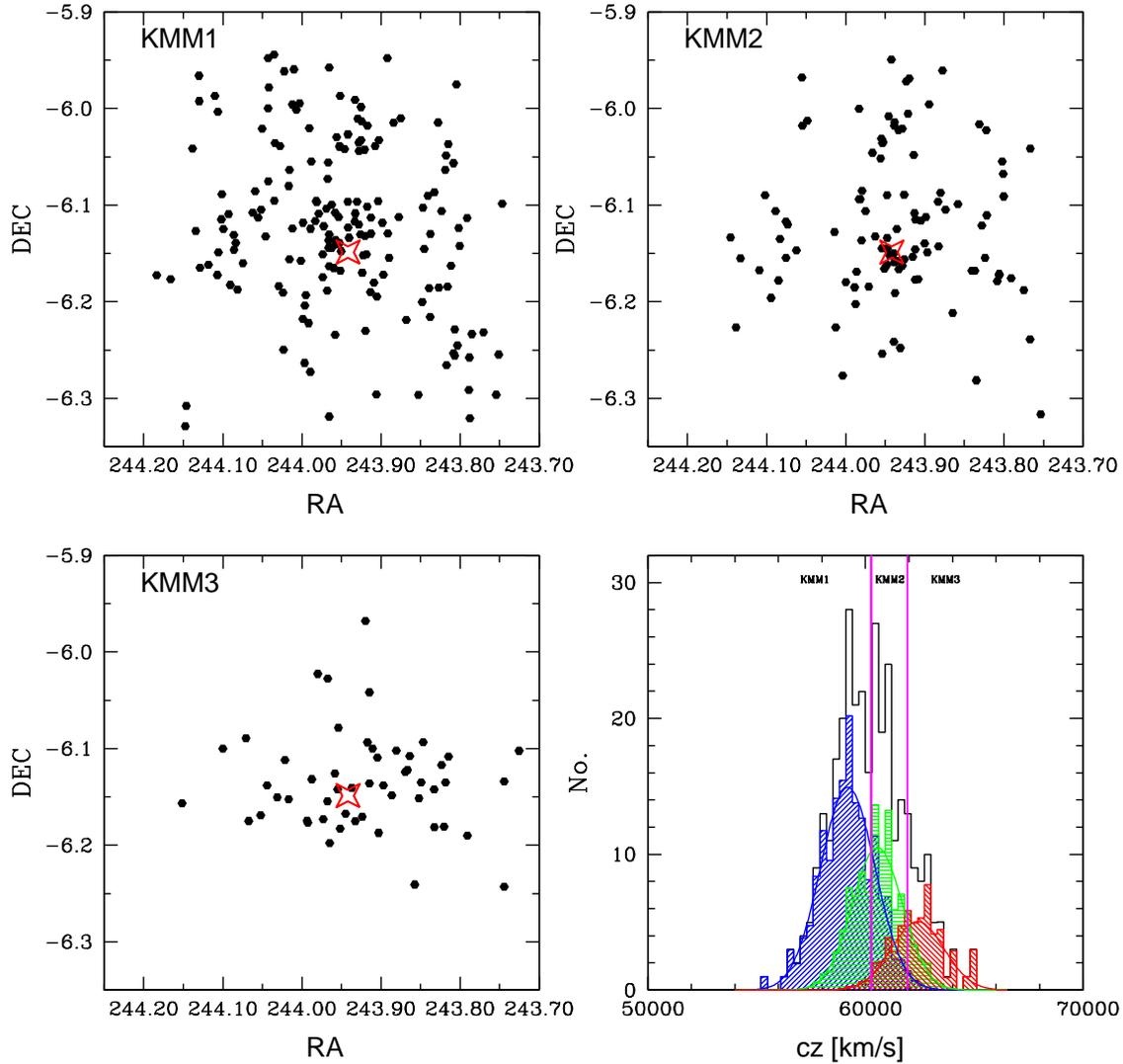}}
\parbox[b]{16cm}{
\caption{\sl Projected distribution of galaxies
and velocity histograms (with a binning of 300 km/s)
for the three KMM partitions of A2163. The open star indicates
the peak of the X-ray emission.
The bottom right panel shows the global velocity histogram (in black)
and the expected histograms for KMM1 (blue) KMM2 (green) and KMM3 (red)
with their respective Gaussian fits.
The two magenta vertical lines represent the velocity limits $V_1$ and $V_2$
of the three partitions. Galaxies with velocity $< V_1$ are associated
to KMM1; galaxies with velocity $> V_2$ are associated to KMM3; galaxies
in the intermediate range are associated to KMM2 (see text).}
\label{fig:kmmD}
}
\end{figure*}

When excluding emission line galaxies, 
the cluster velocity dispersion slightly decreases but still has 
a high value: $1403^{+63}_{-63}$ km/s.
However, our estimate could be affected by substructures. 
For example, a visual
inspection of the histogram in Fig.~\ref{histoV} 
shows a bimodality in the central part of the velocity distribution,
with two peaks at $\sim 59000$km/s and $\sim 60500$ km/s; it also
suggests a lack of galaxies (with respect to the Gaussian fit) at 
$\sim 61000$~km/s, and the presence of a third peak at $\sim 63000$ km/s.
It is not clear how much
these deviations from a Gaussian distribution are real or 
artifacts of sampling, and in the following section 
we will address the details of the
velocity distribution of the cluster with the appropriate statistical tools.

The velocity histogram suggests the presence of a
background structure at $\sim 74000$ km/s.
Selecting all galaxies between
$71000$ km/s and $77000$ km/s (18 objects, 
marked with blue triangles in Fig. \ref{A2163_spectro}), 
and applying ROSTAT, we
find $C_{BI}=73724 \pm 182$ km/s and
$S_{BI}=848_{-242}^{+243}$ km/s.
These values would be consistent with
a background cluster seen on the same line--of--sight of A2163,
but no significant concentration is seen in the projected distribution
of the 18 galaxies (see Fig. \ref{A2163_spectro}).
Moreover, most of these galaxies (13 over 18)
show emission lines. These facts suggest that
the background overdensity is probably due to field galaxies within
a large--scale structure and not to a cluster: maybe a large scale
filament spreading from the SE to the NW of the field.

Finally, there is another small peak in the velocity histogram 
at $\sim 90000$ km/s, for which we have 11 redshifts.
As in the previous case, the projected distribution of these galaxies
(green diamonds in Fig. \ref{A2163_spectro}) 
does not show any significant density concentration. The objects seem to
populate a strip crossing the field along the N-S direction, and half 
of them
present emission lines, suggesting again a population of field galaxies. 
More redshifts would be necessary to identify possible structures.

\subsection {Statistical analysis of the velocity distribution}\label{sec:KMM}

In order to characterize the dynamical state of A2163, we performed
quantitative tests which check if the velocity distribution 
can be reproduced by one or a combination of Gaussian functions,
applying them to the high precision spectroscopy
sample (326 objects). Among the 15 normality 
tests of the ROSTAT program (Beers et al. 1990), only two
(the B2 and KS tests) reject the Gaussian hypothesis at a 
significance level $\lesssim$10\%.

\begin{figure*}
\centering
\resizebox{6cm}{!}{\includegraphics{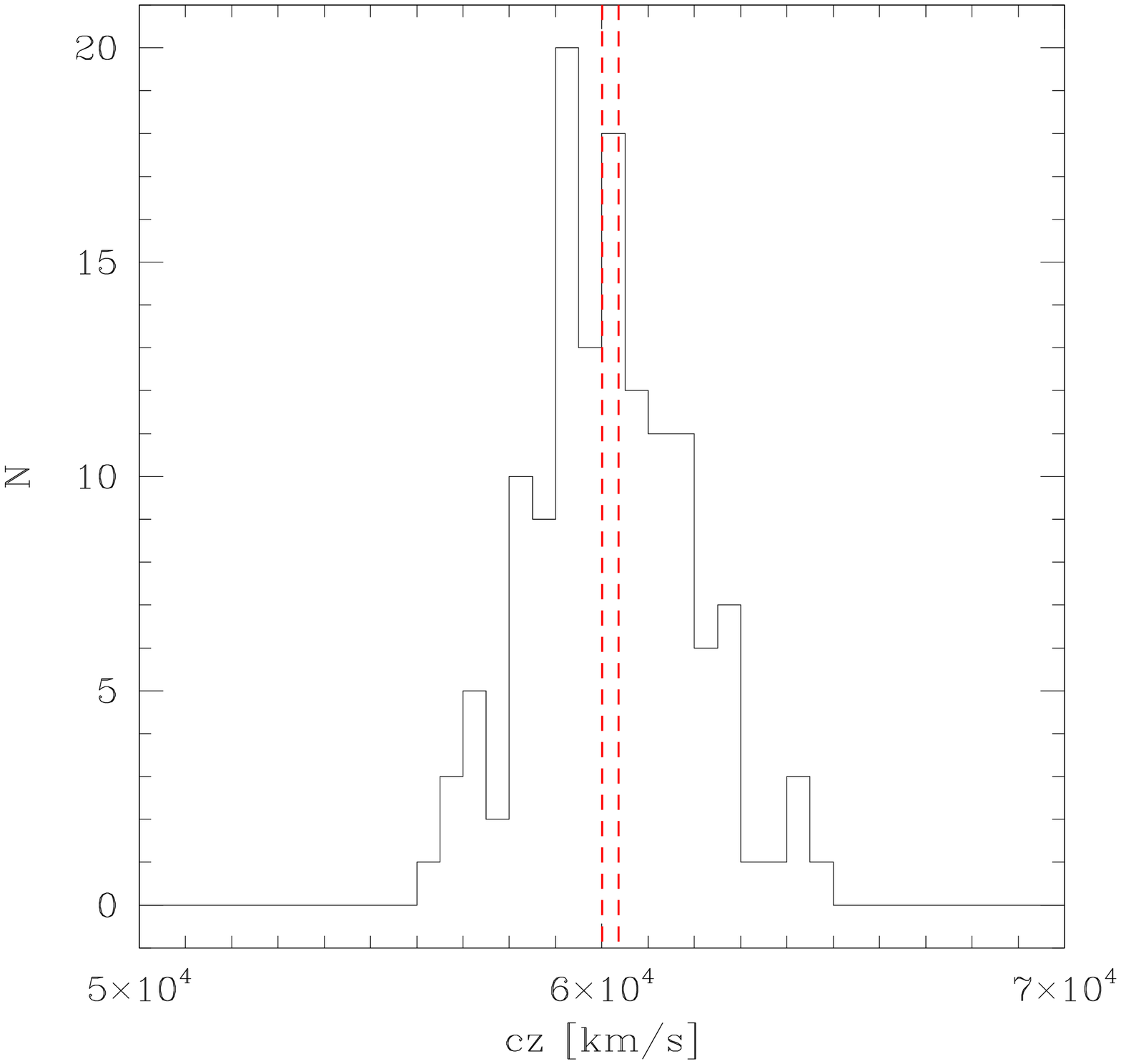}}
\resizebox{6cm}{!}{\includegraphics{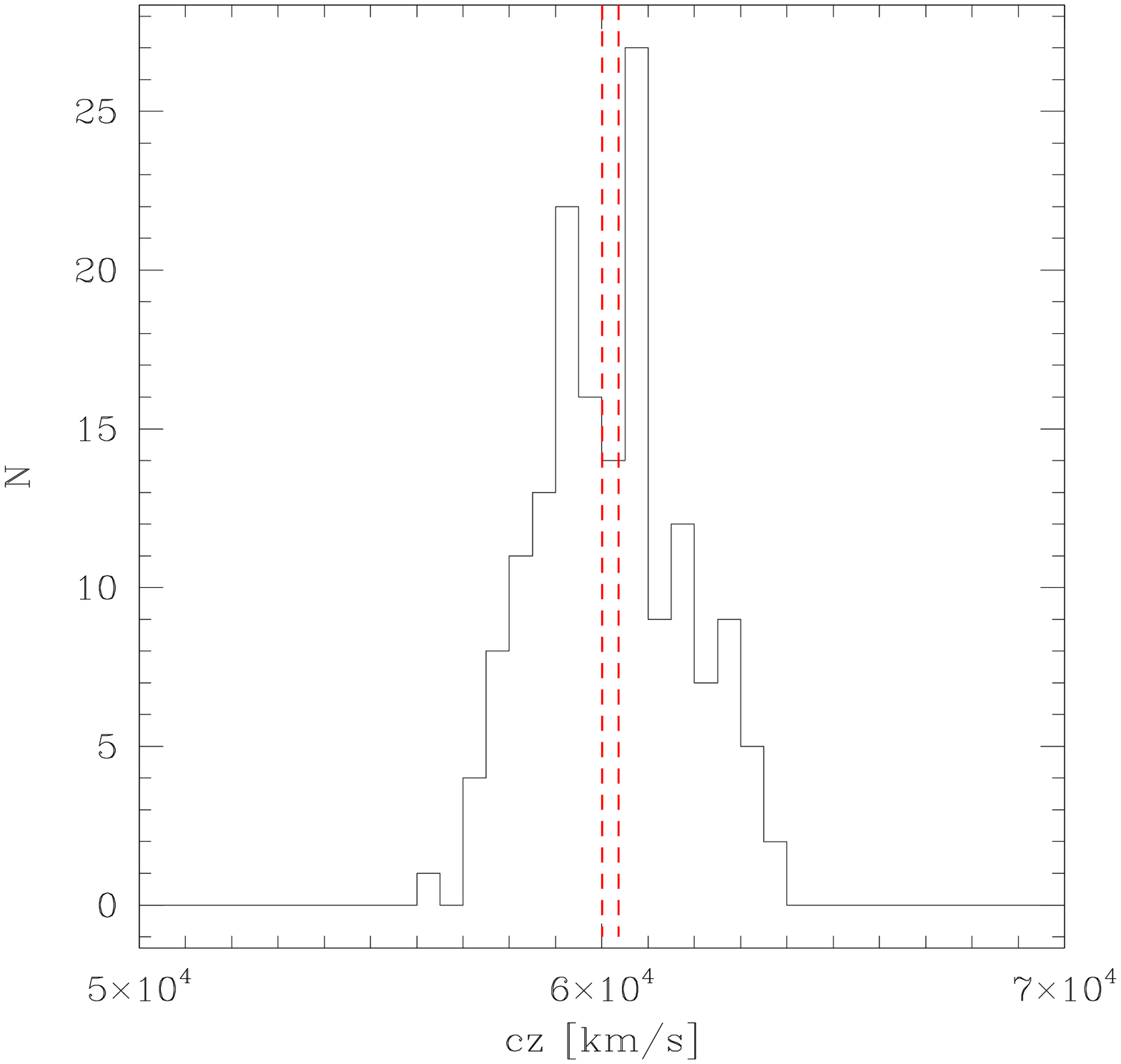}}
\resizebox{6cm}{!}{\includegraphics{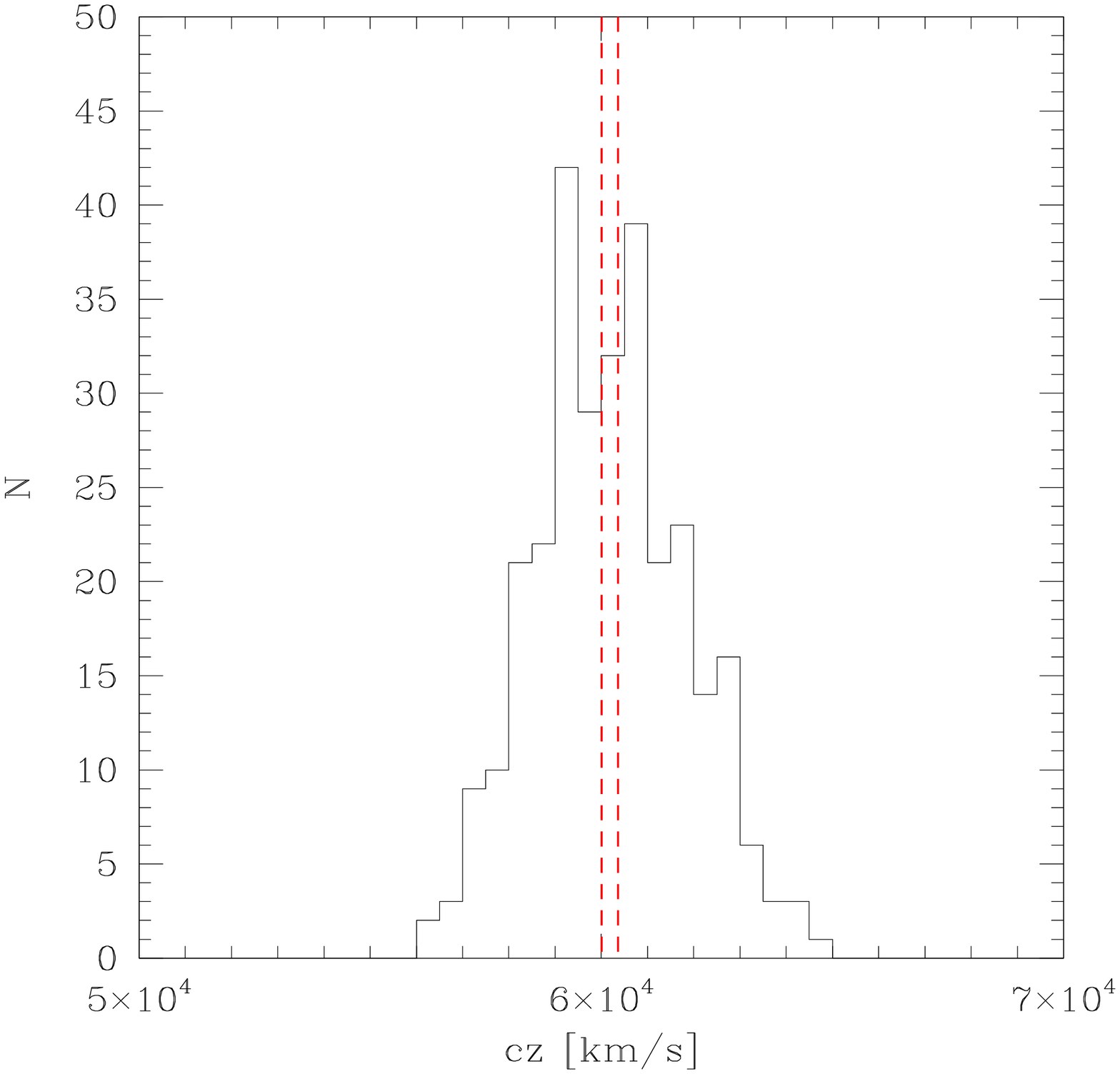}}
\resizebox{6cm}{!}{\includegraphics{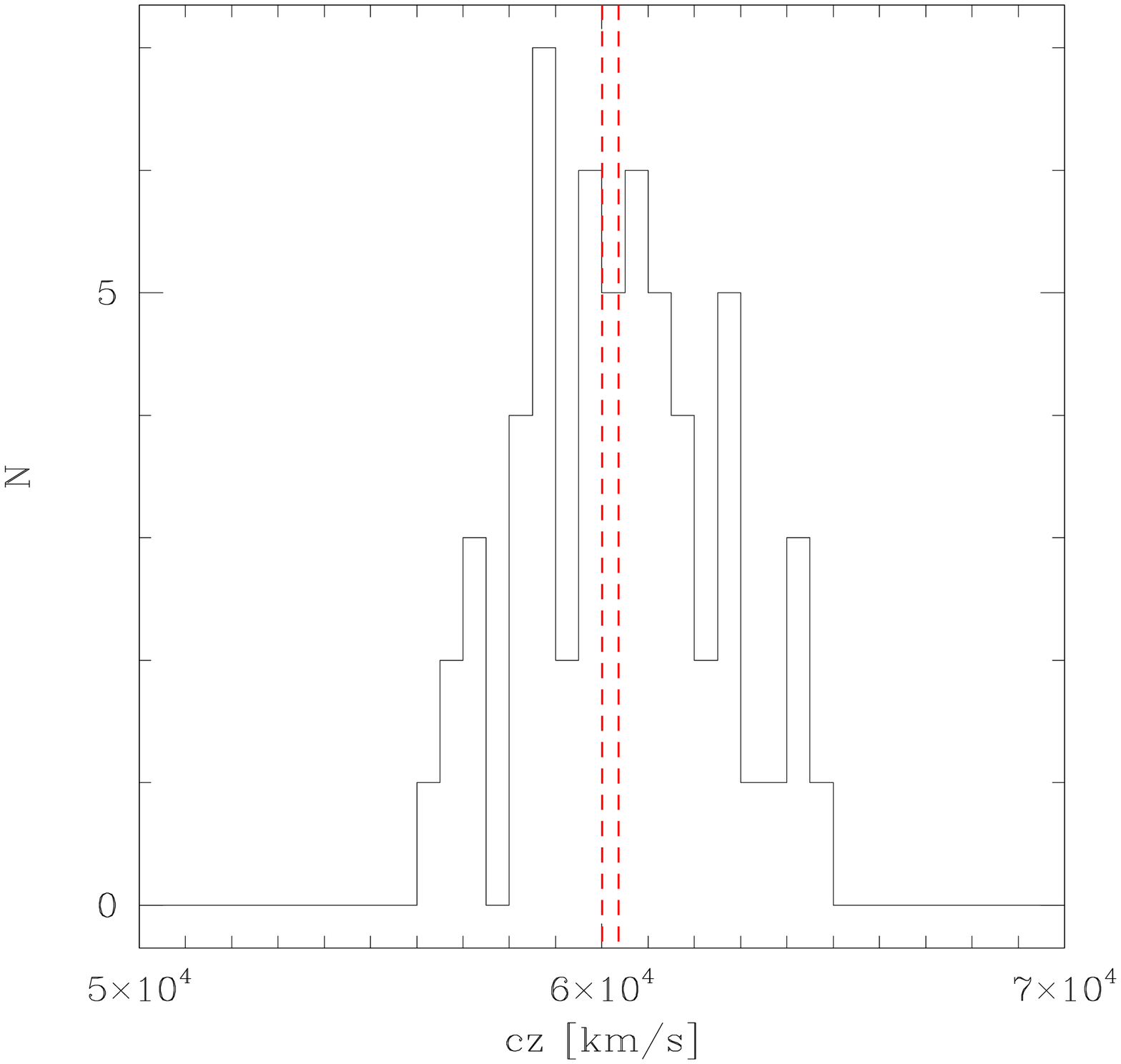}}
\resizebox{6cm}{!}{\includegraphics{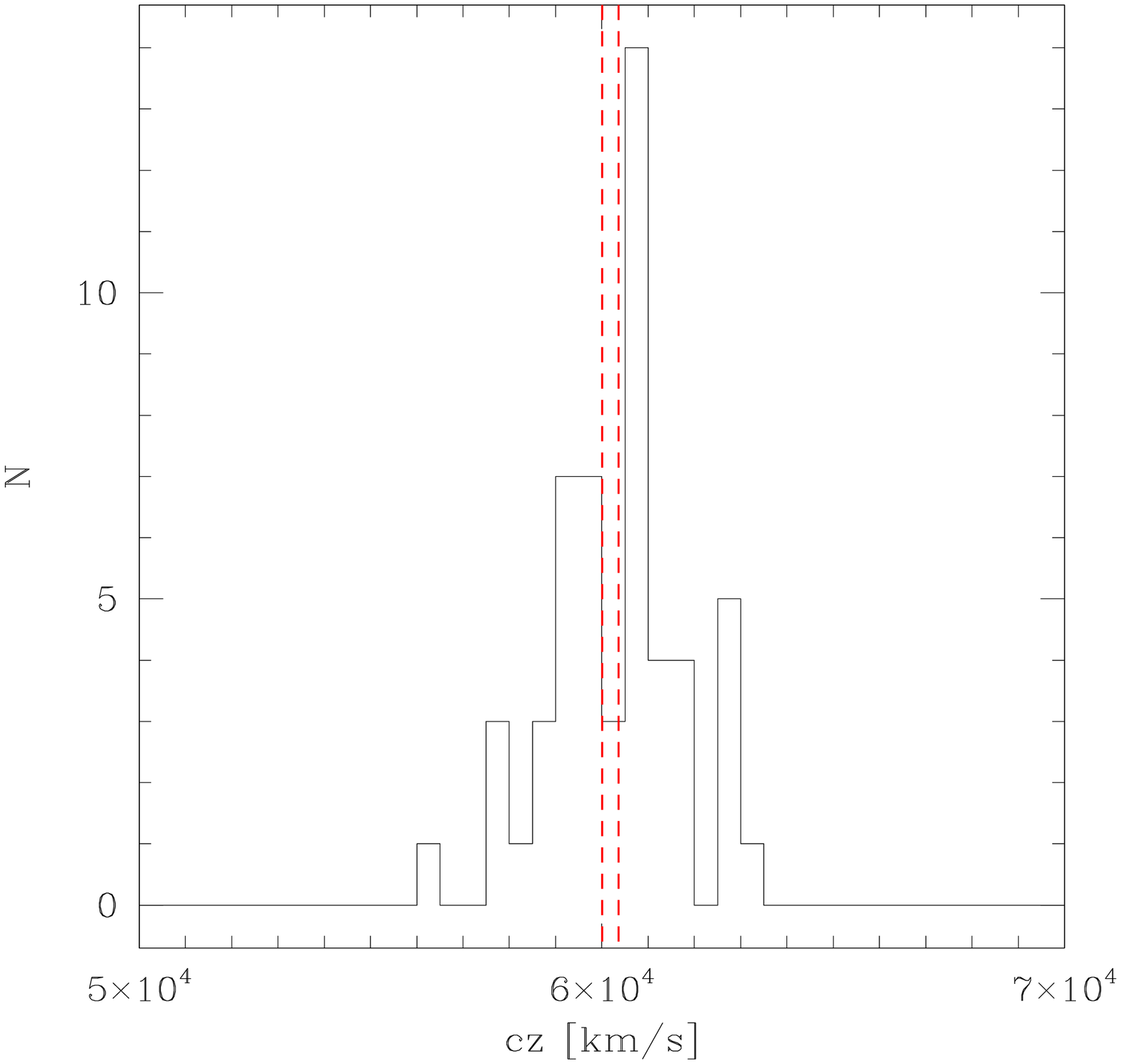}}
\resizebox{6cm}{!}{\includegraphics{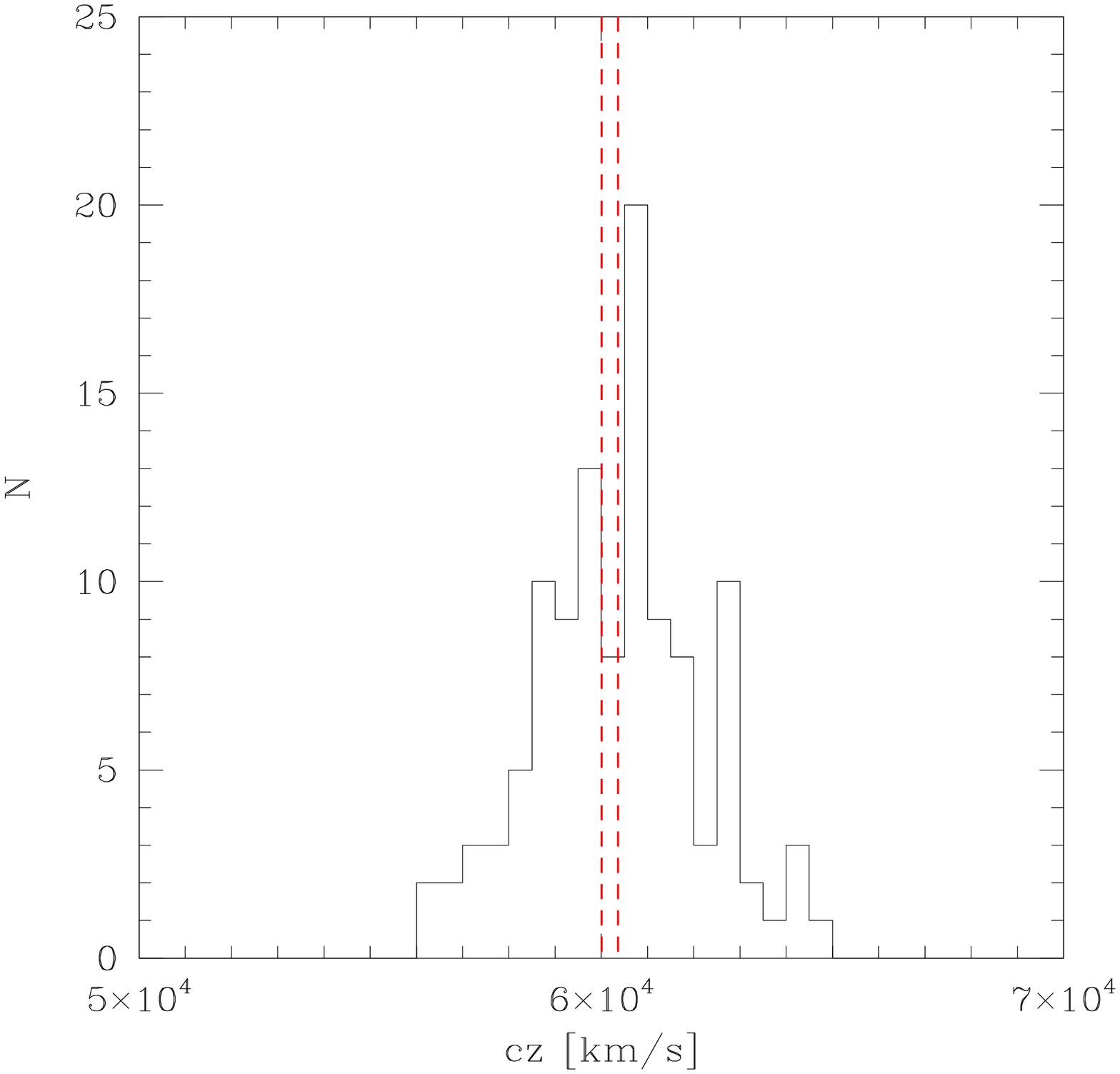}}

\caption{\sl Velocity histograms as functions of luminosity and field size.
Top row: whole field; bottom row: 
circular region within a 5 arcmin radius centered on A2163-A. 
Left column: ``Bright'' ($R \le 19$) galaxies; central column: 
``faint'' ($19 < R \le 21$) galaxies; right column: the whole
magnitude range $R \le 21$.
The velocity of the two BCGs are plotted as red dashed lines.
}
\label{histo_lum}
\end{figure*}

We also used two kinds of shape estimators (the traditional third and
fourth moments, i.e. skewness and kurtosis, and the asymmetry and tail
indices (Bird \& Beers 1993) in order to test the null hypothesis
of a Gaussian velocity distribution. None of them shows strong
deviation from a Gaussian velocity distribution, with the exception of
a slight indication of positive (i.e. above the mean velocity)
asymmetry in the whole sample given by the skewness parameter.  
An indication of possible deviation from Gaussianity is the 
presence of
gaps in the observed velocity distribution (Beers et al. 1991). 
We detected five significant weighted gaps; two of them are
around the mean radial velocity of the cluster (60131 km/s).

In order to quantify the deviation of the velocity distribution
from a single Gaussian,
we have applied the Kaye's Mixture Model algorithm (KMM, McLachlan
\& Basford 1988) in the implementation of Ashman et al. (1994).
This algorithm fits a given number of Gaussian distributions to the data,
calculating the maximum likelihood values for the mean and the variance,
and evaluating the improvement with respect to a single Gaussian fit.
In particular, it gives as output the $P$--value, which
is the probability of measuring the observed value of the 
likelihood ratio test statistic for a sample drawn from a single 
Gaussian parent population: the null hypothesis of a single Gaussian
parent population is conventionally rejected if $P < 0.05$ and marginally
inconsistent if $0.05 < P < 0.1$.
On the basis of the previous discussion, we have tried to fit two and three
Gaussian components to our data.
In both cases, the KMM does not reject the null hypothesis. 
For example,
the two--component fit identifies the first group at a mean velocity
of $59517$ km/s and the second one at 61729 km/s, with a $P$--value of 0.17,
assuming the homoskedastic case, i.e. the same velocity dispersion for the 
different components. 
Moreover, the two-Gaussian fit fails to detect the peak at $\sim 60500$ km/s.
This is not surprising, as the velocity separation between the
first two peaks is comparable to the velocity
dispersion of the groups (theoretically two equal Gaussian distributions can
be separated if their peaks have a separation $\ge 2 \sigma$, Everitt \& 
Hand 1981).
However, when applying a three--component Gaussian fit to the
data (assuming as first guesses the peaks at 59000 km/s, 60500 km/s 
and 63000 km/s), 
the KMM identifies the two central peaks around the mean cluster 
velocity, with a third component at $\sim 62400$ km/s
(see Table \ref{ROST_Tab}). While such groups are not statistically 
significant in the velocity space, we use the KMM results with three Gaussians 
as an objective 
partition of the data and a starting point for the following combined analysis,
which makes use of the projected positions.

The mean velocities of the three fitted groups are shown with
thick lines in the left panel of Fig.~\ref{histoV}, and the velocity
distribution properties of the three partitions are listed in Table 
\ref{ROST_Tab} (notice that
the relative richness of the KMM groups is not very robust;
for example, in the homoskedastic case, i.e. imposing the same velocity
dispersion for the three components, KMM2 becomes the richest one).
The two main partitions KMM1 and KMM2 correspond
to the central peaks in the velocity histogram and have an 
offset in velocity of $\sim 1360$ km/s 
(corresponding to a physical $\Delta V \sim 1130$ km/s 
in the reference system of the cluster), while KMM3 includes the 63000 km/s 
peak but is not centered on it.

The spatial and velocity distributions of the three KMM partitions are
shown in Fig.~\ref{fig:kmmD}. For each galaxy, 
the KMM algorithm estimates the probability of belonging to the
different partitions, associating the galaxy to the partition with
the highest probability. As a result,
all galaxies below a ``critical''
velocity $V_1$ are associated to KMM1, all galaxies above
a critical velocity $V_2$ are associated to KMM3,
and those between $V_1$ and $V_2$ are associated to KMM2.
In order to reconstruct the velocity distributions 
without artificial cutoffs,
for each partition we have summed up the corresponding probabilities 
of all the galaxies in each velocity bin. 
The fourth panel of Fig.~\ref{fig:kmmD} shows $V_1$ and $V_2$
as vertical lines and the reconstructed velocity distributions
of the three partitions with the corresponding Gaussian fits.

\begin{figure*}
\begin{center}
\hspace{6mm}
\epsfxsize=15.0cm \epsfverbosetrue \epsfbox{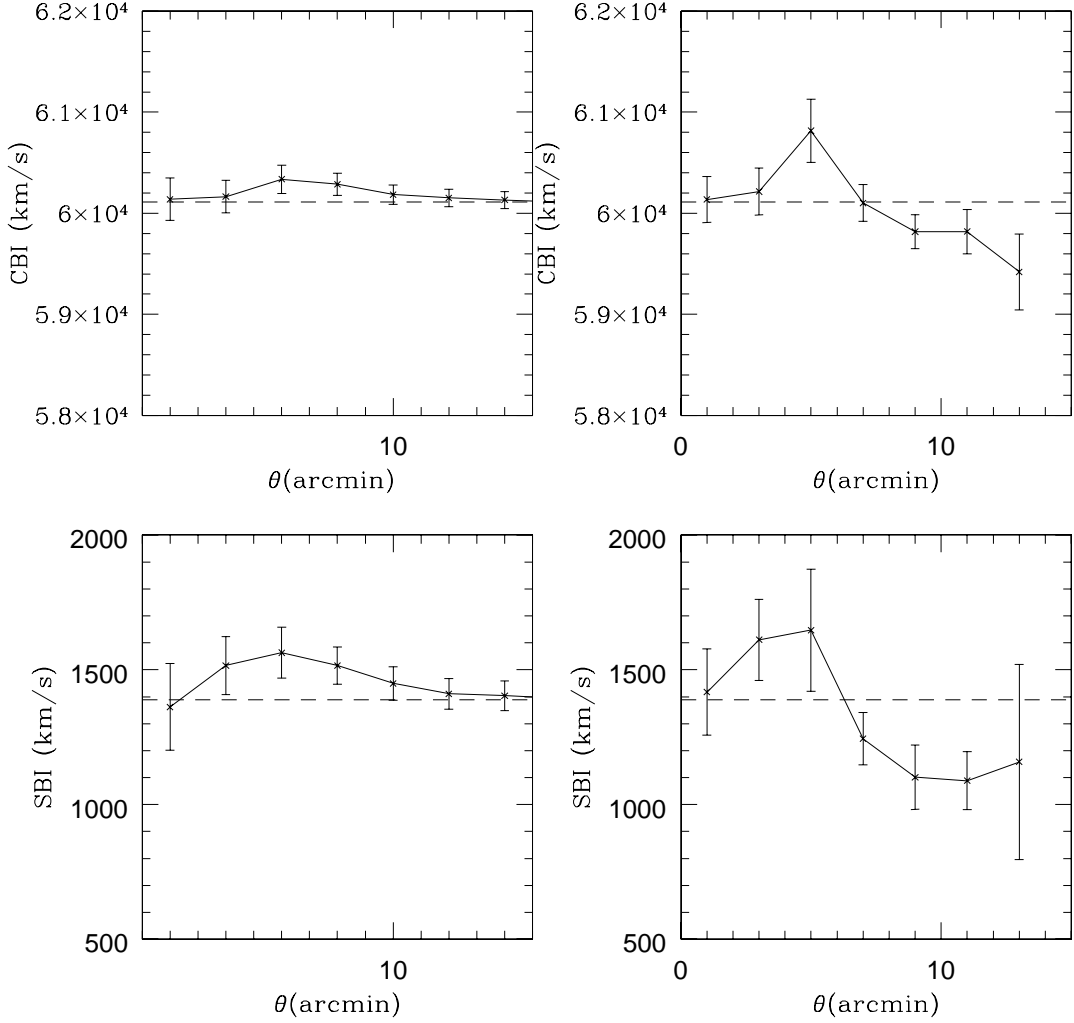}
\end{center}
\caption[]
{\sl Mean velocity (top) and velocity dispersion (bottom) profiles. The 
 plots on the left correspond to integrated profiles, while the ones on the
 right to differential profiles. 
The error bars correspond to $1\sigma$ errors calculated on 
10000 bootstrap catalogs. 
Horizontal short dashed lines mark the estimated values 
of the mean velocity and velocity dispersion, respectively, for the
whole sample.}
\label{velprof}
\end{figure*}

\begin{figure*}
\begin{center}
\hspace{6mm}
\epsfxsize=8.0cm \epsfverbosetrue \epsfbox{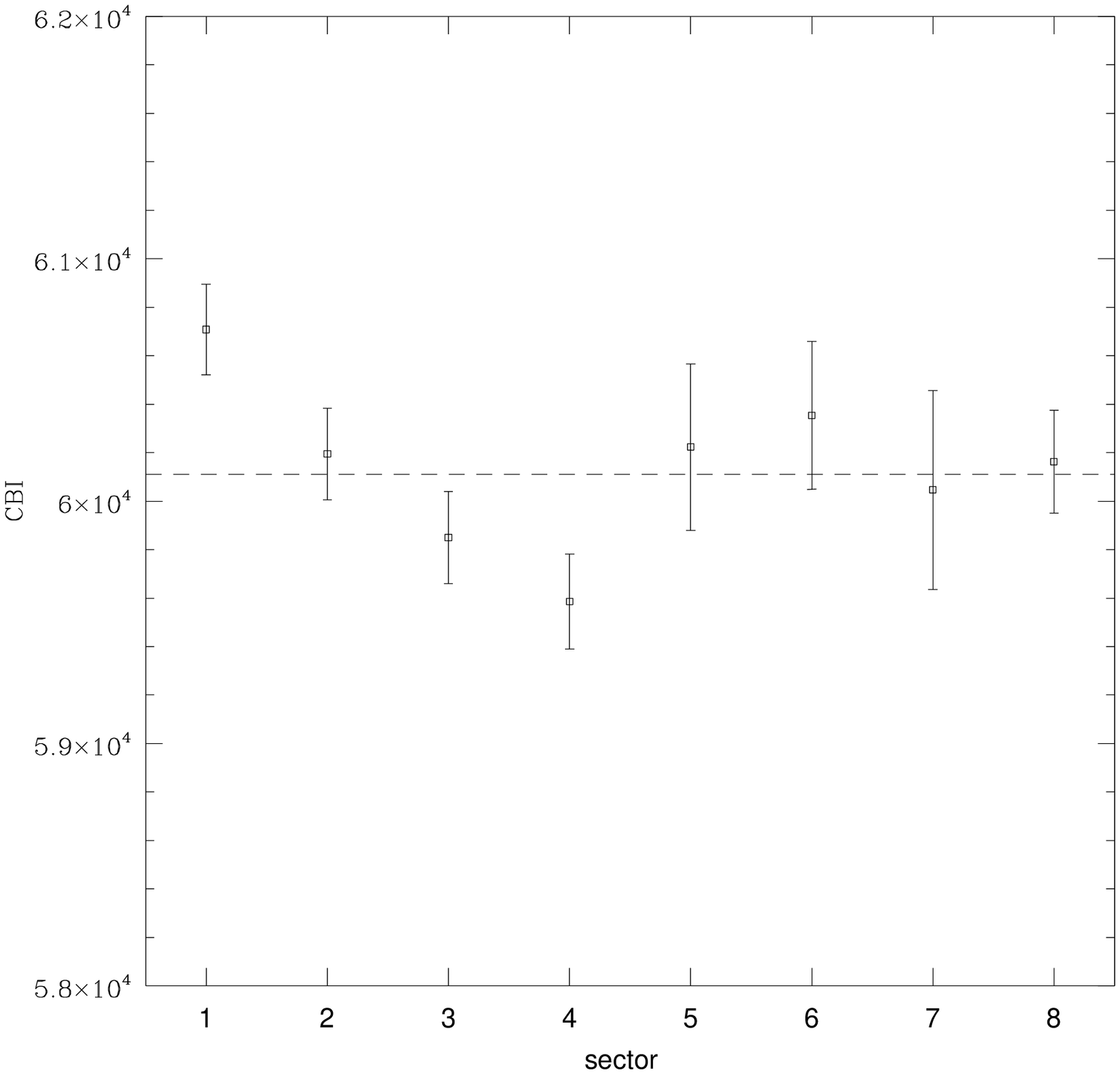}
\epsfxsize=8.0cm \epsfverbosetrue \epsfbox{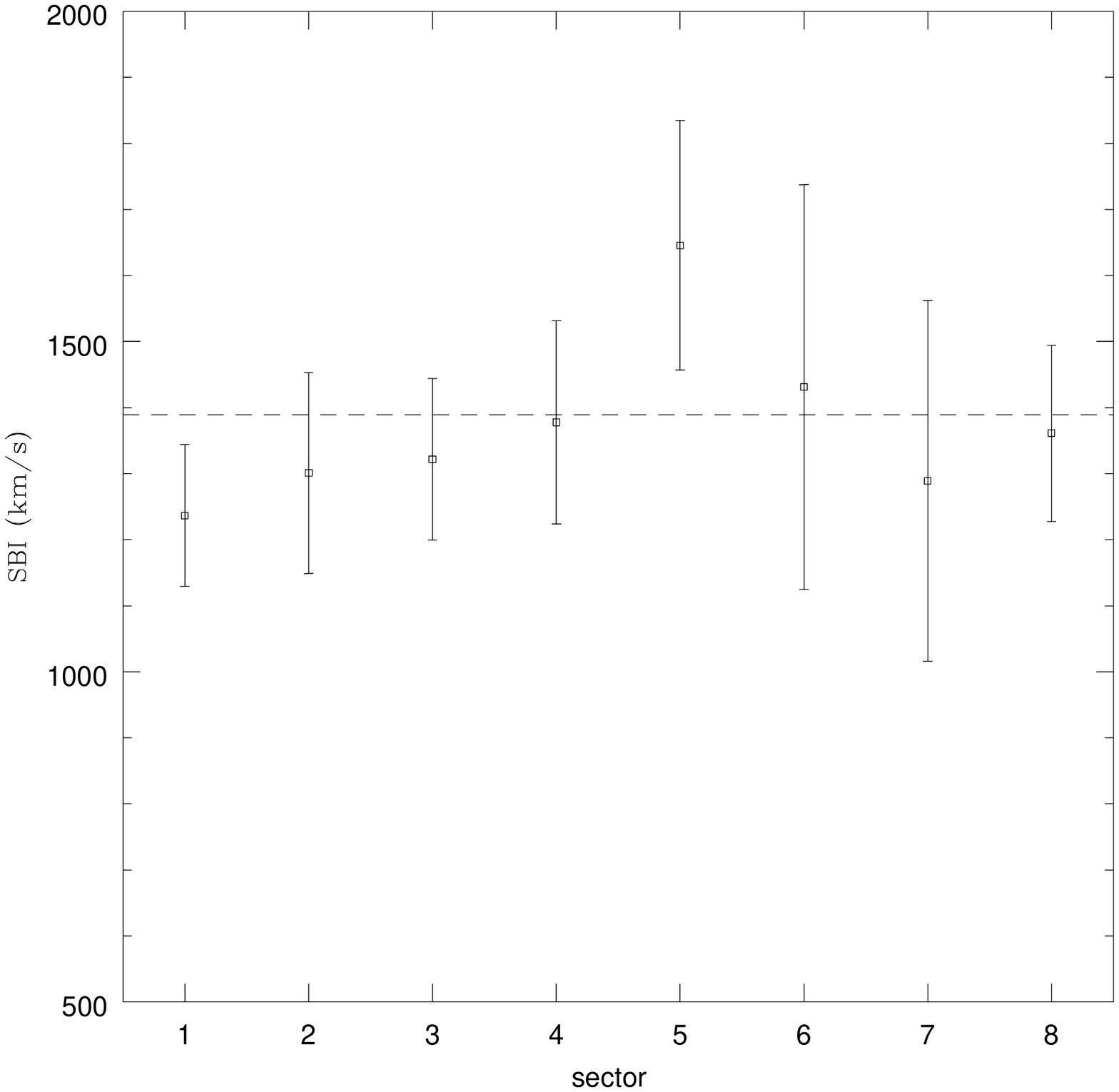}
\end{center}
\caption[]
{\sl Mean velocity (left) and velocity dispersion (right) estimated in 
45 degrees sectors. These are labeled from 1 to 8 anti-clockwise. 
With this notation, sector 1 corresponds to 0-45 degrees, 
sector 2 to 45-90 degrees, etc...}
\label{velsect}
\end{figure*}

From the projected distribution of the galaxies displayed in 
the first three panels of Fig.~\ref{fig:kmmD}, 
an offset between the spatial distributions of the
two main velocity partitions is apparent along the NE/SW direction  
: KMM1 objects are more concentrated
towards the NE side of the peak of X-ray emission (open star in
Fig.~\ref{fig:kmmD}), while KMM2 objects show a more clumpy
distribution, the main concentration being located more towards the central-SW
region of the cluster compared to KMM1 galaxies. Applying the
Kolmogorov-Smirnov test indicates that the probability that the two
distributions are issued from the same one is excluded at more than
$99 \%$ confidence level. A more detailed comparison of the spatial
distributions of KMM1 and KMM2 objects will be addressed in
Sect.~\ref{sec:slice}. Galaxies in the KMM3 partition are more 
concentrated in the central region of the cluster, tracing the elongation
in the EW direction. This could be due to
projection effects, since infalling field galaxies, characterized by
radial orbits towards the cluster center, have a higher radial
component of their proper velocity when observed in projection at the
cluster center. Indeed galaxies in the lower tail of the velocity
distribution of A2163 show a very similar spatial distribution.
Nevertheless, we cannot exclude that the small excess of galaxies
in KMM3 is a group at a higher radial velocity with respect to the cluster
mean.

\begin{table*}
\begin{center}
\begin{tabular}{ccccrrr}
\hline
\hline
Field &Population  & Flag &R  & ${C}_{\rm BI}$ & ${S}_{\rm BI}$  
& $N_{\rm gal}$\\
 & &  &[$km~s^{-1}$] &  [$km~s^{-1}$]\\
\hline
\multicolumn{7}{c}{ }\\
30'x30' &all &0+1 &$R<22$ & $60217 ^{+82} _{-82}$ 
& $1437 ^{+56} _{-56}$ & 361 \\
30'x30'&all &0 &$R<22$ & $60131 ^{+86} _{-86}$ 
& $1434 ^{+59} _{-59}$ & 326 \\
30'x30'&no emission &0 &$R<22$ & $60080 ^{+89} _{-89}$ 
& $1409 ^{+63} _{-63}$ & 284\\
30'x30'&only emission &0 &$R<22$ & $60491 ^{+266} _{-266}$ 
& $1564 ^{+152} _{-151}$ & 42\\
30'x30'&all &0 &$R<19$ &$60153  ^{+135}_{-135}$ 
& $1529 ^{+106} _{-106}$ & 137 \\
30'x30'&all &0 &$19<R<21$&$60190  ^{+115}_{-115}$ 
& $1360 ^{+69} _{-69}$ & 161 \\
$\theta<5'$ &all &0 &$R<22$ & $60386 ^{+155} _{-155}$ 
& $1636 ^{+111} _{-112}$ & 123\\
$\theta<5'$ &all &0 &$R<19$ & $60548  ^{+258}_{-258}$
& $1862 ^{+156} _{-156}$ &  61 \\
$\theta<5'$ &all &0 &$19<R<21$ &$60416 ^{+169}_{-169}$ 
& $1240 ^{+141} _{-141}$ &  54 \\
\hline
\hline
Field &Partition  & Flag &R  & $<V>$ & $\sigma$  & $N_{\rm gal}$\\
 & &  &[$km~s^{-1}$] &  [$km~s^{-1}$]\\
\hline
30'x30'&KMM1 &0 &$R<22$ & $ 59186 ^{+317}_{-317} $ & $ 1039 \pm 230 $  & 172 \\
30'x30'&KMM2 &0 &$R<22$ & $ 60545 ^{+484}_{-484} $ & $  951 \pm 283 $  & 102 \\
30'x30'&KMM3 &0 &$R<22$ & $ 62410 ^{+292}_{-292} $ & $ 1016 \pm 212 $  &  52 \\
\hline
\hline
\multicolumn{7}{c}{ }\\
\end{tabular}
\caption{\sl ${C}_{\rm BI}$ and ${S}_{\rm BI}$, the biweight estimators 
for location and scale 
and the associated errors computed with ROSTAT (Beers et al. 1990) for
different subsamples, with the exception of 
the KMM partitions (see section \ref{sec:KMM}), for which
we give the standard mean velocity and velocity dispersion 
and the corresponding bootstrap errors.}
\label{ROST_Tab}
\end{center}
\end{table*}

\subsection{Velocity distribution as a function of 
luminosity}\label{sec:vellum}

We have discussed the variation of the projected density distribution of
galaxies as a function of galaxy luminosity in the central component A2163-A. 
Here we will investigate the dependence of the velocity distribution 
on the galaxy luminosity. 
For this purpose, we have divided our high quality cluster spectroscopic sample
in two luminosity classes, bright ($R \le 19$, 137 galaxies), 
and faint ($19 \le R \le 21$, 161 galaxies), and we have analyzed
the two subsamples with ROSTAT.
Their velocity distributions are shown in
Fig. \ref{histo_lum} and the values of location and scale
are listed in Table \ref{ROST_Tab}. A similar analysis was
performed in the central region of the cluster, corresponding to A2163-A, 
within a radius of 5 arcmin from the X-ray center. 

Concerning the $30' \times 30'$ field,
the estimates of location 
for the two luminosity classes are consistent and also in good agreement 
with the estimate for the whole sample.
The velocity dispersion of the bright sample is $\sim 90$ km/s higher than
that of the total sample, while the velocity dispersion of the faint sample is 
$\sim 80$ km/s lower than that of the total sample, but at the
$1 \sigma$ level. 
The histograms (Fig. \ref{histo_lum}) 
show that the bimodality previously detected 
around 60000 km/s for the whole sample is still present for both the 
bright and faint subsamples.
Faint galaxies appear to be
slightly more numerous in the $\sim 60500$ km/s peak. 

Restricting the analysis to the inner 5
arcmin field (bottom row in Fig.\ref{histo_lum}), 
we find that the estimates of location for the two luminosity
classes are stable and consistent, with a value $\sim 200$ km/s 
higher than the estimate for the whole field but still consistent
taking into account the large error bars.
On the other hand, as in the case of the $30' \times 30'$ field,
the velocity dispersion of the faint sample 
($1239 \pm 141$ km/s) is
smaller than that of the brightest ones ($1862 \pm 156$ km/s),
but now the effect appears statistically significant ($3\sigma$). 
This might indicate that the population of faint objects in the core 
is more relaxed.
The peak at $\sim 60500$ km/s becomes more dominant when analyzing 
the distribution of faint objects in the core. 
The comparison between the velocity
distribution in the core and that in the whole field suggests that the 
velocity peak at $\sim 59000$ km/s 
is mostly composed of faint and bright objects outside the inner
5 arcmin, while the one at $\sim 60500$ km/s is dominated 
by the faint population within the core. 
The third peak at $\sim 62500$ km/s is particularly apparent
in the core (as it was apparent in the projected
distribution of KMM3 objects shown in Fig. \ref{fig:kmmD}) 
and composed both of faint and bright galaxies.
In the following sections we present 
a more detailed analysis of the relative projected 
density distribution of these objects.

\subsection{Spatial variation of the velocity distribution}\label{sec:velspa}

We have analyzed the velocity distribution in different cluster regions,
addressing the variation of the global
quantities, location and scale, as a function of the angular distance 
to the cluster center and of the position angle. 
As for the projected density profiles, we
have chosen the position of the X-ray centroid as the cluster center. 
For our first purpose, we have measured 
the location $C_{BI}$ and scale $S_{BI}$ in increasing circular annuli.
The differential and integrated profiles of these quantities are shown 
in Fig. \ref{velprof}. 
The velocity profile reaches a maximum $V \sim 60800$ km/s at an
angular scale of $\sim 5$ arcmin, then declines to $\sim 59400$ km/s. 
While a determination of the exact profile is difficult due to the
large error bars, the mean velocity
in the inner cluster region is significantly higher than in the outer region. 

The velocity dispersion profile has very high values 
($\sim 1400$ to $1600$ km/s) in the
inner region (up to $\sim 6$ arcmin), then drops to lower values  (1100
km/s) at larger scales. Here again, the error bars are 
large,
but when comparing the value in the 2-4 arcmin bin to that in the 8-10
and 10-12 arcmin bins, the effect is $3 \sigma$ significant.

Given the complexity of the system, we analyzed the variations
of the velocity distribution not only as a function of the distance 
to the center, but also in sectors at different position angles. 
We have estimated $C_{BI}$ and $S_{BI}$ in eight
angular sectors, each one 45 degrees wide, centered on the X-ray centroid and
rotating counterclockwise starting from West on the right ascension axis. 
These measurements are shown in Fig. \ref{velsect}. 
The mean velocity shows a systematic trend, reaching
a maximum value ($C_{BI}\sim 60700$ km/s)
in the 0-45 degrees sector, and a minimum nearly in the opposite direction,
with values lower than the mean in the North-East 
quarter (90-180 degrees). The velocity dispersion is quite high in all 
sectors, with a particularly high value of $S_{BI} \sim 1700$ km/s in
sector 5 (180-225 degrees). Unfortunately our sampling is not sufficient 
to consider this high value as statistically significant.

We conclude that there are variations of the velocity distribution 
between different field regions; 
the velocity dispersion decreases with radius, 
and the value of the mean velocity is
lower in the NE region than in the NW (and also lower
with respect to the cluster mean). 

In what follows we will investigate how the features of
the velocity distribution are connected to the overdensities identified in the 
projected density distribution.

\section{Combined velocity-density analysis of subclustering}

\subsection{Kinematical indicators of subclustering}

We have applied three classical methods that quantify the amount of
substructure in galaxy clusters using a combined analysis of the
velocity and spatial distributions of confirmed cluster members,
i.e. the $\Delta$, $\alpha$ and $\epsilon$ tests by Dressler \&
Shectman (1988), West \& Bothun (1990) and Bird (1994) respectively.
The actual values of the $\Delta$, $\alpha$ and $\epsilon$ parameters
and their significance levels are summarised in
Table~\ref{tab:3D}. Significance levels were obtained using the
bootstrap technique and normalizing with 1000 Monte Carlo
simulations.

\begin{table}
\begin{center}
\caption{\label{tab:3D}{\small {\sl 3-D substructure indicators for
the sample of 326 confirmed cluster members in our dataset}}}
\begin{tabular}{ccc}
\hline
\hline
Indicator & Value & Significance\\
\hline
$\Delta$ & 428.790 & 0.002  \\
$\alpha$ & 0.478~${{\rm h}_{70}}^{-1}$~Mpc & 0.012 \\
$\epsilon$ & ${3.369}{\times}{10}^{27}~{\rm kg}$ & 0.999 \\
\hline
\end{tabular}
\end{center}
\end{table}

Assuming that these tests reject the null hypothesis for significance
levels lower than 10\%, both the $\Delta$ and $\alpha$ tests find
strong evidences of subclustering, with significance levels
$\lesssim$~1\%. Subclustering is not detected by the $\epsilon$
test, which was shown in Pinkney et al. (1996) to be less sensitive than the 
$\Delta$ and $\alpha$ tests. 

\begin{figure}
\resizebox{8cm}{!}{\includegraphics{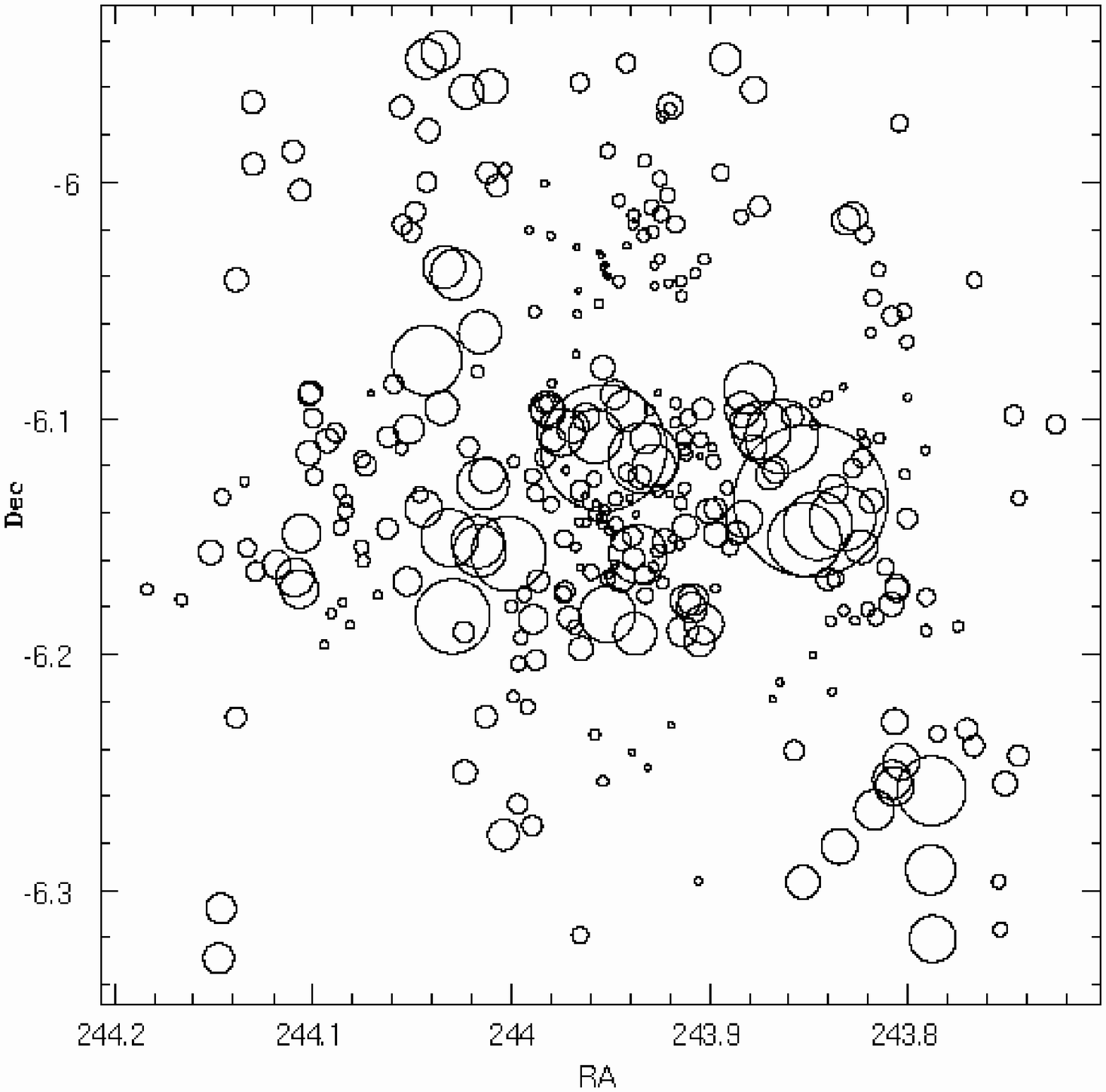}}
\centering
\hfill
\parbox[b]{8cm}{
\caption{\sl Projected positions of the galaxies in our spectroscopic sample,
represented by circles with size weighted by the $\delta$
estimator of Dressler \& Shectman (1988). Concentrations of large
circles indicate a correlated spatial and kinematical variation.}
\label{fig:delta}}
\end{figure}

In Fig.~\ref{fig:delta} we show the results of the Dressler \& Shectman (1988)
test. 
The projected position of each galaxy
is represented by a circle whose radius is weighted by the
corresponding $\delta$ parameter\footnote{The test by Dressler \&
Shectman (1988) associate a $\delta$ value to each galaxy, $\Delta$
being the sum of all $\delta$.}. Large circles indicate local spatial
and/or kinematic variations with respect to the whole cluster properties,
i.e local velocity dispersion and/or mean velocity significantly
different from the global cluster values.  From this plot, we do not
detect any strong concentration of large circles which would 
indicate the presence of a significant substructure.
However, large circles appear to be more frequent
in the northern region of A2163-A, and west of the main cluster. 
Given the highly significant value of the $\Delta$ and $\alpha$ tests 
and the local concentrations of circles, 
we have investigated in more detail 
the correlation of subclustering in velocity and density space.

\subsection{Slicing the density distribution in velocity and 
luminosity space}\label{sec:slice}

As shown in sections \ref{sec:vellum} and \ref{sec:velspa}, 
we have  detected several signatures indicating that the
velocity distribution depends on the luminosity and on the spatial location of
galaxies. Here we investigate the relation between the substructures visible
in the cluster projected density distribution and 
the partitions detected in the velocity distribution, testing 
the dependence of the projected density both 
on the velocity and on the luminosity range. For this purpose, we have built
several sets of density maps for the galaxies with a measured redshift
(spectroscopic sample). 

\begin{figure*}
\begin{center}
\hspace{6mm}
\epsfxsize=17.0cm \epsfverbosetrue \epsfbox{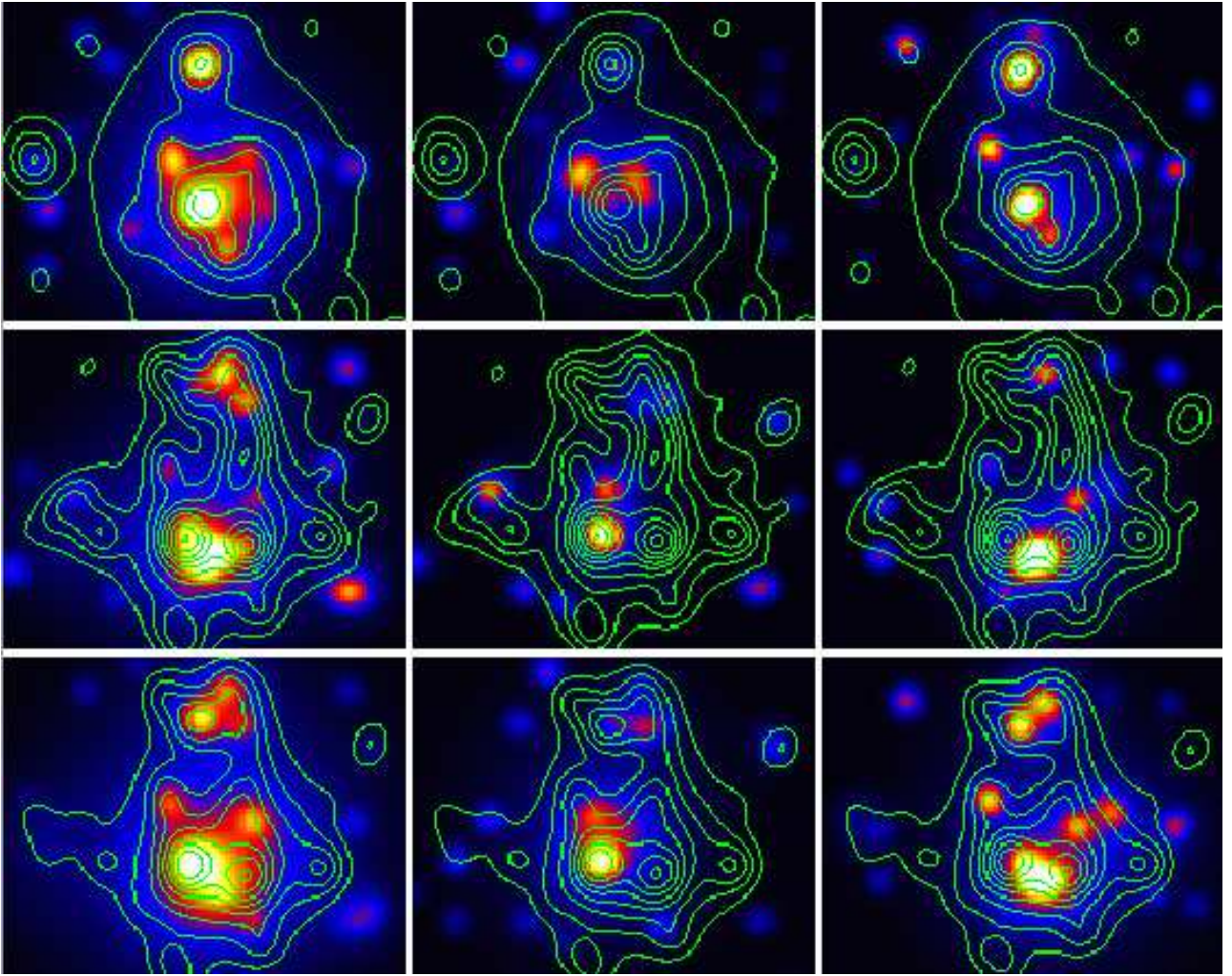}
\end{center}
\caption[] {\sl Density maps of A2163 in velocity and luminosity space. 
Rows correspond to different magnitude limits:
$R<19$ bright subsample (top), 
$19<R<21$ faint subsample (middle), $R<21$ total sample (bottom). 
Columns correspond to different velocity limits:
all spectroscopically confirmed cluster members (left), 
members of the KMM1 partition (middle), members of the KMM2 partition (right). 
Projected density maps previously calculated are superimposed, corresponding
to galaxies with R $< 19$ (top), and to galaxies with $19 < R < 21$ (middle),
and to $R < 21$ (bottom).}
\label{densvelmap}
\end{figure*}

\begin{figure*}
\begin{center}
\hspace{6mm}
\epsfxsize=17.0cm \epsfverbosetrue \epsfbox{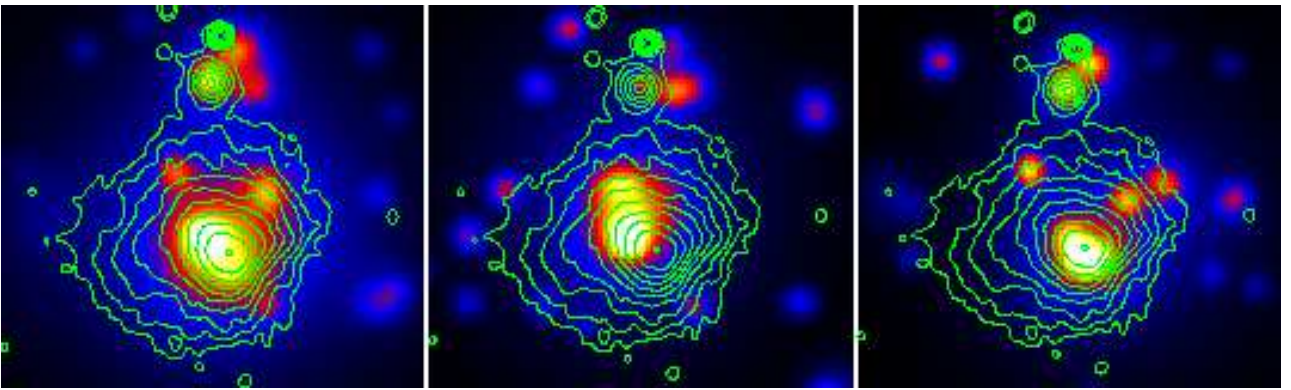}
\end{center}
\caption[] {\sl Density maps of A2163 in velocity space for galaxies with
  $R<21$ (the same as in the bottom row in Fig. \ref{densvelmap}) 
with superimposed X-ray isocontours.
Left panel: all galaxies with measured redshift;
middle panel: galaxies associated to the KMM1 partition;
right panel: galaxies associated to the KMM2 partition. 
X-ray contours have a logarithmic step of 0.2 dex, 
with the lowest contour at 4.65E-3 ct/s/arcmin$^2$ and the
highest at 1.86E-1 ct/s/arcmin$^2$}
\label{Xvelmap}
\end{figure*}

In Fig. \ref{densvelmap} we show the density maps
for all the galaxies which are members of the cluster 
and for the galaxies associated 
to the KMM1 and KMM2 partitions (Sect.~\ref{sec:KMM}). We omit
KMM3 as its number of objects is too small; these objects were shown 
to be located 
preferentially in the central region corresponding to A2163-A. 
We have furthermore selected galaxies in three magnitude ranges,
applying the same criteria as in Sect.~\ref{sec:vellum}: 
the whole magnitude range ($R<21$), and
the bright  ($R \le 19$) and faint
 ($19 < R \le 21$) ranges. In each case, we overplot 
the isocontours of the projected density distribution for
the {\em photometric} sample
corresponding to the same magnitude limits, as it
does not suffer of the sampling inhomogeneities of the
spectroscopic sample. 
We stress that the spectroscopic sampling is homogeneous only in the central
field and for the bright galaxies ($R \le 19$): for this reason the
density maps of fainter galaxies are biased and require a careful 
interpretation.

We have also built density maps for galaxies 
in different velocity ranges. In the case of the bright subsample
(R$<19$, top row), the density map of all the confirmed velocity members of the
cluster (top left) is in very good agreement with the projected 
density map of the photometric catalogue at the same magnitude limit. 
Both subclusters A2163-A and A2163-B are
clearly visible on the map (and to some extent A2163-C, which is however in a
region not very well sampled by spectroscopy). 
In the main cluster A2163-A, the NE/SW orientation is
confirmed, and the deformations of the projected isocontours in its periphery
are shown to be due to the presence of small groups belonging to the cluster.
The density maps of galaxies in the KMM1 and KMM2 partitions 
(top middle and top right) reveal some differences. 
While KMM2 follows the projected high density structure with maxima A1 and A2, 
KMM1 is clearly located in the northern region of A2163-A. 
This is in good agreement with the previous analysis of the velocity 
distribution 
in sectors, 
which showed a lower mean velocity in the NE sectors (Fig. \ref{velsect}).

\begin{figure*}
\begin{center}
\hspace{6mm}
\epsfxsize=8.0cm 
\epsfverbosetrue 
\epsfbox{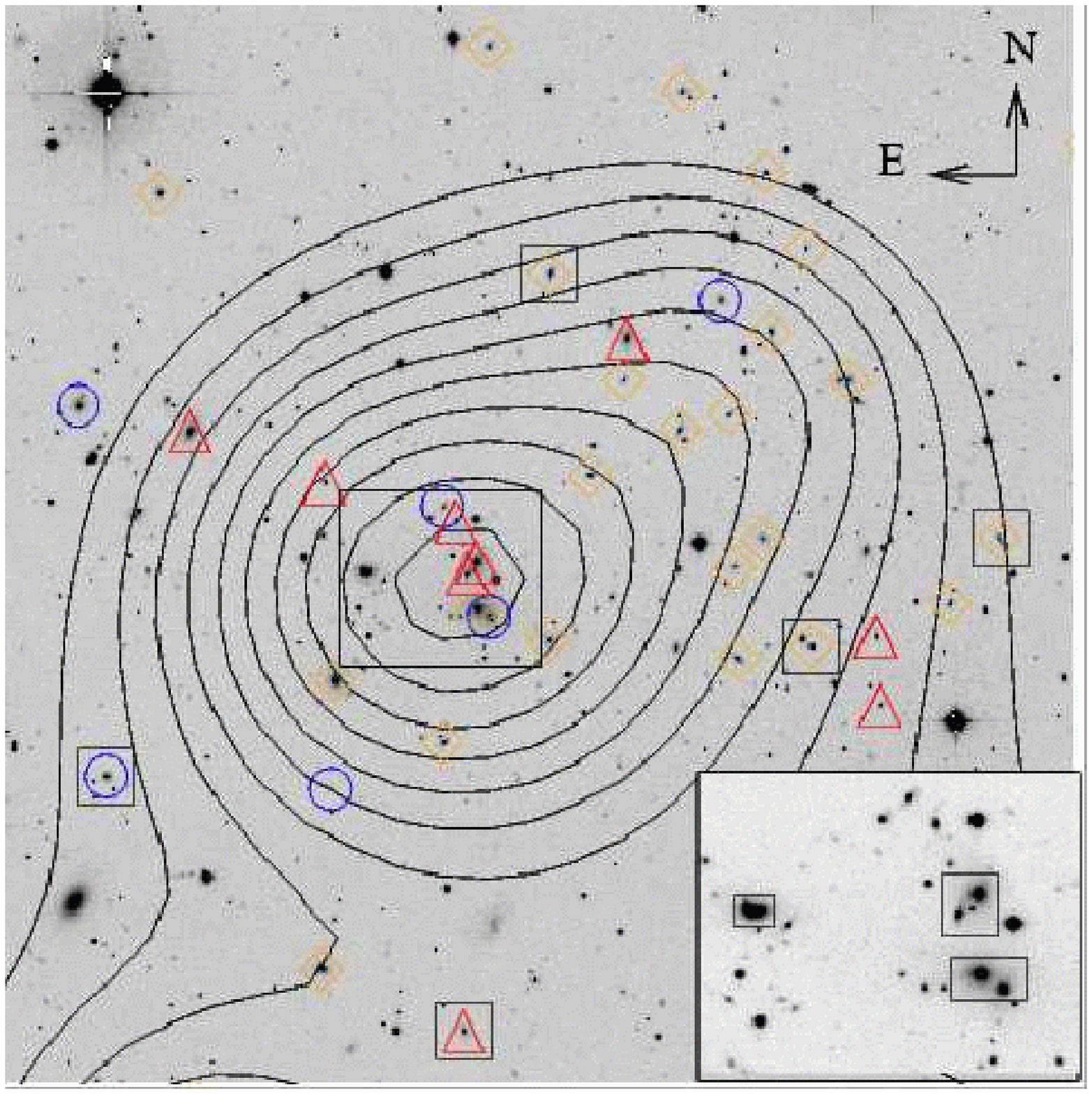}
\epsfxsize=8.0cm 
\epsfverbosetrue 
\epsfbox{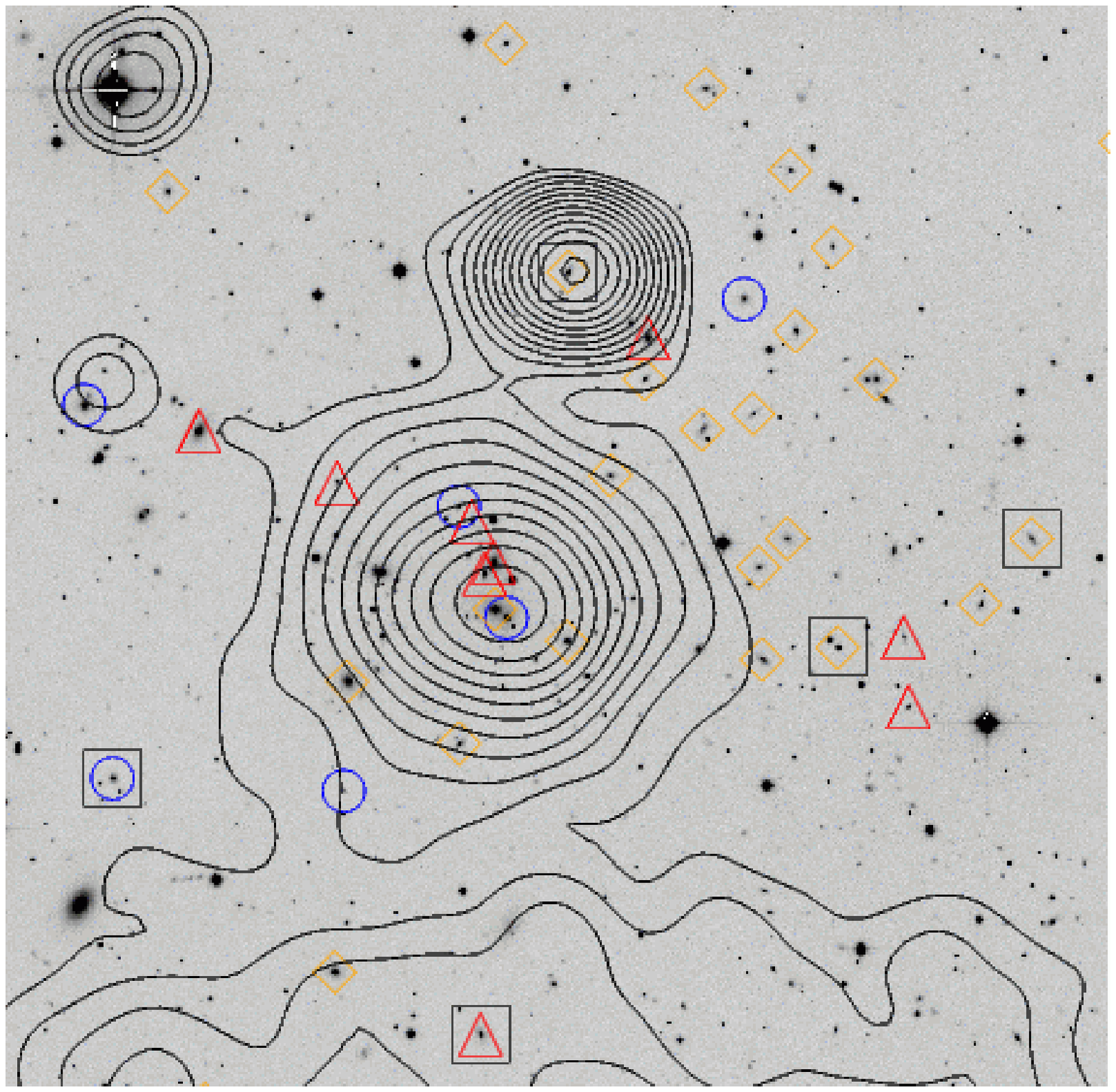}
\end{center}
\caption[]
{\sl A close-up view on a 6'x6' region centered on A2163B (northern group).
Left) The
projected galaxy density contours ($R \leq 19.5$)
are superimposed on the R-band WFI image.  
Galaxies with measured redshift are separated in three velocity intervals, 
corresponding
to the three peaks in the velocity histogram. Blue circles correspond to
the range [57000-59000] km/s, yellow diamonds to [59000-61000] km/s
and red triangles to [61000-63000] km/s. 
Black squares identify galaxies with emission lines. 
A zoom on the central region of the subcluster
is shown in the bottom right corner, where rectangles identify
the northern pair, the southern pair and the dumbbell galaxy.
Right) The same with the projected X-ray contours from XMM.}
\label{isodens_North}
\end{figure*}

\begin{figure}
\begin{center}
\hspace{6mm}
\epsfxsize=8.0cm \epsfverbosetrue \epsfbox{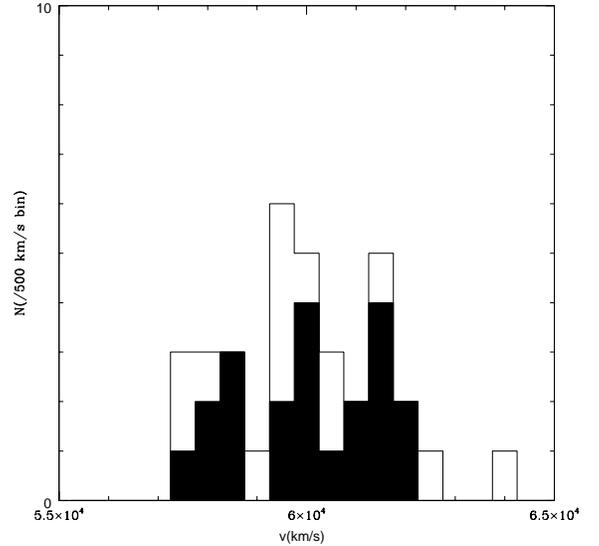}
\end{center}
\caption[]
{\sl Velocity
histogram of A2163B (northern group).
Open and filled histograms include 
galaxies within 3 and 2 arcmin, respectively 
from the density maximum.}
\label{histo_North}
\end{figure}

The density maps
of faint galaxies ($19 < R \le 21$, middle line)
in KMM1 (central panel) and KMM2 (middle row, right column) 
indicate clear differences. The same
is observed when considering all galaxies with $R \le 21$. 
Looking at A2163-A, there is an offset in the core along the NE/SW axis 
between the projected density maxima of KMM1 and KMM2, 
consistently with Fig. \ref{fig:kmmD}. 
In the density maps of the spectroscopic sample there are no clear peaks
in correspondence to the A2, D and E components, while they are apparent
in the density maps of the photometric catalogue: this is
due to the low sampling rate in those regions.

For the whole velocity range within the
cluster (left column), when varying the magnitude range
from bright to faint galaxies (top left to middle left panels), there is
a shift of the main peak within the core of A2163-A 
along the N-NE/S-SW axis. The positions of the density maxima
both for KMM1 and KMM2 (middle and right columns)
show also variations with luminosity. 
We conclude that the galaxies associated to different velocity partitions 
have different projected density distributions, and this is true 
whatever be their luminosity range.
Moreover, the projected density distributions of all the
velocity partitions vary systematically with the luminosity range. 

In Fig. \ref{Xvelmap} we compare the density maps of the spectroscopic
sample including the whole magnitude range ($R<21$) to the X-ray isocontours. 
The X-ray maximum of the main
component A2163-A is located in between the density peaks 
corresponding to KMM1 and KMM2, indeed
very close to the KMM2 peak. KMM1 is clearly offset to the NE 
with respect to the X-ray contours. More strikingly,
the  compressed X-ray contours in the SW region are roughly perpendicular to
the axis joining KMM1 and KMM2.

\subsection{The peripheral subclusters}

 The projected galaxy density distribution (Fig. \ref{isodens_mag}) has shown
the presence of two particularly significant subclusters,
North and East of the main component (A2163-B and A2163-C), and of
smaller ones, West (A2163-D) and South (A2163-E) respectively.
We have used our spectroscopic data to establish
if these components belong to the  A2163 complex. 
Unfortunately the regions of A2163-D and A2163-E are sparsely
sampled. 
In the case of A2163-C, 15 redshifts are available within a radius
of 2 arcmin from its center: the mean location is 
$C_{BI}= 59960$ km/s, thus confirming its association to the A2163 complex.

In Fig.\ref{isodens_North} we show
the projected density distribution of galaxies in A2163-B 
($R \leq 19.5$), which  is centered on 
a dense group of bright objects with two galaxy pairs. 
Contours are roughly elliptical in the center and progressively 
elongated along a tail extending North-West. 
This tail becomes more prominent and extended when
including fainter magnitudes (see Fig.\ref{densvelmap}, left panel).

The right panel of Fig.\ref{isodens_North} 
shows the X-ray contours from XMM data. 
A secondary peak in the X-ray emission, North of A2163B, is due
to a galaxy which is a cluster member: it is
a type 1 AGN, and its spectrum is shown in Fig.\ref{spectra}. 
The velocity histograms of galaxies within projected
separation of 2 arcmin and 
3 arcmin respectively from the position of the 
main density maximum of A2163-B are shown in Fig.\ref{histo_North}. 
Values of location and scale have been calculated with ROSTAT, 
leading to $C_{BI}=60272$ km/s and $S_{BI}= 1223$ km/s (2 arcmin, 22
galaxies) and $C_{BI}= 60190$ km/s and $S_{BI}= 1323$ km/s (3 arcmin, 35
galaxies). The value of the mean velocity is in good agreement with the global
one for the whole cluster (Table \ref{ROST_Tab}). 
However the velocity dispersion
is high, typical of a rich cluster and not of a group.

The mean velocity of A2163-B confirms that 
it is member of the same complex as A2163-A, 
as already indicated by the density map of galaxies within the cluster
redshift range displayed in Fig. \ref{densvelmap} (bottom left). 
However, the velocity distribution
is complex, with three peaks at different velocities.
One difficulty, due to the small angular
separation of the two components (7 arcmin), is to disentangle the 
contribution of A2163-B from that of the
main underlying cluster A2163-A. 
In order to better understand the spatial location of the galaxies belonging
to  the different peaks in velocities, 
in Fig. \ref{isodens_North} (left)
we have marked them 
with different symbols according to the velocity range.
A significant component of galaxies belonging to the
intermediate
velocity peak ($\sim 60000$ km/s)
extends all over A2163-B, including the
central elliptical galaxy on which the X-ray contours are centered
(Fig. \ref{isodens_North}, right) . This
suggests that A2163-A and A2163-B have approximately the same redshift.
However, both low an high velocity components are present, and strongly
contribute to the measured high velocity dispersion.   
There is no BCG in the center but a couple of galaxy  pairs and 
a dumb-bell galaxy.
The two galaxies of the  northern pair lie at 61804 km/s (for
the brightest) and
61920 km/s, revealing a ``physical'' pair at high velocity as compared to the
main cluster. However, the two galaxies of the southern pair (a bright
elliptical on which are centered the X-ray contours and a smaller object) 
show very
different velocities (59914 km/s  and 58192 km/s), implying that their
pair-like aspect probably results from a projection effect.
Unfortunately, the Eastern region of A2163-B is poorly sampled. In particular,
there is a concentration of galaxies around a bright dumb-bell galaxy
whose redshift is unknown. With the present incomplete velocity sampling 
of A2163-B, it is difficult to separate its velocity distribution 
from that of the main cluster. 
Present evidence might suggest a complex dynamical state 
for A2163-B with the presence of several components in velocity space, but  
the low spectroscopic sampling does not allow us to reach a definitive 
conclusion. The relation of A2163-B with the main complex
will be further discussed in Sect.7.

\section{The mass of A2163}

It is well known that estimating the mass of a merging system 
is a difficult problem, as the general
assumption on which mass calculation is based is that the system
is at equilibrium (virial theorem for optical analysis, and hydrostatic
equilibrium for X-ray). However, the simple presence of
substructures in a system does not automatically imply 
that such an estimate is unreliable: the crucial point is how much
the system is far from equilibrium.
 
In the case of A2163, we have shown that clear signs of merging are present;
on the other hand, the general regularity of
its density and velocity profiles suggests 
that the system is not too far for equilibrium. 
In fact, the projected galaxy distribution of the main cluster is 
reasonably regular when smoothed on a sufficiently large scale.
Moreover the velocity distribution, in spite of the existing bimodality, is 
well fitted by a Gaussian distribution. Finally, 
there is an excellent agreement between the X--ray temperature of the cluster
(T=12.4 keV, see Govoni et al. 2004 and Pratt et al. 2001) and its global 
velocity dispersion:
using for example the relation $\sigma (km/s) = 10^{2.52} (kT)^{0.6}$ 
(Lubin \& Bahcall 1993) we obtain $\sigma = 1490$ km/s, 
to be compared to our estimate of $\sigma = \sim 1400$ km/s. 
There are also mass estimates from weak lensing, which 
do not rely on hydrostatic equilibrium, 
virialization or the symmetry of the system.
Unfortunately, the two weak lensing velocity dispersions
presently available for A2163 are quite uncertain: for a singular 
isothermal model Squires et al. find $\sigma_{SIS} = 740$ km/s,
but note that a value of $1000$ km/s would still be consistent
with observations; with the same model, Cypriano et al. (2004)
find $\sigma_{SIS} = 1021 \pm 146$ km/s. 
These values correspond to mass
estimates systematically lower than that calculated from X-ray approaches.
Squires et al. (1997) note that the mass
estimate can be biased downwards if the cluster extends to the control annulus,
and applying a tentative correction based on the X-ray derived mass profile
they obtain consistent X-ray and weak lensing mass estimates. 
However, up to now weak lensing analysis have been restricted to relatively
small and central fields (7'x7'), while the cluster extends at least to angular
separations of 10'. A weak lensing analysis using our multi-band wide-field
imaging of A2163 is under progress (Soucail et al. in prep.). Another
alternative,
when the merging clusters are well defined, is to estimate the mass for each 
clump separately and add the two components. This is not possible
here, as there is too much spatial overlap between the two main velocity
components identified as KMM1 and KMM2. 

We have  applied a standard approach to estimate the virial mass from
spectroscopic and photometric data (see
e.g. Ferrari et al. 2005); under the usual assumptions of
spherical symmetry we have $r_{vir} = (\pi/2) R_{vir}$,
where $r_{vir} $ and $R_{vir}$ are respectively the spatial and 
projected virial radii 
(see Limber \& Mathews 1960), and $\sigma_{vir} = 3 \sigma_r$, 
where
$\sigma_{vir}$ is the spatial virial velocity dispersion and
$\sigma_r$ is the radial velocity dispersion of the system. 
The virial mass $M_{vir}$ is then:

\begin{equation}
 M_{vir} = \frac{3}{2} \frac{\pi}{G} R_{vir} \sigma_{r} ^2
 \label{eq:vir}
\end{equation}

$R_{vir}$ is given by: 

\begin{equation}
\label{eq:rvir}
 R_{vir} = \frac{2 N}{N-1}  R_h
 \label{eq:radii}
\end{equation}

\noindent where $R_h$ is the projected harmonic radius

\begin{equation}
\label{eq:rh}
R_h = \frac{N(N-1)}{2} \left( \sum _{i<j} ^{N-1} \sum _{j=i+1} ^{N}
\frac{1}{R_{ij}} \right) ^{-1}
 \label{eq:rhpw}
\end{equation}

\noindent $R_{ij}$ is the projected separation between 
the $i$th and $j$th galaxies, 
and $N$ is the total number of objects in the system.

For the estimate of the harmonic radius we apply the ring-wise estimator
(Carlberg et al. 1996).
As in Ferrari et al. 2005,
we have estimated the harmonic radius selecting from 
our photometric catalogue the galaxies belonging to
the red sequence. Using the photometric catalog we are able to have 
an homogeneous field coverage, and the red sequence guarantees that we are
selecting early--type cluster members, excluding interlopers and therefore
optimizing the mass estimate (see Biviano et al. 2006).

From our photometric catalogue we have selected all galaxies with 
$16.5 \le R \le 20$ which belong to the red sequence,
and are within 10 arcmin of the cluster center (corresponding
to 2 Mpc --and to $\sim 1$ Abell radius-- in our concordance cosmology).

We find a projected virial radius $R_{vir} = 1.9$ Mpc, and a virial mass
$ M_{vir} = 4.1 \times 10^{15} M_{\odot}$.

However, we have
to take into account that our observations cover a large part 
of the cluster, but not the whole cluster: the value of
$M_{vir}$ represents a good estimate of the mass only within
our observational window, but not of the total mass.
In a given cosmology, the total virialized mass of a cluster is expected
 to be correctly estimated if the ratio 
$\Delta_{vir} \equiv \bar{\rho}(r_{vir})/\rho_c(z)$
of the mean density within $r_{vir}$ to the critical density is lower than 
a given value $\Delta_c$,
which depends on the cosmological model and the cluster redshift. 
The value of $\Delta_c$ can be derived in the spherical collapse model 
assuming that the cluster has just virialized: 
in the case of an Einstein - de Sitter model, 
$\Delta_c = 18\pi^2$, usually approximated with $\Delta_c = 200$, while
in our cosmology $\Delta_c = 97$ at $z=0$ 
(see e.g. Eke et al. 1996) and $\Delta_c=118$ at $z=0.2$ 
(using the fitting formula in Bryan \& Norman 1998); for sake of
comparison we will give values both for $\Delta_c = 100$ and $\Delta_c = 200$.

Given that $\rho_c(z) = 3 H(z)^2 / 8 \pi G$ and
$H(z) = H_0 [\Omega_M (1+z)^3 + \Omega_k (1+z)^2 + \Omega_\Lambda]^{1/2}$,
$\Delta_{vir}$ can be related to the virial radius according to the relation:

\begin{equation}
 \frac{\bar{\rho}(r_{vir})}{\rho_c(z)} = 
 \frac{1}{\rho_c(z)} \frac{3 M_{vir}}{4 \pi r_{vir} ^3} =
 \frac{\sigma_r ^2}{r_{vir} ^2} \frac{6}{H^2 (z)}
\label{eq:ratio}
\end{equation}

Introducing the mean cluster redshift and 
the estimated value of the three-dimensional virial radius into the above
equation, with $r_{vir} = \pi R_{vir} / 2 = 2.8$, we find 
$\Delta_{vir} \sim 250$. 
As expected, this value is higher than $\Delta_c$. 
In order to measure the total mass, we have therefore to 
extrapolate $M_{vir}$ to larger radii.

Assuming that at radii larger than $r_{vir}$ 
the profile goes as $\rho(r) \propto r^{-\alpha}$,  
from equation (\ref{eq:ratio}) we find
the radius $r_{\Delta_c}$ within which the mean density is
$\Delta_c \rho_c(z)$:

\begin{equation}
 r_{\Delta_c} = \frac{\sqrt{6} \sigma_r}{H(z)} 
 \frac{\Delta_{vir}^{\frac{1}{\alpha}-\frac{1}{2}}}
{\Delta_c ^{\frac{1}{\alpha}}}
\end{equation}

The mass goes as 
$M(r) \propto r^{3-\alpha}$ if $\alpha \ne 3$, or $M(r) \propto \ln r$
if $\alpha = 3$. For $\alpha \ne 3$ we obtain

\begin{equation}
 M_{\Delta_c} = M_{vir} \left( \frac{\Delta_{vir}}{\Delta_c} \right)^{1/\alpha}
\end{equation}

Assuming $\alpha = 4$, which corresponds to an Hernquist profile 
at large radii (Hernquist 1990) and gives a convergent mass, we find
$r_{100} = 3.6$ Mpc and $r_{200} = 3.1$ Mpc,
while $M_{100} = 5.0$ and  $M_{200} = 4.2 \times 10^{15} \ M_\odot $.

Until now we have ignored the surface pressure term (The \& White 1986).
This is an additional term to the standard
virial theorem which must be taken into account 
when the volume used to estimate the virial mass does not include 
the whole system. 
Neglecting it, the virial mass is overestimated by a factor depending on the 
observed fraction of the cluster and its profile;
this factor cannot be larger than 50\%, which is
the case of an isothermal sphere
(see Carlberg et al. 1996, Girardi et al. 1998).
In our case, the field is quite large and the expected surface pressure
term should be small, but in order to 
quantitatively assess the effect of the field size
we have used numerical simulations.
We have generated 1000 mock clusters, each one
following an Hernquist distribution, with a density profile truncated
at 20 Mpc and $R_e \sim 1.8153 a = 1$ Mpc.
We have chosen this value as
Lanzoni et al. (2004) have found that for cluster dark haloes
the half--mass projected radius is in the range of
0.2 to 0.4 $r_{100}$: the estimated value of $r_{100} = 3.6$ Mpc for A2163 
implies that $R_e$ should be between $\sim 0.7$ and $ 1.4$ Mpc. 
The total mass has been fixed to
$3.5 \times 10^{15} M_\odot$, a value slightly smaller than the estimated one
and reproducing the observed velocity dispersion; the average number
of simulated galaxies has been chosen to reproduce --on the average--
the observed number of cluster members.
We have estimated the virial radii and masses of the simulated
clusters as a function of the field radius, 
applying the same analysis as for the real data.
The dispersion of the mass values around the mean gives
an estimate of the error due to the poissonian sampling, while the systematic
offset of the measured mean mass with respect to the prediction for a 
theoretical Hernquist profile gives an estimate of the surface pressure term.
The average measured virial mass of the simulations within 10 arcmin
is $2.75 \pm 0.25   \times 10^{15}\ M_\odot$ , i.e. 79\% of the total mass,
with a $1 \sigma$ error of 9\%.
According to the theoretical Hernquist profile with a cutoff at 20
Mpc, we would expect 71\% of the total mass within 10 arcmin, i.e.
on average we overestimate the mass of the simulated cluster by 8\%.

Taking into account the pressure term, 
we finally have
$M_{vir} = 3.8 \pm 0.4 \times 10^{15}\ M_\odot $, 
while the extrapolated total masses are
$M_{200} =  3.9 \pm 0.4 \times 10^{15} \ M_\odot$ 
and $M_{100} = 4.6 \pm 0.5 \times 10^{15}\ M_\odot$.
Our values are in good agreement with the X--ray
estimate of Elbaz et al. which, rescaled to $H_0=70$ km/s/Mpc,
corresponds to $3.3 \times 10^{15} M_\odot$ within $r = 3.3$ Mpc,
and confirm that A2163 is one of the most massive clusters known.

\section{Discussion and conclusions}

Multiple signatures of merging have been detected in A2163. 
Comparing this optical analysis to previous results at other
wavelengths allows  important clues about the merging history 
of this cluster to be derived.

A2163 appears to be composed of a main component (A2163-A), of a
subcluster 7' North of its center (A2163-B), both identified at optical and
X-ray wavelengths, and of various clumps: A2163-C, A2163-D, and A2163-E, East,
West and South of
the main component respectively, detected only in 
the optical. A2163-A, A2163-B and A2163-C are spectroscopically confirmed. 
A large scale elongation along the E--W direction appears at faint magnitudes,
embedding A2163-A, A2163-C and A2163-D and extending over 20 arcmin (4 Mpc). 
A bridge of faint galaxies seems to connect A2163-A to A2163-B
along the North-South direction. A2163-E also lies on this N--S axis.  
The central cluster A2163-A shows a
strong luminosity segregation in its projected density distribution. 
At bright magnitudes, there is a NE/SW structure,   
with a bright maximum in the NE part and a secondary maximum 
in the SW part. When including galaxies at fainter
magnitudes, the orientation of the inner subclustering changes, showing two
maxima aligned in the E-W direction. 
We find that there are relative offsets in the positions of:
a) the BCGs; b) the peaks in the projected galaxy density maps; 
c) the X-ray density peak.
Restricting the analysis to the members of the red-sequence 
shows the same properties, while enhancing the density contrast 
of the bimodal structure in the central part of A2163-A.

Another signature of merging is the presence of multi-modality in the velocity
distribution of the whole cluster. 
With a KMM analysis we identify two partitions very close in velocity, 
centered at $59186$ km/s (KMM1) and $ 60545$ km/s (KMM2), respectively,
corresponding to the two main peaks in the velocity histogram,
and a third partition peaked at $62410$ km/s (KMM3). 
The spatial distribution of the galaxies belonging to KMM1 and KMM2 partitions 
is different; in particular in the central region
KMM1 galaxies are preferentially in the NE part while KMM2 galaxies 
rather populate the central region. Galaxies 
in the KMM3 partition do not show any particular concentration
and are distributed along an elongated structure, 
following the main E--W axis of the cluster.

A comparison of the optical and X-ray density maps 
indicates a strong segregation between gas and galaxies, 
and interesting alignments effects.
In A2163-A, the main NE/SW axis joining the two maxima in the 
isodensity map of bright galaxies is nearly aligned with
the inner direction in the gas density contours, and perpendicular to the
compression of the X-ray isophotes in the SW region. 
This can be expected in the case of
merging between these two clumps along a NE/SW axis.
On the other hand, at larger scales the gas density
distribution is mostly oriented along an East-West axis, which is 
also the case for the large--scale over-density
embedding A2163-A, A2163-C and A2163-D, and for the density distribution of
faint galaxies of A2163-A. 

Several facts argue that we are witnessing a post-merger
event: for instance the
relatively mixed velocity distribution (Schindler and B\"ohringer 1993), and
the location of the density maximum in the gas distribution traced by X-ray
observations, in between the two peaks of the galaxy density distribution.
  The most likely scenario for the main cluster 
is that a collision has already occurred within A2163-A,  
as suggested by Elbaz et al. (1995) and Squires et al. (1997). 
The two density clumps existing in the core of the galaxy distribution 
would be the fossils of the previously colliding subclusters.
Another signature of  a post-merger event is the luminosity
segregation detected in A2163-A. 
A luminosity segregation  in projected density has
been observed in several observed clusters (Biviano et al. 1996, 
Barrena et al. 2007). The most luminous cluster galaxies are then supposed to 
trace the remnants of the pre-merging substructures, while the distribution
of fainter objects, elongated in the same direction as the ICM distribution,
traces the large-scale structure and morphology of the recently formed
cluster. 
Moreover, in A2163-A the velocity distribution of bright
objects ($R<19$) is more dispersed than that of fainter ones ($19<R<21$),
varying from $\sim 1800$ km/s to $\sim 1200$ km/s. The density distribution of
those faint objects with measured redshift is also roughly coinciding with
that of the gas shown by X-ray isocontours, but this result may  be affected by
incomplete spectroscopy sampling at faint magnitudes. 
These facts converge to the conclusion that the distribution of faint objects
is quietly settling within the cluster potential, while that of bright ones is
still substantially disturbed, as expected in the early period after a merger.
 
The comparison with the dark matter and gas density maps derived 
from numerical simulations is essential in order to constrain the merging
scenario, in particular the epoch and the axis of the merger event. 
This is due to the difference in relaxation times between
collisionless and collisional components of the cluster.
Roettiger et al. (1997) have followed the evolution of gas density and dark
matter density distributions during the merging process.  While gas and dark
matter contours are coincident before the merging, they become quite
different after core passage, with the gas ``sloshing'' about within the
gravitational potential. 
The observed relative distribution of gas towards galaxies in A2163-A, with
a single peak in the gas contours located in between the two peaks of the
galaxy distribution (supposed to trace the dark matter one) is quite
similar to that observed in simulations immediately ($\sim 0.5$ Gyr) after
the merger (see Figure 23c of Roettiger et al. 1997). 
Chandra observations (Markevitch et al. 2001, Govoni et al. 2004) at higher
spatial resolution indicate a possible double peak structure in the very
center of A2163. However the separation of these peaks ($\leq 5$ arcsec)
is quite smaller than that shown by the galaxy distribution, and their
positions are in between the galaxy peaks. 

Subclustering tests can also bring information on the merging axis.
From numerical simulations Pinkney et al. (1996) 
conclude that if the
angle between the merger axis and the line of sight is greater than 30
degrees, little substructure will be detected in the 1D velocity
distribution during the 2 Gyr after the merging occurred. 
We detect subclustering in A2163
with 2D or 3D tests, but also from the analysis 
of the radial velocity distribution (strong velocity gradient). 
This implies that the merging axis (whose projection on the plane of the sky 
follows the NE/SW direction) has also a component along the line of sight and 
that the angle of the merger axis with the line of sight should be less than 
30 degrees.  
It is interesting to note that in correspondence to 
the SW clump A2 revealed at bright magnitudes, one can find 
the cold core detected by Govoni et al. (2004) 
in the Chandra temperature map, near the
region of compressed inward X-ray contours. 

Some facts still remain to be explained. {\em A priori}, one
would have expected a correlation between the positions of the BCGs and the
two density peaks A1 and A2 defined by bright galaxies.
The Eastern one, BCG1, is located near the NE bright clump A1.
However, the location of BCG2, completely offset to the West side of
A2163-A, is quite puzzling. However, in case of a post-merger event,
such apparent discrepancies can be expected. 
The velocities of the two BCGs
are comparable to the cluster mean velocity
and are within the range defined by the two central velocity peaks:
they have respective offsets of +230 km/s and -125 km/s with respect to the 
cluster velocity derived from the high precision
sample, which correspond to physical velocity
differences of +192 km/s and -104 km/s 
in the reference system of the cluster.
 
Another issue is the geometry of the merging event. 
While the projected density distribution of bright galaxies ($R < 19$)
suggests a NE/SW axis, the two BCGs are aligned along a E--W axis, 
as well as the projected density distribution of the fainter galaxies. 
The analysis of the velocity distribution of bright galaxies 
seems to confirm the NE/SW orientation for the merger axis, 
and that of fainter galaxies appears to be consistent. 
However, one has to note that the velocity distribution of the fainter
objects is very poorly sampled, especially in the western subcluster
of A2163-A visible in the projected density map at R$<21$, 
and responsible of the twist of the isophotes from NE/SW to E--W. 
Important issues on the velocity distribution of faint
galaxies may therefore be hidden by this inhomogeneous sampling.
A detailed analysis of the density and temperature distribution of the gas
from XMM observations, and of the further constraints 
on the merging scenario which can be derived from a comparison to
numerical simulations will be presented in
a forthcoming paper.
   
The northern component A2163-B is shown from our spectroscopy 
analysis to lie at a  redshift comparable to that of the
main cluster, so it is part of the same complex. 
This confirms the photometric redshift analysis of La Barbera
et al. 2004. However, the velocity distribution is complex, with three
velocity peaks and a high velocity dispersion, more typical of a 
rich cluster than of a group. Part of this high dispersion might
be due to contamination from galaxies of the main cluster. 

A possible scenario is that A2163-B is infalling towards
A2163-A. This pre-merger hypothesis is consistent with 
the coincident and round shape of the gas and bright galaxies 
density distributions. 
However, two features are not easily explained: the fact that at fainter
magnitudes the projected density distribution of A2163-B shows an 
important Western trail (confirmed by the redshift space density maps) 
which seems to extend towards A2163-A through a N--S bridge of galaxies, 
and that the velocity distribution looks highly dispersed. While the high
velocity dispersion may be an artifact due to an insufficient spectroscopic
sampling, the N--S bridge of galaxies
is not explained in this scenario.  

This bridge could in fact have a tidal origin. Another scenario
is that A2163-B has already undergone a high impact parameter merging
event with A2163-A. Crossing the very peripheral region
of A2163-A, the gas core would have been preserved, but
the tidal interaction would have left the
N--S bridge of faint galaxies. Moreover,
the velocity dispersion of A2163-B would have been strongly 
increased by the merging event. This scenario is however difficult to reconcile
with present X-ray observations, since whatever the impact parameter, one
should in principle detect a trailing emitting X-ray gas, which is not 
the case.
Deeper study of the X-ray observations, in combination
with the present data, should help to resolve the issue.

The Eastern and Western clumps, A2163-C, and 
A2163-D less dense than A2163-B and not visible in X-ray, are likely to be
groups infalling on A2163-A along the E--W structure. 
The presence of a significant excess
of emission line galaxies in the Western periphery of A2163-D could also be
related to the merging process, which has been suggested to 
trigger a burst of star formation in the galaxy population (Bekki 1999), or to
accretion on the E--W filament. The southern clump A2163-E is probably
accreted on the N--S direction. 

Although the central region of the cluster exhibits clear signatures
of recent merging, the velocity distribution on the whole sampled region
 is not statistically deviant from a Gaussian
distribution, and the averaged density profiles are reasonably regular. 
Therefore it appears reasonable to estimate the mass
applying the virial theorem: the resulting value, 
$M_{vir} = 3.8 \pm 0.4 \times 10^{15} M_\odot$,
is among the largest measured for a cluster. 
This would remain true even taking into account the uncertainties
which could cause an overestimate of the mass 
(by $\sim 25$\% if for example the true velocity dispersion were as low as 
as found for the subsample of the central, faint galaxies, i.e. 
$\sim 1200$ km/s).

In conclusion, the observed properties of A2163 result
from multiple merging and accretion processes.
We should probably speak of an ``A2163 complex'',  
a very massive structure composed of a main component having recently 
undergone a strong merger phase along the NE/SW or E--W axis with a velocity 
gradient $\sim 1500$ km/s,
and accreting a variety of groups aligned along a large scale filament
spreading over 4Mpc in the E-W direction. The secondary northern 
subcluster A2163-B belongs to the same complex and according to optical
observations alone has probably already crossed the periphery of A2163-A.

\acknowledgements{}
The data published in this paper have been reduced using VIPGI,
designed by the VIRMOS Consortium and developed by INAF Milano.
We thank Bianca Garilli for her generous help, together with Marco Scodeggio,
Paolo Franzetti and Luigi Paioro for numerous interactions in the VIPGI data
reduction. 
We thank the Programme National de Cosmologie of CNRS for his constant support 
on this program, the Observatoire de la C\^ote d'Azur and the
Laboratoire Cassiop\'ee, CNRS, for funding this project. 
We are very grateful to Luigina Feretti for fruitful discussions.
We also thank Eric Slezak and Albert Bijaoui for providing the density
estimator based on the Multiscale Vision Model used in this paper. We 
thank Andrea Biviano, Jos\'e de Freitas Pacheco and Heinz Handernach 
for their reading of the manuscript and for their
comments. We thank the referee for his/her useful 
comments which helped to improve the quality of this paper.
CF acknowledges financial support of Austrian Science Foundation
(FWF) through grant number P18523, and Tiroler Wissenschaftsfonds
(TWF) through grant number UNI-0404/156. GWP acknowledges support from DFG
Transregio Programme TR33.
H.B. acknowledges the financial support from contract ASI--INAF I/023/05/0.


\begin{thebibliography}{}

\bibitem[Adami et al.(2001)]{2001A&A...371...11A} 
Adami C., Mazure A., Ulmer M.P., Savine C., 2001, A\&A 371, 11 

\bibitem[Arnaud et al.(1992)]{1992ApJ...390..345A} 
Arnaud M., Hughes J.P., Forman W., Jones C., Lachieze-Rey M., Yamashita K., 
Hatsukade I., 1992, ApJ 390, 345 

\bibitem[]{} 
Arnaud M., Maurogordato S., Slezak E., Rho J., 2000, A\&A 355, 461

\bibitem[Ashman et al.(1994)]{1994AJ....108.2348A} Ashman K.M., Bird 
C.M.,  Zepf S.E., 1994, AJ 108, 2348 

\bibitem[Barrena et al.(2007)]{2007astro.ph..1833B} 
Barrena R., Boschin W., Girardi M.,  Spolaor M., 2007, 
A\&A 469, 861

\bibitem[Bardelli et al.(2001)]{2001MNRAS.320..387B} 
Bardelli S., Zucca E., Baldi A., 2001, MNRAS 320, 387 

\bibitem[]{} Beers T.C., Flynn K., Gebhardt K., 1990, AJ 100, 32

\bibitem[]{} Beers T.C., Forman W., Huchra J.P., Jones C.,
Gebhardt K., 1991, AJ 102, 1581

\bibitem[]{} Bekki K., 1999, ApJ 510, L15

\bibitem[]{} Bertin E.,  Arnouts S., 1996, A\&AS 117, 393

\bibitem[]{} Bird C.M., Beers T.C., 1993, AJ 105, 1596

\bibitem[Bird(1994)]{1994ApJ...422..480B} 
Bird C., 1994, \apj 422, 480 

\bibitem[]{} Biviano A., Murante G., Borgani S., Diaferio A., Dolag K., 
Girardi M., 2006, A\&A 456, 23

\bibitem[Boschin et al.(2004)]{2004A&A...416..839B} 
Boschin W., Girardi M., Barrena R., Biviano A., Feretti L.,  Ramella M., 
2004, A\&A 416, 839 

\bibitem[Boschin et al.(2006)]{2006A&A...449..461B} 
Boschin W., Girardi M., Spolaor M.,  Barrena R., 2006, A\&A 449, 461 

\bibitem[]{} Bottini D., Garilli B., Maccagni D., Tresse L.,
            Le Brun V. et al., 2005, PASP 117, 996

\bibitem[]{} Bourdin H, Slezak E., Bijaoui A., Arnaud M., 2001, Proceedings
  of the XXXVIth Rencontres de Moriond , XXIst Moriond Astrophysics Meeting,
  March 10-17, 2001 Savoie, France. Edited by D.M. Neumann 
  J. Tran Thanh Van (astro-ph/0106138)

\bibitem[]{} Brunetti G., 2003, ASP Conf. Series 301, p.349 (astro-ph/0208074)

\bibitem[]{} Bruzual G., Charlot S., 2003, MNRAS 344, 1000

\bibitem[]{} Bryan G.L., Norman M.L., 1998, ApJ 495, 80

\bibitem[]{} 
Carlberg R.G., Yee H.K.C., Ellington E., Abraham R., Gravel P., Morris S., 
Pritchet C.J., 1996, ApJ 462, 32

\bibitem[]{} Coleman G., Wu C., Weedman D., 1980, ApJS 43, 393

\bibitem[]{} Czoske O., Moore B., Kneib J.-P., Soucail G., 2002, 
A\&A 386, 31

\bibitem[]{} 
Cypriano E.S., Sodr\'e L.Jr., Kneib J.P., Campusano L.E.,
 2004, ApJ 613, 108

\bibitem[den Hartog  Katgert]{1996MNRAS.279..349D} den Hartog R., 
 Katgert P., 1996, MNRAS 279, 349 

\bibitem[]{} Dressler A.,  Shectman S., 1988, AJ 95, 985

\bibitem[]{} Elbaz D., Arnaud M., B\"ohringer H., 1995, A\&A 293, 337

\bibitem[]{} Eke V.R., Cole S., Frenk C.S., 1996, MNRAS 282, 263

\bibitem[]{} Everitt B.S., Hand D.J., 1981, 
{\em Finite mixture distributions}, Chapman \& Hall

\bibitem[]{} Fadda D., Slezak E.  Bijaoui A. 1998, A\&AS 127, 335

\bibitem[]{} Ferrari C., Maurogordato S., Cappi A., Benoist C., 
2003, A\&A 399, 813

\bibitem[]{} Ferrari C., Benoist C., Maurogordato S., Cappi A., Slezak E.,
   2005, A\&A 430, 19

\bibitem[]{} Ferrari, C., Govoni, F., Schindler, S., Bykov, A.M., Rephaeli, Y., 2008, SSR, in press

\bibitem[]{} Feretti L., Fusco-Femiano R., Giovannini G., Govoni F., 2001,
     A\&A 373, 106

\bibitem[]{} Feretti L., Orru E., Brunetti G., Giovannini G., Kassim N.,
     Setti G., 2004, A\&A 423, 11

\bibitem[]{} 
Feretti L., 2006, proceeding of XLIst Rencontres de Moriond, XXVIth
Astrophysics Moriond Meeting: "From dark halos to light", L.Tresse, S.
Maurogordato and J. Tran Thanh Van, Eds, astro-ph/0612185

\bibitem[]{} Flores R. A., Quintana H., Way M. J., 2000, ApJ 532, 206

\bibitem[]{} Hernquist L., 1990, ApJ 356, 359

\bibitem[]{} 
Girardi M., Giuricin G., Mardirossian F., Mezzetti M., Boschin W., 1998,
ApJ 505, 74

\bibitem[Girardi et al.(2005)]{2005A&A...442...29G} Girardi M., Demarco 
R., Rosati P.,  Borgani S., 2005, A\&A 442, 29 

\bibitem[Girardi et al.(2006)]{2006A&A...455...45G} 
Girardi M., Boschin W.,  Barrena R., 2006, A\&A 455, 45 

\bibitem[]{} Giovannini G.  Feretti L., 2002, 
Merging Processes in Galaxy Clusters. Edited
by L. Feretti, I.M. Gioia, G. Giovannini. Astrophysics and Space Science
Library, Vol. 272. Kluwer Academic Publishers, Dordrecht, 2002, p.197-227

\bibitem[Gonzalez-Casado et al.(1994)]{1994ApJ...433L..61G} 
Gonzalez-Casado G., Mamon G.A.,  Salvador-Sole E., 1994, ApJ 433, 
L61 

\bibitem[]{} Govoni F., Markevitch M., Vikhlinin A., VanSpeybroeck L.,
     Feretti L., Giovannini G., 2004, ApJ 605, 695

\bibitem[Henriksen et al.(2000)]{2000ApJ...529..692H} 
Henriksen M., Donnelly R.H., Davis D.S., 2000, ApJ 529, 692 

\bibitem[Hernquist(1990)]{1990ApJ...356..359H} 
Hernquist L., 1990, ApJ 356, 359 

\bibitem[]{} La Barbera, F., Merluzzi, P., Busarello, G., Massarotti, M.,
     Mercurio, A., 2004, A\&A, 425, 797

\bibitem[]{} Lanzoni B., Ciotti L., Cappi A., Tormen G., Zamorani G., 
2004, ApJ 600, 640

\bibitem[Ledlow et al.(2005)]{2005AJ....130...47L} 
Ledlow M.J., Owen F.N.,  Miller N.A., 2005, AJ 130, 47 

\bibitem[]{} Le F\`evre O. et al., 2000, SPIE Proc. 4008, 546

\bibitem[L{\'o}pez-Cruz et al.(2004)]{2004ApJ...614..679L} 
L{\'o}pez-Cruz O., Barkhouse W.A., Yee H.K.C., 2004, ApJ 614, 679 

\bibitem[]{} Lubin L.M., Bahcall N.A., 1993, ApJ 415, L17

\bibitem[]{} Markevitch M., and Vikhlinin A., 2001, ApJ 563, 95

\bibitem[]{} {Maurogordato S., Proust D., Beers T. C., Arnaud M., Pell\'o R.,
   Cappi A., Slezak E., Kriessler J. R., 2000, A\&A, 355, 848}

\bibitem[]{} {McLachlan G. J., Basford K. E., 1988, Mixture Models (Marcel
     Dekker, New York)}

\bibitem[]{} {Miller N. A., Owen F. N., Hill J. M., Keel W. C.,
   Ledlow M. J., Oegerle W. R., 2004, ApJ 613, 841}

\bibitem[]{} {Miller N. A., Oegerle W., Hill J. M., 2006, ApJ 131, 2426}

\bibitem[]{} {Pinkney J., Roettiger K., Burns J.O., 1996, ApJSS 104, 1}

\bibitem[]{} {Popesso P., Bohringer H., Romaniello M. et al., 2005, A\&A
   433, 415}

\bibitem[]{} {Pratt G., Arnaud M., Aghanim N., 2001, astro-ph/0105431}

\bibitem[]{} {Quintana H., Ramirez A., and Way M.J., 1996, AJ 112, 36}

\bibitem[]{} {Ricker P.M.,  Sarazin C.L., 2001, ApJ 561, 621R}

\bibitem[]{} {Roettiger K., Loken C., Burns J.O., 1997, ApJ 109, 307}

\bibitem[]{} {Schindler S., B\"ohringer H., 1993, A\&A 269, 83}

\bibitem[Schlegel et al.(1998)]{1998ApJ...500..525S} 
Schlegel D.J.,  Finkbeiner D.P., Davis M., 1998, ApJ 500, 525 

\bibitem[]{} Scodeggio M., Franzetti P., Garilli B., Zanichelli A.,
            Paltani S. et al., 2005, PASP 117, 1284

\bibitem[]{} Slezak et al. 2007, in preparation

\bibitem[]{} Squires G., Neumann D., Arnaud M., Babul A., Bohringer H.,
     Fahlman G., Woods D., 1997, ApJ 482, 648

\bibitem[]{} Tonry J., Davis M., 1981, ApJ 246, 666

\bibitem[]{} Vandame B., 2002, SPIE Proc. 4847, 123 (astro-ph/0208230)

\bibitem[]{} Yahil A., Vidal N.V., 1977, ApJ 214, 347

\bibitem[]{} Yasuda N. et al. 2001, AJ, 122, 1104

\bibitem[]{} West M.J., Bothun G.D., 1990, ApJ 350, 36

\end{thebibliography}
\end{document}